\documentclass[12pt,a4paper]{article}
\usepackage{amsmath,amssymb,cite,comment,mathtools}
\usepackage[dvipdfmx]{graphicx}
\usepackage[usenames]{color}
\usepackage[bookmarksopen,colorlinks=true,linkcolor=dark_green,citecolor=dark_red,urlcolor=dark_red,linktocpage=false]{hyperref}
\usepackage{afterpage}
\usepackage[normalem]{ulem}

\definecolor{dark_blue}{rgb}{0,0,0.6}
\definecolor{dark_green}{rgb}{0,0.4,0}
\definecolor{dark_red}{rgb}{0.6,0,0}

\usepackage[height=22.5cm,width=16.5cm,centering]{geometry}

\def\lsim{\mathrel{\rlap{\lower4pt\hbox{\hskip1pt$\sim$}} \raise1pt\hbox{$<$}}}
\def\gsim{\mathrel{\rlap{\lower4pt\hbox{\hskip1pt$\sim$}} \raise1pt\hbox{$>$}}}
\def\thefootnote{\fnsymbol{footnote}}

\renewcommand{\thefootnote}{\fnsymbol{footnote}}
\newcommand{\paren}[1]{\left(#1\right)}
\newcommand{\sqbr}[1]{\left[#1\right]}

\newcommand{\ab}[1]{\left|#1\right|}
\newcommand{\fn}[1]{\!\left(#1\right)}
\newcommand{\ov}{\over}
\newcommand{\tx}{\text}

\newcommand{\J}{\text{J}}
\newcommand{\al}[1]{\begin{align}#1\end{align}}

\DeclareMathOperator{\arcsinh}{arcsinh}
\DeclareMathOperator{\arctanh}{arctanh}

\pdfoutput=1
\begin{document}
%%%%%%%%%%%%%%%%%%%%%%%%%%%%%%%%%%%%%%%%%%%%%%%%%%

%%%%%%%%%%%%%%%%%%%%%%%%%%%%%%%%%%%%%%%%%%%%%%%%%%
\begin{titlepage}

\begin{center}

\hfill DESY 19-058 \\
\hfill CTPU-PTC-19-10 \\
\hfill OCHA-PP-354 \\
\hfill OU-HET-1005 \\
\hfill LDU2019-03 \\

\vskip .5in
\begin{center}
{\fontsize{16pt}{0pt} \bf
Higgs inflation in metric and Palatini formalisms:
}
\\ \vspace{4.5mm} 
{\fontsize{16pt}{0pt} \bf
Required suppression of higher dimensional operators
}
\end{center}
\vskip .5in

{\large
Ryusuke Jinno,$^{a,b}$ 
Mio Kubota,$^{c}$
Kin-ya Oda,$^{d}$
and
Seong Chan Park$^{e}$
}

\vskip .3in

\begin{tabular}{ll}
$^{a}$ &\!\! 
{\em Deutsches Elektronen-Synchrotron DESY, 22607 Hamburg, Germany} \\[.3em]
$^{b}$ &\!\! 
{\em Center for Theoretical Physics of the Universe, Institute for Basic Science (IBS),} \\[.2em]
&{\em Daejeon 34126, Korea} \\[.3em]
$^{c}$ &\!\! 
{\em Department of Physics, Ochanomizu University, Tokyo 112-8610, Japan} \\[.3em]
$^{d}$ &\!\! 
{\em Department of Physics, Osaka University, Osaka 560-0043, Japan} \\[.3em]
$^{e}$ &\!\! 
{\em Department of Physics \& IPAP, Yonsei University, Seoul 03722, Korea}
\end{tabular}

\end{center}
\vskip .5in

\begin{abstract}
We investigate the sensitivity of Higgs(-like) inflation 
to higher dimensional operators in the nonminimal couplings and in the potential,
both in the metric and Palatini formalisms.
We find that, while inflationary predictions are relatively stable against the higher dimensional operators 
around the attractor point in the metric formalism,
they are extremely sensitive in the Palatini one:
for the latter, inflationary predictions are spoiled 
by $\ab{\xi_4} \gtrsim 10^{-6}$ 
in the nonminimal couplings $\paren{\xi_2 \phi^2 + \xi_4 \phi^4 + \cdots}R$,
or by $\ab{\lambda_6} \gtrsim 10^{-16}$ 
in the Jordan-frame potential $\lambda_4 \phi^4 + \lambda_6 \phi^6 + \cdots$
(both in Planck units).
This extreme sensitivity results from the absence of attractor in the Palatini formalism.
Our study underscores the challenge of realizing inflationary models with the nonminimal coupling in the Palatini formalism.
\end{abstract}

\end{titlepage}

\tableofcontents
\thispagestyle{empty}

\renewcommand{\thepage}{\arabic{page}}
\setcounter{page}{1}
\renewcommand{\thefootnote}{$\diamondsuit$\arabic{footnote}}
\setcounter{footnote}{0}
%%%%%%%%%%%%%%%%%%%%%%%%%%%%%%%%%%%%%%%%%%%%%%%%%%

\newpage
\setcounter{page}{1}

%%%%%%%%%%%%%%%%%%%%%%%%%%%%%%%%%%%%%%%%%%%%%%%%%%
\section{Introduction}
\setcounter{equation}{0}
\label{sec:Intro}
%%%%%%%%%%%%%%%%%%%%%%%%%%%%%%%%%%%%%%%%%%%%%%%%%%

The inflationary paradigm~\cite{Starobinsky:1980te,Sato:1980yn,Guth:1980zm} 
is one of the fundamental elements of modern cosmology,
providing an elegant solution to the horizon and flatness problems~\cite{Guth:1980zm} 
and  the dilution of possible unwanted relics~\cite{Sato:1980yn}. Furthermore, the slow-roll inflation~\cite{Linde:1981mu,Albrecht:1982wi,Linde:1983gd} successfully produces the primordial density perturbations 
by quantum fluctuations of the inflaton field~\cite{Mukhanov:1981xt,Kodama:1985bj}.
The inflation has become precision science, 
in which a few parameters of the model beautifully fits hundreds of data points 
in the observational data~\cite{Akrami:2018odb}.
We expect further experimental improvements in near future,
with the tensor-to-scalar ratio $r$ being explored by the LiteBIRD~\cite{Matsumura:2013aja}, POLARBEAR-2~\cite{Inoue:2016jbg}, and CORE~\cite{Delabrouille:2017rct} down to $r\sim 10^{-3}$.\footnote{
The observation of $r$ is important as it will be an indirect evidence of the quantum graviton fluctuation during the inflation, namely the first evidence of quantum gravity. Even direct observation of the cosmic graviton background might be possible in the future space interferometers such as the (Ultimate) DECIGO experiment~\cite{Seto:2001qf,Kawamura:2011zz}.
}

The Higgs field is the only elementary scalar field in the Standard Model (SM) of particle physics, and it is tempting to consider a possibility that it plays the role of the inflaton field~\cite{Salopek:1988qh,vanderBij:1993hx,CervantesCota:1995tz,vanderBij:1994bv}.
In these early studies, the possibility of induced gravity has been mainly pursued where the Einstein-Hilbert action $S_\tx{EH}$ is absent, and the nonminimal coupling $\xi_2$ between the Higgs field and the Ricci scalar is required to be of order $\xi_2\sim 10^{34}$.\footnote{
See also Refs.~\cite{Lucchin:1985ip,Futamase:1987ua} for inflation with the nonminimal coupling.
}
The modern version of the Higgs inflation~\cite{Bezrukov:2007ep} allows $S_\tx{EH}$, 
which is not prohibited by any symmetry, and the nonminimal coupling becomes of order 
$\xi_2\sim 10^{5\tx{--}6}$.\footnote{
This possibility has also been commented in the earlier Ref.~\cite{Salopek:1988qh} 
with essentially the same parameters: $\xi_2\sim 10^4$ and the Higgs quartic coupling (around the Planck scale) 
$\lambda_4\sim \paren{\xi_2/10^5}^2\sim10^{-2}$.
}
This modern version is indeed one of the best fit models of the current observational data~\cite{Akrami:2018odb}.\footnote{
We note that the value of the running Higgs quartic coupling $\lambda_4$ has to be positive 
near the Planck scale for the successful Higgs inflation
(see e.g. Refs.~\cite{Bezrukov:2014ipa,Rubio:2018ogq} 
for an uncertainty due to the lack of our knowledge of the ultraviolet completion),
requiring the pole mass of the top quark to be $m_t^\tx{pole}\lesssim 171.3\,\tx{GeV}$ 
(see e.g.\ Ref.~\cite{Hamada:2014wna}). 
The frequently quoted constraint $m_t^\tx{MC}=173.0\pm0.4$\,GeV~\cite{PDG2018} is on the Monte Carlo mass,
a parameter in the Monte Carlo code, whose relation to the pole mass is unknown. 
Currently the pole mass is best deduced from the cross-section measurements, 
and a simple combination of experiments gives 
$m_t^\tx{pole}=173.1\pm0.9$\,GeV~\cite{PDG2018}, which is 2$\sigma$-consistent with the Higgs inflation. 
Note that this bound may be too stringent 
since it assumes totally uncorrelated systematic errors among experiments.
We also note that the inclusion of the Higgs-portal dark matter greatly relaxes the constraint 
even under the existence of the right-handed neutrinos~\cite{Hamada:2017sga}.
}
There are also variations in addition to this vanilla model: 
the critical~\cite{Hamada:2014iga,Bezrukov:2014bra,Hamada:2014wna,Hamada:2017sga}, 
the hill-climbing~\cite{Jinno:2017jxc,Jinno:2017lun}, and the hill-top~\cite{Enckell:2018kkc} Higgs inflations. 
Sharing the virtue of the vanilla model, these variations predict different values of $r$
along with other observables such as $d n_s/d\ln k$, $d^2 n_s/{d\ln k}^2$ as well as $n_t$, $dn_t/d\ln k$, 
which can possibly be used to distinguish between these models in near-future experiments.

However, such various Higgs inflations have been studied mainly in the so-called metric formalism for past decades.
It is known that, even in the Einstein gravity,
there are two different ways to formulate gravity: metric and Palatini formalisms.\footnote{
It has been noted~\cite{Ferraris1982} that the original Palatini's paper~\cite{Palatini1919} `was rather far from what is usually meant by ``Palatini's method,'' which was instead formulated ...\ by Einstein'~\cite{Einstein1925}. Here we follow the convention and call it Palatini anyway.
}
The former assumes the Levi-Civita connection from the beginning by imposing the metricity and torsion free conditions,
while the latter regards the metric and connection as independent variables,
with respect to which we take the variation of the action.\footnote{
Throughout this paper we assume vanishing torsion. 
In stringy context, this corresponds to assuming the trivial background for any higher-rank gauge fields. 
We note that while a three-form gauge-field background can be treated as a torsion, five- and higher-form 
gauge-field background cannot.
}
Though these two formalisms give the same dynamics within the Einstein gravity,
their predictions differ once we introduce nontrivial couplings with matter and gravity.
This indeed happens in the Higgs inflation, and their difference has been attracting considerable interest 
recently~\cite{Bauer:2008zj,Bauer:2010jg,Rasanen:2017ivk,Tenkanen:2017jih,
Racioppi:2017spw,Markkanen:2017tun,Jarv:2017azx,
Racioppi:2018zoy,Enckell:2018kkc,Carrilho:2018ffi,Enckell:2018hmo,Antoniadis:2018ywb,Rasanen:2018fom,
Kannike:2018zwn,Almeida:2018oid,Takahashi:2018brt,Jinno:2018jei,
Tenkanen:2019jiq}.

One may worry about the theoretical issues arising in the Palatini formalism, namely, strong coupling and uniqueness, which are pointed out in Ref.~\cite{Iglesias:2007nv} for $f(R)$ theories. 
However, we found that such issues do not directly apply to Palatini formalism in the case of Higgs inflation, in contrast to f(R) theories (see Appendix~\ref{app:Referee}).

In this paper, we study the (vanilla) Higgs inflation both in metric and Palatini formalisms.
From the viewpoint of effective theory, the nonminimal coupling in the Higgs inflation
$\xi_2 \phi^2 R$ should be regarded as a truncation in the infinite series $(1 + \xi_2 \phi^2 + \xi_4 \phi^4 + \cdots) R$.
The potential term $\lambda_4 \phi^4$ should also be regarded as a truncation in
$\lambda_4 \phi^4 + \lambda_6 \phi^6 + \cdots$.
While there are extensive discussion on the scale at which these higher-order couplings 
appear~\cite{Burgess:2009ea,Barbon:2009ya,Hertzberg:2010dc,Bezrukov:2010jz,Bezrukov:2011sz},\footnote{
Note that the cutoff-scale issue has turned out to be more severe at the preheating regime
in the metric formalism~\cite{Ema:2016dny} compared with 
earlier estimations~\cite{Bezrukov:2008ut,GarciaBellido:2008ab,Repond:2016sol}.
This is due to explosive production of the longitudinal gauge bosons (see also Ref.~\cite{He:2018mgb}).
On the other hand, such an effect is reported to be absent in the Palatini formalism~\cite{Rubio:2019ypq}.
}
we at least expect that new physics effects appear at the Planck scale $M_\text{P}$.
Therefore, if the inflationary predictions significantly depend on the higher-order terms 
suppressed by the Planck scale, it means that the model construction is challenging in such setups.
This is indeed what we find for the Higgs inflation with the Palatini formalism.

The organization of the paper is as follows.
In Sec.~\ref{sec:Setup} we first introduce basic ingredients 
and review the inflationary predictions in the standard setup.
In Sec.~\ref{sec:Nonminimal}, we investigate the sensitivity of the Higgs-like inflation to the higher-order corrections in the Weyl rescaling factor.
In Sec.~\ref{sec:Potential}, we investigate the sensitivity to the potential.
Finally, we give summary and discussion in Sec.~\ref{sec:DC}.
In Appendix~\ref{app:Equations}, we show the full expressions for the relevant formulae.

%%%%%%%%%%%%%%%%%%%%%%%%%%%%%%%%%%%%%%%%%%%%%%%%%%
\section{Setup}
\setcounter{equation}{0}
\label{sec:Setup}
%%%%%%%%%%%%%%%%%%%%%%%%%%%%%%%%%%%%%%%%%%%%%%%%%%

In this section, we summarize two independent formalisms of Higgs-like inflation, namely the metric and Palatini formalisms, 
and review the standard predictions without higher order corrections for each case. 

%%%%%%%%%%%%%%%%%%%%%%%%%%%%%%%%%%%%%%%%%%%%%%%%%%
\subsection{Basic ingredients: metric and Palatini formalisms}
%%%%%%%%%%%%%%%%%%%%%%%%%%%%%%%%%%%%%%%%%%%%%%%%%%

We start from the action in the Jordan frame:
\begin{align}
S
&= 
\int d^4x \sqrt{-g_\J}
\left[
\frac{1}{2} \Omega^2\fn{\phi} g_\J^{\mu \nu} R_{\J \mu \nu}
- \frac{1}{2} g_\J^{\mu \nu} \partial_\mu \phi\, \partial_\nu \phi - V_\J(\phi)
\right],
\label{eq:S}
\end{align}
where $\phi$ is the (Jordan-frame) inflaton and
\begin{align}
\Omega^2\fn{\phi}
&=
1+\xi_2\phi^2+\xi_4\phi^4+\cdots,
~~~~~~
V_\J\fn{\phi}
=
\lambda_4\phi^4+\lambda_6\phi^6+\cdots,
\label{infinite series}
\end{align}
are the Weyl-rescaling factor and the Jordan-frame potential, respectively;
here we have assumed the $Z_2$ symmetry: $\phi \to  -\phi$.\footnote{
For the Higgs field, this is a natural assumption since the gauge invariance allows only a pair $H^\dagger H$ as a basic building block. If we relax this condition, there are other potentially interesting cases without the $Z_2$ symmetry such as $\Omega^2 =1+\xi_2 \phi^2+\xi_3 \phi^3$ and $V_\J =\lambda_4 \phi^4 + \lambda_6 \phi^6$. In this case, the potential in Einstein frame has a plateau at a large field limit, $\phi\gg\max\left({\xi_2\ov\xi_3}, \sqrt{\lambda_4\ov\lambda_6}\right)$ as
$
 V\sim  \frac{V_\J}{\Omega^4} \sim \frac{\lambda_6}{\xi_3^2}\left(1-\frac{\xi_2}{\xi_3 \phi}+\frac{\lambda_4}{\lambda_6 \phi^2}\right),
$
which is universal as long as $\lim_{\phi \to \infty} {V_\J\ov\Omega^4} \to\tx{const.}$~\cite{Park:2008hz}. In this case  inflationary dynamics would happen at a lower regime $\phi < {\xi_2\ov\xi_3}, \sqrt{\lambda_4\ov\lambda_6}$. 
}
The Ricci tensor is 
$R_{\J \mu \nu} = R_{\J \mu \nu}(g_{\J})$ in the metric formalism
while $R_{\J \mu \nu} = R_{\J \mu \nu}(\Gamma_\J)$ in the Palatini formalism,
with the arguments $g_\J$ and $\Gamma_\J$ symbolically denoting the metric and connection in the Jordan frame, respectively.
The subscript $\J$ refers to the Jordan frame.
We use the Planck unit $M_\text{P} = 1$ throughout the paper,
with $M_\text{P} \equiv 1/\sqrt{8\pi G} = 2.4 \times 10^{18}$ GeV being the reduced Planck mass. 

A metric redefinition by a Weyl transformation $g_{\mu \nu}=\Omega^2 g_{\J \mu \nu}$ gives the transformation  of the Ricci scalar as
\begin{align}
R_\J
=
\left\{
\begin{matrix}
\displaystyle
\Omega^2
\left[
R
+ 3\Box \ln \Omega^2
- \frac{3}{2}(\partial \ln \Omega^2)^2
\right]
~~~~
&{\rm (metric)},
\\[2ex]
\Omega^2
R(\Gamma)
~~~~
&{\rm (Palatini)},
\end{matrix}
\right.
\end{align}
where the result for the Palatini formalism is trivially obtained as the Ricci tensor is independent of the metric transformation. 
After this, we obtain the Einstein-frame action
\begin{align}
S
&= 
\int d^4x \sqrt{-g}
\left[
\frac{1}{2} g^{\mu \nu} R_{\mu \nu}
- \frac{1}{2} (\partial \chi)^2 - V(\phi)
\right],
\end{align}
where $V$ is the Einstein-frame potential $V = V_\J/\Omega^4$ and $\chi$ is the canonical inflaton field  in this frame.
The relation between $\chi$ and the Jordan-frame inflaton $\phi$ depends on the formalism:
\begin{align}
\frac{d\chi}{d\phi}
&=
\left\{
\begin{matrix}
\displaystyle
\sqrt{
\frac{1}{\Omega^2}
+ 
\frac{3}{2}
\left(
\frac{d\ln \Omega^2}{d\phi}
\right)^2
}
~~~~
&{\rm (metric)},
\\[2ex]
\displaystyle
\frac{1}{\Omega}
~~~~
&{\rm (Palatini)}.
\end{matrix}
\right.
\label{eq:dchidphi}
\end{align}
The slow-roll parameters and the $e$-folding are calculated through
\begin{align}
\epsilon
&= 
\frac{1}{2} 
\left(
\frac{dV/d\chi}{V}
\right)^2
= 
\frac{1}{2} 
\left(
\frac{dV/d\phi}{V}
\right)^2
\frac{1}{(d\chi/d\phi)^2},
\label{eq:epsGeneral}
\\
\eta
&= 
\frac{d^2V/d\chi^2}{V}
= 
\left[
\frac{d^2V/d\phi^2}{V}
\frac{1}{(d\chi/d\phi)^2}
-
\frac{dV/d\phi}{V}
\frac{d^2\chi/d\phi^2}{(d\chi/d\phi)^3}
\right],
\label{eq:etaGeneral}
\\
N
&= 
\int 
\frac{d\chi}{\sqrt{2\epsilon}}
=
\int 
\frac{d\phi}{d\ln V/d\phi}
\left(
\frac{d\chi}{d\phi}
\right)^2,
\label{eq:NGeneral}
\end{align}
and the resulting inflationary predictions are expressed by
\begin{align}
A_s
&= 
\frac{1}{24\pi^2}\frac{V}{\epsilon},
~~~~~~
n_s
= 
1 - 6\epsilon + 2\eta,
~~~~~~
r 
= 
16\epsilon.
\label{eq:Pzeta_ns_r}
\end{align}
We use $A_s = 2.1 \times 10^{-9}$
(Planck TT,TE,EE+lowE+lensing+BK14+BAO~\cite{Akrami:2018odb}) 
throughout our analysis.

%%%%%%%%%%%%%%%%%%%%%%%%%%%%%%%%%%%%%%%%%%%%%%%%%%
\subsection{Standard predictions without higher dimensional operators}
%%%%%%%%%%%%%%%%%%%%%%%%%%%%%%%%%%%%%%%%%%%%%%%%%%

Before including higher dimensional operators in Secs.~\ref{sec:Nonminimal} and \ref{sec:Potential},
let us review the inflationary predictions in the simplest setup.
We truncate the Weyl-rescaling factor and the Jordan-frame potential as
\begin{align}
\Omega^2
&= 
1+\xi_2 \phi^2,
~~~~~~
V_\J
=
\lambda_4
\phi^4,
\label{eq:original_truncation}
\end{align}
for both metric and Palatini formalisms.
The Einstein-frame potential becomes
\begin{align}
V
&= 
\frac{\lambda_4 \phi^4}{\paren{1 + \xi_2 \phi^2}^2}.
\end{align}
%%

%%%%%%%%%%%%%%%%%%%%%%%%%%%%%%%%%%%%%%%%%%%%%%%%%%
\subsubsection*{Metric formalism}
%%%%%%%%%%%%%%%%%%%%%%%%%%%%%%%%%%%%%%%%%%%%%%%%%%
In the metric formalism, the relation (\ref{eq:dchidphi}) between $\chi$ and $\phi$ becomes
\begin{align}
\frac{d\chi}{d\phi}
&=
\frac{\sqrt{1 + \xi_2 \paren{1 + 6\xi_2} \phi^2}}{1 + \xi_2 \phi^2},
\label{eq:Metric_dchidphi}
\end{align}
and the integral of the $e$-folding (\ref{eq:NGeneral}) can be exactly performed:
\begin{align}
N
&= 
\frac{1}{8}
\left[
\paren{1 + 6\xi_2} \phi^2
-
6\ln
\left(
1 + \xi_2 \phi^2
\right)
\right],
	\label{e-folding in metric formalism}
\end{align}
where we have defined $N$ as the number of $e$-folding from $\phi = 0$.
Note that generically the difference between this definition and the usual one measured from $\max\fn{\ab{\epsilon},\ab{\eta}} = 1$ 
is suppressed by $1/N$ in the expression of $\phi(N)$.

Eq.~\eqref{eq:Metric_dchidphi} can be solved as\footnote{
The following formulae are useful:
for $-1<x<1$,
\begin{align}
\arcsinh x
&=
\ln \left( x + \sqrt{1+x^2} \right),
~~~~~~
\arctanh x
=
\frac{1}{2} \ln \left( \frac{1+x}{1-x} \right).
\end{align}
}
\begin{align}
\chi(\phi)
&= 
\sqrt{\frac{1 + 6\xi_2}{\xi_2}}
~
\arcsinh
\left[
\sqrt{\xi_2 \paren{1 + 6\xi_2}} \phi
\right]
-
\sqrt{6}
~
\arctanh
\left[
\frac{\sqrt{6}\xi_2 \phi}{\sqrt{1 + \xi_2 \paren{1 + 6\xi_2} \phi^2}}
\right].
\label{eq:Metric_chiphi}
\end{align}
Although this is an exact formula, it is hardly inverted in general.  
However, in the limit $\xi_2 \gg 1$, 
the new kinetic term $\sim \paren{\partial \ln \Omega^2}^2$ originating from the Ricci scalar dominates over 
the original kinetic term $\sim \paren{\partial \phi}^2$ of the
inflaton field, and thus the canonical inflaton field is well approximated as
\begin{align}
\chi
&\simeq
\sqrt{\frac{3}{2}}
\ln
\Omega^2
=
\sqrt{\frac{3}{2}}
\ln
\left(
1 + \xi_2 \phi^2
\right).
\label{eq:approximated_chi}
\end{align}
Substituting this back to the potential, we get
\begin{align}
V
&\simeq
\frac{\lambda_4}{\xi_2^2}
\left(
1 - e^{-\sqrt{\frac{2}{3}}\chi}
\right)^2.
\end{align}
The slow-roll parameters with this potential are calculated as
\begin{align}
\epsilon
&\simeq
\frac{4}{3} e^{-2\sqrt{\frac{2}{3}}\chi}
\simeq
\frac{3}{4 N^2},
~~~~~~
\eta
\simeq
- \frac{4}{3} e^{-\sqrt{\frac{2}{3}}\chi}
\simeq
- \frac{1}{N},
\end{align}
with the $e$-folding given by
\begin{align}
N
&\simeq
\frac{3}{4} e^{\sqrt{\frac{2}{3}}\chi}.
\label{eq:N_chi}
\end{align}
The inflationary predictions become
\begin{align}
A_s
&\simeq
\frac{N^2}{32\pi^2}
\frac{\lambda_4}{\xi_2^2},
~~~~~~
n_s
\simeq
1 - \frac{2}{N},
~~~~~~
r
\simeq
\frac{12}{N^2}.
~~~~~~
\tx{(Metric)}
\end{align}
In the expression of $n_s$, the deviation from unity is dominated by
the contribution from $\eta$ in Eq.~(\ref{eq:Pzeta_ns_r}).

Calculations above are performed in terms of the canonical inflaton $\chi$ in the Einstein frame.
From Eq.~\eqref{e-folding in metric formalism}, we can read off the behavior of $\phi$ for $\xi_2 \gg 1$ as
\begin{align}
\phi
&\simeq
\sqrt{\frac{4N}{3\xi_2}},
\label{eq:Metric_phiN}
\end{align}
which is consistent with Eqs.~\eqref{eq:approximated_chi} and \eqref{eq:N_chi}.
We see that the value of $\phi$ for a fixed $e$-folding scales as $\propto 1/\sqrt{\xi_2}$.\footnote{In general, for monomial functions $\Omega^2=1+\xi_m \phi^{m}$ and $V_\J \propto \phi^{2m}$,  the scaling behavior is $\phi \simeq \left(\frac{4N}{3\xi_m}\right)^{1/m}$ with $m\geq 2$~\cite{Park:2008hz}.}
This scaling behavior turns out to be important in interpreting our results later.

%%%%%%%%%%%%%%%%%%%%%%%%%%%%%%%%%%%%%%%%%%%%%%%%%%
\subsubsection*{Palatini formalism}
%%%%%%%%%%%%%%%%%%%%%%%%%%%%%%%%%%%%%%%%%%%%%%%%%%

In Palatini formalism, the relation (\ref{eq:dchidphi}) becomes
\begin{align}
\frac{d\chi}{d\phi}
&=
\frac{1}{\sqrt{1 + \xi_2 \phi^2}},
\label{eq:Palatini_dchidphi}
\end{align}
and the $e$-folding~\eqref{eq:NGeneral} is obtained as
\begin{align}
N
= 
\frac{\phi^2}{8}
~~~~~~
\leftrightarrow
~~~~~~
\phi
=
\sqrt{8N},
\label{eq:Palatini_phiN}
\end{align}
where we have again defined $N$ as the number of $e$-folding from $\phi = 0$.
We see that the relation is independent of $\xi_2$ in contrast to the metric formalism.
We also see that the value of $\phi_{\rm Palatini}\sim \sqrt{N}$ at a given $e$-folding $N$ is 
significantly larger than that in metric one, $\phi_{\rm metric} \sim \sqrt{N/\xi_2}$ for $\xi_2 \gg 1$ in Eq.~(\ref{eq:Metric_phiN}).

Again one can exactly solve Eq.~\eqref{eq:Palatini_dchidphi}:
\begin{align}
\chi 
= 
\frac{1}{\sqrt{\xi_2}}
\arcsinh
\left[
\sqrt{\xi_2}\phi
\right]
=
\frac{1}{\sqrt{\xi_2}} \ln
\left[
\sqrt{\xi_2}\phi + \sqrt{1 + \xi_2 \phi^2}
\right]
~~~~~~
\leftrightarrow
~~~~~~
\phi 
= 
\frac{1}{\sqrt{\xi_2}}
\sinh 
\left[ 
\sqrt{\xi_2}\chi 
\right],
\label{eq:Palatini_chiphi}
\end{align}
and the Einstein-frame potential becomes
\begin{align}
V
&= 
\frac{\lambda_4}{\xi_2^2} 
\tanh^4 
\left[ 
\sqrt{\xi_2}\chi
\right],
\end{align}
In the limit $\sqrt{\xi_2}\chi\gg1$, we see
\begin{align}
\phi
&\simeq
{1\ov2\sqrt{\xi_2}}e^{\sqrt{\xi_2}\chi},
~~~~~~
N
\simeq
\frac{1}{32\xi_2} e^{2\sqrt{\xi_2}\chi},
\end{align}
and the slow-roll parameters become
\begin{align}
\epsilon
&\simeq 
128 \xi_2 e^{-4\sqrt{\xi_2}\chi}
\simeq \frac{1}{8N^2 \xi_2},
~~~~~~
\eta
\simeq 
-32 \xi_2 e^{-2\sqrt{\xi_2}\chi}
\simeq -\frac{1}{N}.
\end{align}
From the expression of $\eta$ we see that the inflation ends at 
$\chi \simeq {\ln\paren{32\xi_2}\ov2\sqrt{\xi_2}}\ll 1$ for $\xi_2 \gg 1$,
which corresponds to $\phi \simeq 2\sqrt{2}$.
The curvature perturbation and other inflationary observables become
\begin{align}
A_s
&\simeq
\frac{N^2}{3\pi^2}\frac{\lambda_4}{\xi_2},
~~~~~~
n_s
\simeq
1 - \frac{2}{N},
~~~~~~
r
\simeq 
\frac{2}{N^2 \xi_2}.
~~~~~~
\tx{(Palatini)}
\end{align}
In the expression of $n_s$, the deviation from unity is dominated by
the contribution from $\eta$ in Eq.~(\ref{eq:Pzeta_ns_r}), similarly to the metric formalism.
Note that the scaling $A_s \propto 1/\xi_2$ in Palatini formalism is different from $A_s \propto 1/\xi_2^2$ in metric one. 
Also, the tensor-to-scalar ratio $r$ is multiplicatively suppressed by a factor of $1/\xi_2 \ll 1$ in contrast to the metric formalism.

%%%%%%%%%%%%%%%%%%%%%%%%%%%%%%%%%%%%%%%%%%%%%%%%%%
\section{Sensitivity to corrections in the Weyl-rescaling factor}
\label{sec:Nonminimal}
%%%%%%%%%%%%%%%%%%%%%%%%%%%%%%%%%%%%%%%%%%%%%%%%%%

The original truncation~\eqref{eq:original_truncation} takes into account up to the next-to-leading and leading terms in $V_\J$ and $\Omega^2$, respectively.
We naturally expect further terms arising from radiative corrections and from new physics effects. 
Without knowing all sources of corrections, we phenomenologically parametrize the corrections 
by introducing higher dimensional operators as in Eq.~\eqref{infinite series}.

To be specific, we first consider the correction to the Weyl rescaling factor~$\Omega^2$ 
in both metric and Palatini formalisms in detail 
and see their effects to the inflationary dynamics and observables.
Correction to the Jordan-frame potential will be discussed in Sec.~\ref{sec:Potential}.
Throughout both the analysis, we assume that $|\xi_{p>2}| \ll \xi_2$ and $|\lambda_{q>4}| \ll \lambda_4$ so that 
the success of the original Higgs-like inflation with nonminimal coupling $\xi_2\gg1$ is maintained. 

The first case we consider is a correction to the Weyl rescaling factor:
\begin{align}
\Omega^2
&= 
1+\xi_2 \phi^2+\xi_4 \phi^4,
~~~~~~
V_\J
=
\lambda_4
\phi^4.
\end{align}
Here we take into account the correction to $\Omega^2$ only, and leave the correction to $V_\J$ for the later section.
We fix their signs of $\xi_2$ and $\lambda_4$ to be positive,
while we allow both positive and negative values for $\xi_4$.
Now the Einstein-frame potential becomes
\begin{align}
V
&= 
\frac{\lambda_4 \phi^4}{\paren{1 + \xi_2 \phi^2 + \xi_4 \phi^4}^2}.
\end{align}
In this setup we have three parameters
\begin{itemize}
\item[]
\begin{center}
$(\lambda_4,\xi_2,\xi_4)$,
\end{center}
\end{itemize}
 but we can eliminate one of them by fixing the observed value of $A_s$,
which we take to be 
$A_s = 2.1\times 10^{-9}$ (Planck TT,TE,EE+lowE+lensing+BK14+BAO~\cite{Akrami:2018odb}) 
as mentioned above.

%%%%%%%%%%%%%%%%%%%%%%%%%%%%%%%%%%%%%%%%%%%%%%%%%%
\subsection{Predictions}
%%%%%%%%%%%%%%%%%%%%%%%%%%%%%%%%%%%%%%%%%%%%%%%%%%

%%%%%%%%%%%%%%%%%%%%%%%%%%%%%%%%%%%%%%%%%%%%%%%%%%
\subsubsection*{Metric formalism}
%%%%%%%%%%%%%%%%%%%%%%%%%%%%%%%%%%%%%%%%%%%%%%%%%%

In metric formalism, the relation (\ref{eq:dchidphi}) between $\chi$ and $\phi$ becomes
\begin{align}
\frac{d\chi}{d\phi}
&=
\frac{\sqrt{1 + \xi_2 \phi^2 + \xi_4 \phi^4 + 6\paren{\xi_2 \phi + 2\xi_4 \phi^3}^2}}{1 + \xi_2 \phi^2 + \xi_4 \phi^4}.
\end{align}
The slow-roll parameters and $e$-folding are calculated through Eqs.~(\ref{eq:epsGeneral})--(\ref{eq:NGeneral}):
\begin{align}
\epsilon
&=
\frac{8(1 - \xi_4 \phi^4)^2}
{\phi^2 \left[ 1 + \paren{\xi_2 + 6\xi_2^2} \phi^2 + \paren{\xi_4 + 24\xi_2 \xi_4} \phi^4 + 24\xi_4^2 \phi^6 \right]},
\label{eq:Nonminimal_Metric_eps}
\\[0.6ex]
\eta
&= 
\frac{
4
\left[
3 + \paren{\xi_2 + 12 \xi_2^2} \phi^2
+ (- 2 \xi_2^2 - 12 \xi_2^3 + 24 \xi_2 \xi_4 - 11 \xi_4) \phi^4 
+ \cdots
+ 96 \xi_4^4 \phi^{14}
\right]
}
{\phi^2 \left[1 + \paren{\xi_2 + 6 \xi_2^2} \phi^2 + \paren{\xi_4 + 24 \xi_2 \xi_4} \phi^4 + 24 \xi_4^2 \phi^6\right]^2},
\label{eq:Nonminimal_Metric_eta}
\\[0.6ex]
N
&= 
\frac{1 + 6\xi_2}{16\sqrt{\xi_4}}
\ln
\left[
\frac{1 + \sqrt{\xi_4}\phi^2}{1 - \sqrt{\xi_4}\phi^2}
\right]
-
\frac{3}{4}
\ln
\left[
\left(
1 - \xi_4 \phi^4
\right)
\left(
1 + \xi_2 \phi^2 + \xi_4 \phi^4
\right)
\right].
\label{eq:Nonminimal_Metric_N}
\end{align}
Here we do not show a complete expression for $\eta$ just to avoid complications; see Appendix~\ref{app:Equations} for it.
Note that the $e$-folding diverges for $\xi_4 > 0$ at $\phi = \xi_4^{-1/4}$, 
which corresponds to the point where the potential derivative vanishes.
The scalar power spectrum amplitude reads
\begin{align}
A_s
&=
{\lambda_4\phi^6\ov192\pi^2}
{
{1+\xi_2\paren{1+6\xi_2}\phi^2+\paren{1+24\xi_2}\xi_4\phi^4+24\xi_4^2\phi^6}
\ov
\paren{1-\xi_4\phi^4}^2
\paren{1+\xi_2\phi^2+\xi_4\phi^4}^2
}.
\end{align}
%%

%%%%%%%%%%%%%%%%%%%%%%%%%%%%%%%%%%%%%%%%%%%%%%%%%%
\subsubsection*{Palatini formalism}
%%%%%%%%%%%%%%%%%%%%%%%%%%%%%%%%%%%%%%%%%%%%%%%%%%

In Palatini formalism, the relation (\ref{eq:dchidphi}) becomes
\begin{align}
\frac{d\chi}{d\phi}
&=
\frac{1}{\sqrt{1 + \xi_2 \phi^2 + \xi_4 \phi^4}},
\end{align}
which can be explicitly solved as shown in Appendix~\ref{app:Equations}.
Also, from Eqs.~(\ref{eq:epsGeneral})--(\ref{eq:NGeneral}), the slow-roll parameters and $e$-folding become
\begin{align}
\epsilon
&= 
\frac{8(1 - \xi_4 \phi^4)^2}{\phi^2 \left[ 1 + \xi_2 \phi^2 + \xi_4 \phi^4 \right]},
\label{eq:Nonminimal_Palatini_eps}
\\
\eta
&= 
\frac{4(3 - 2\xi_2 \phi^2 - 14 \xi_4 \phi^4 - 2\xi_2 \xi_4 \phi^6 + 3 \xi_4^2 \phi^8)}
{\phi^2 \left[ 1 + \xi_2 \phi^2 + \xi_4 \phi^4 \right]},
\label{eq:Nonminimal_Palatini_eta}
\\
N
&= 
\frac{1}{16\sqrt{\xi_4}} \ln \left[
\frac{1 + \sqrt{\xi_4}\phi^2}{1 - \sqrt{\xi_4}\phi^2}
\right]
~~~~~~
\leftrightarrow
~~~~~~
\phi
= 
\xi_4^{-1/4}
\sqrt{\tanh \left[ 8\sqrt{\xi_4}N \right]},
\label{eq:Nonminimal_Palatini_N}
\end{align}
where we inverted the relation in the last equality.
Note that the $e$-folding diverges at the same point $\phi = \xi_4^{-1/4}$ as the metric formalism
for $\xi_4 > 0$ because the potential derivative vanishes there.
The scalar power spectrum amplitude reads
\begin{align}
A_s
&=
{\lambda_4\phi^6\ov192\pi^2}
{1
\ov
\paren{1-\xi_4\phi^4}^2
\paren{1+\xi_2\phi^2+\xi_4\phi^4}
}.
\end{align}
%%

%%%%%%%%%%
\begin{figure}
\begin{center}
\includegraphics[width=0.42\columnwidth]{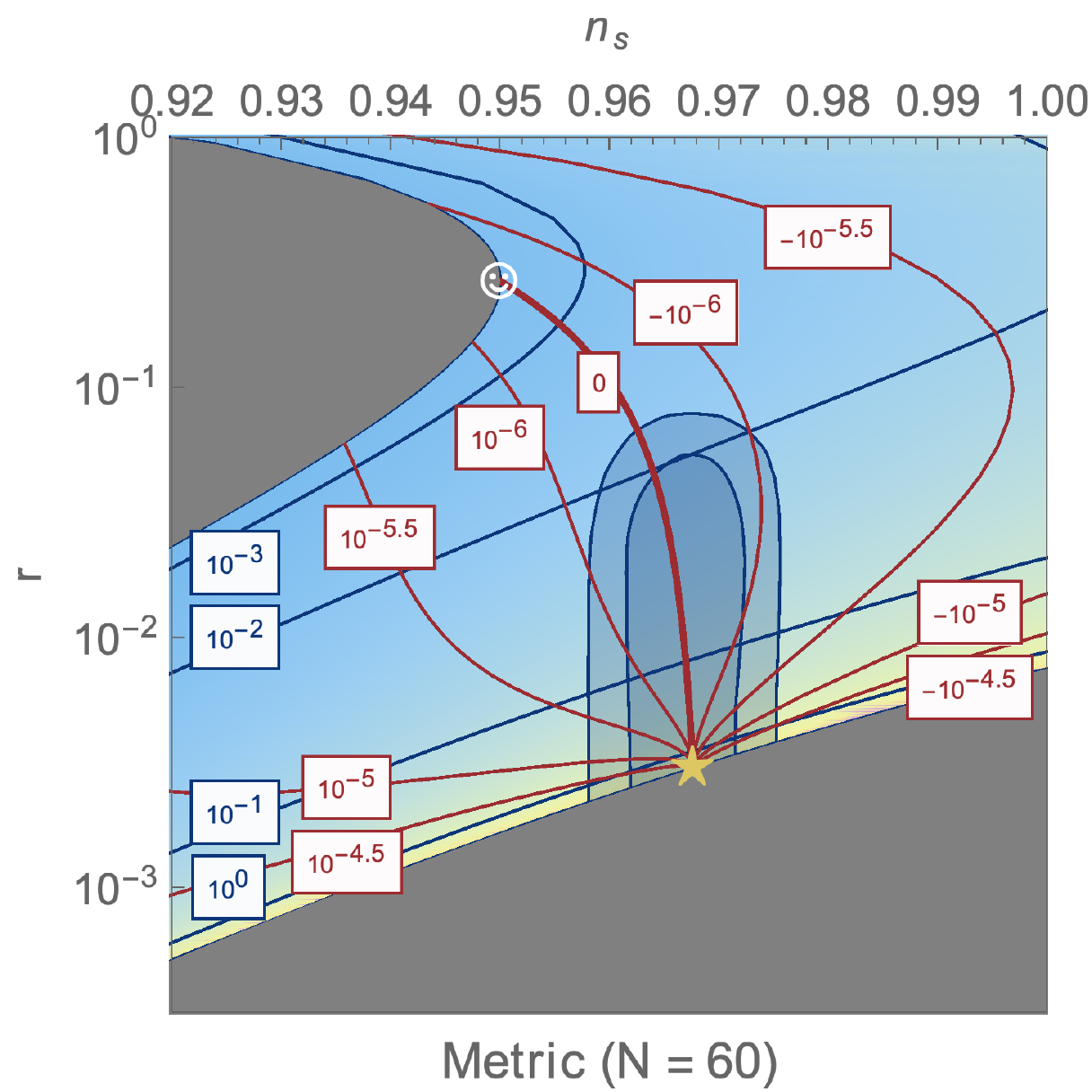}
\hskip 3mm
\includegraphics[width=0.42\columnwidth]{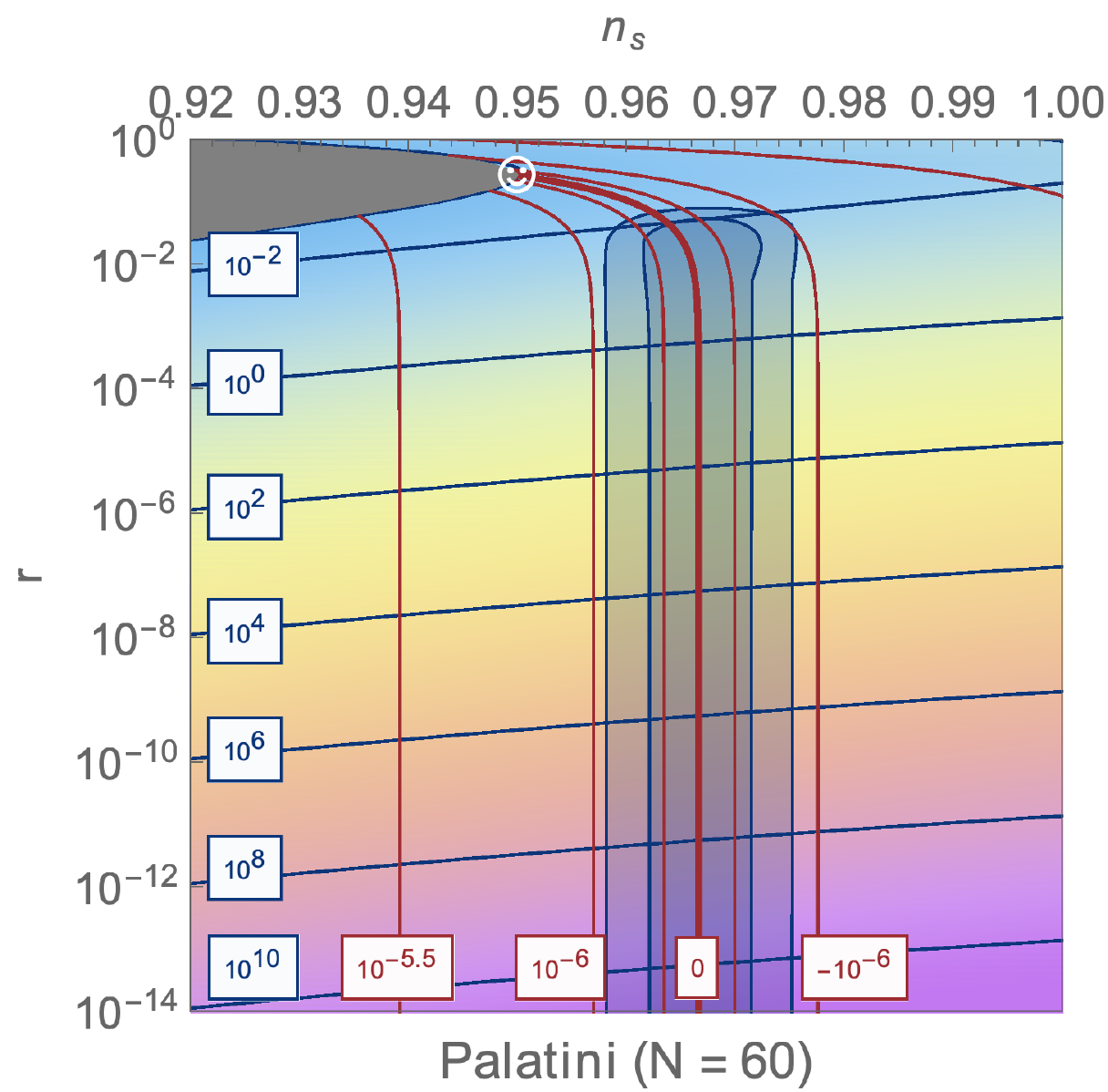}
\hskip 3mm
\includegraphics[width=0.085\columnwidth]{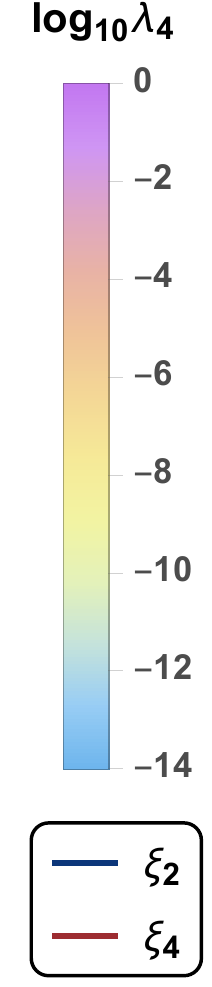}
\\
\vskip 4mm
\includegraphics[width=0.42\columnwidth]{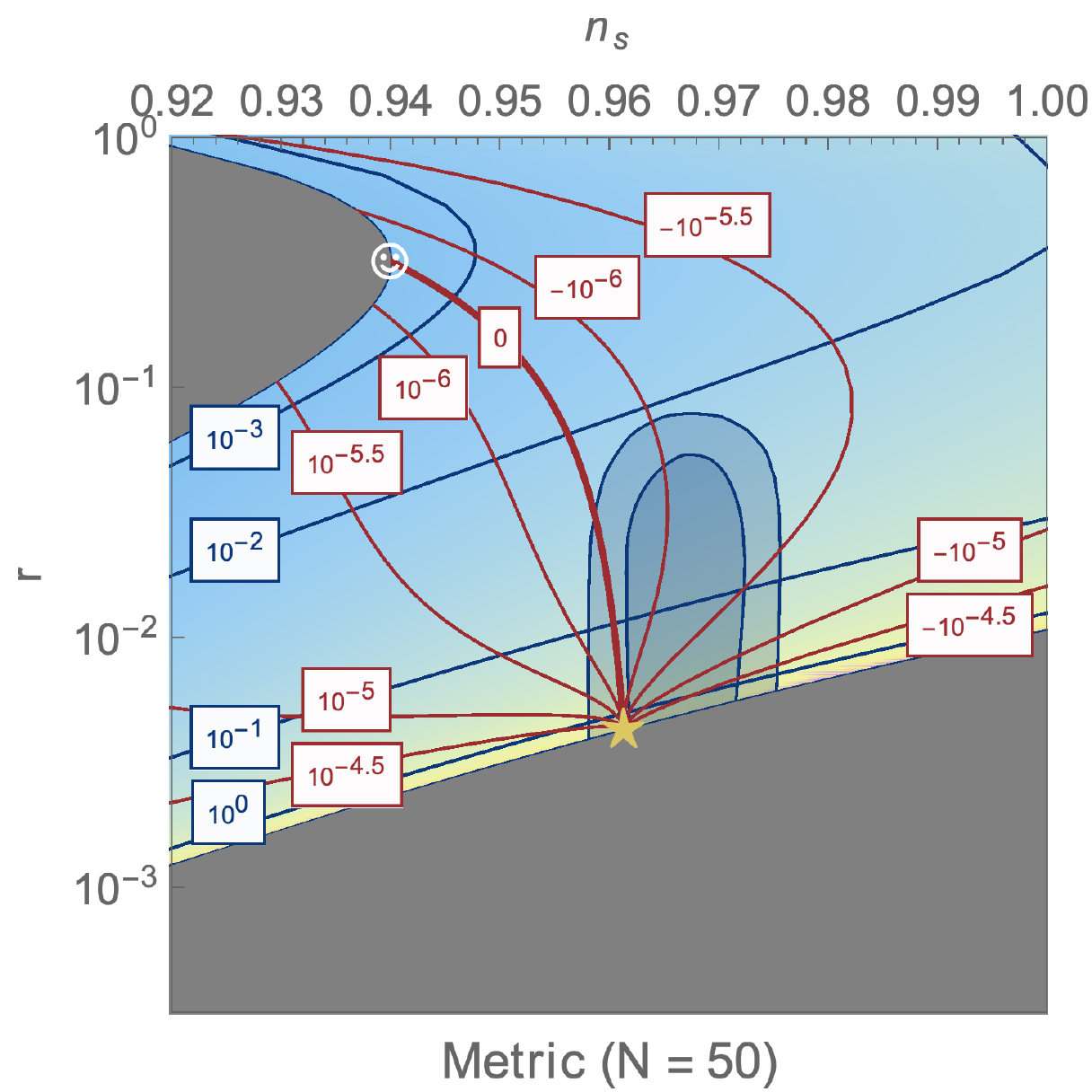}
\hskip 3mm
\includegraphics[width=0.42\columnwidth]{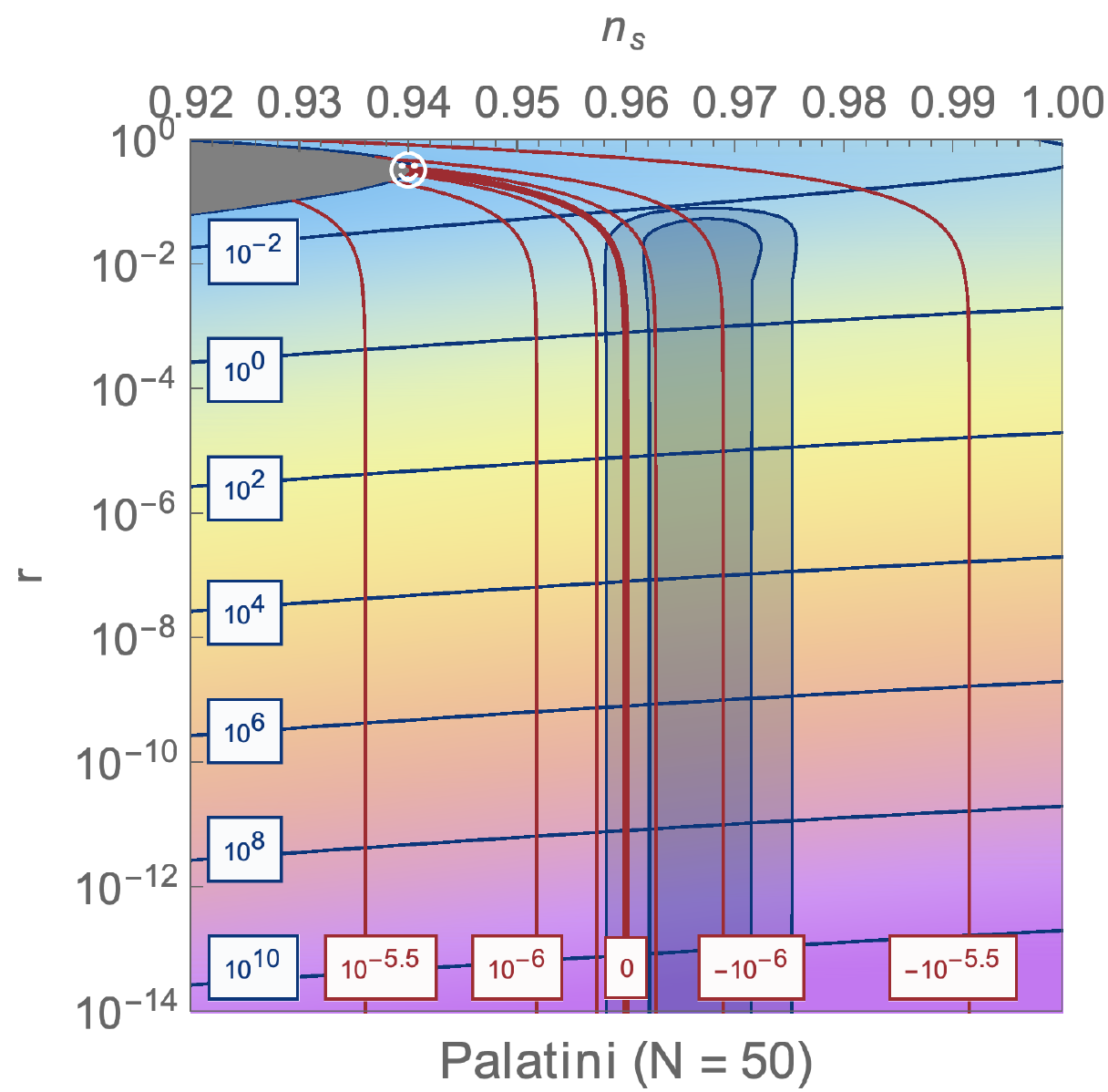}
\hskip 3mm
\includegraphics[width=0.085\columnwidth]{./figs/legNonminimal.pdf}
\caption{\small
Effects of the higher dimensional operator $\xi_4\phi^4R_\J$ in the metric (left) and Palatini (right) formalisms with $N=60$ (top) and 50 (bottom).
We plot contours of fixed $\xi_2$ (blue, horizontal) and of fixed $\xi_4$ (red, vertical) in the $n_s$-$r$ plane.
The value of $\lambda_4$ is also shown as a density plot.
Allowed regions of $1\sigma$ and $2\sigma$ from the Planck experiment~\cite{Akrami:2018odb}
(TT,TE,EE+lowE+lensing+BK14+BAO) are also shown in the center.
}
\label{fig:Nonminimal_nsr}
\end{center}
\end{figure}
%%%%%%%%%%

%%%%%%%%%%
\begin{figure}
\begin{center}
\small Metric ($N=60$)\smallskip\\
\fbox{
\begin{minipage}{0.32\textwidth}
\includegraphics[width=\columnwidth]{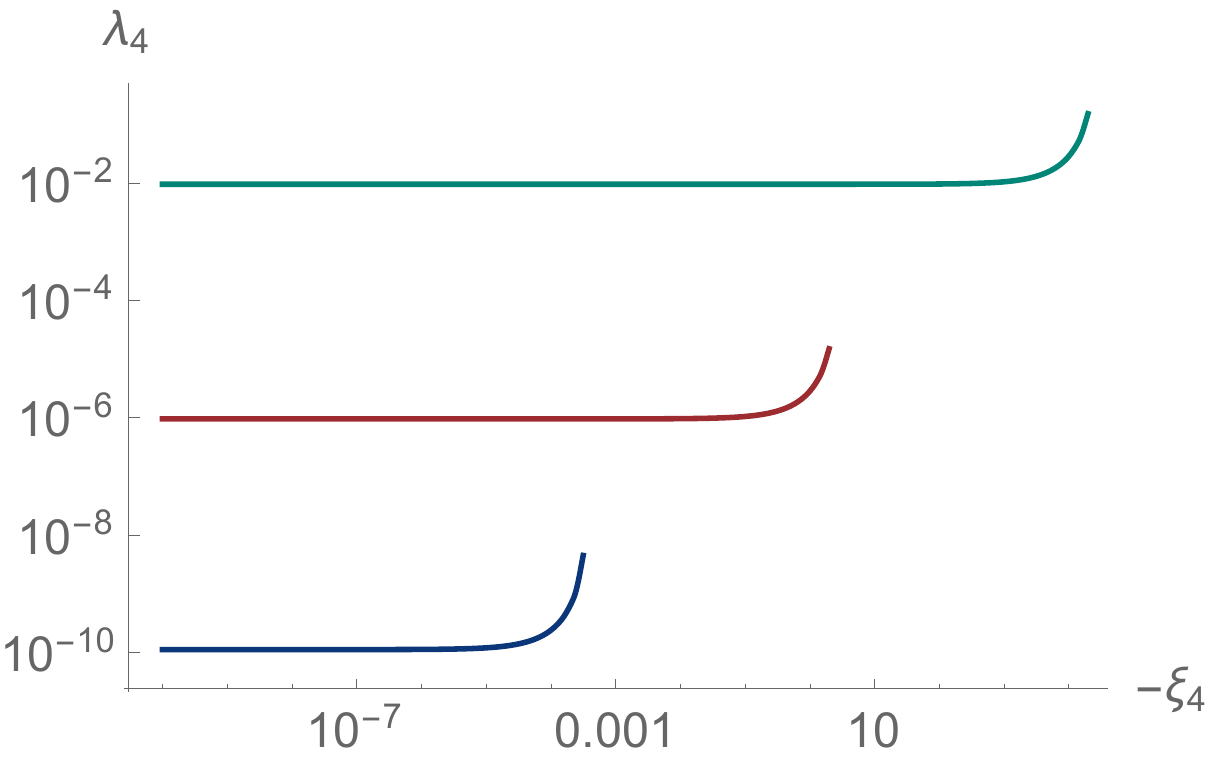}
\end{minipage}
\hskip 3mm
\begin{minipage}{0.32\textwidth}
\includegraphics[width=\columnwidth]{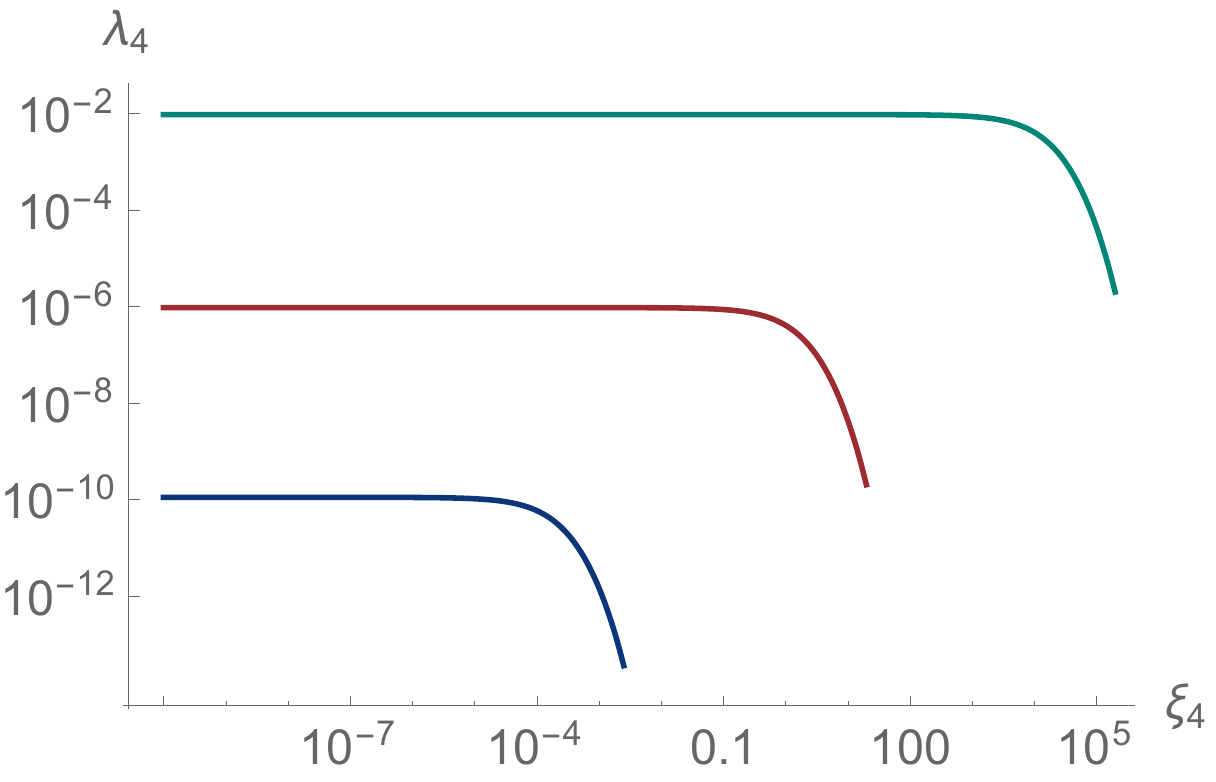}
\end{minipage}
\hskip 3mm
\begin{minipage}{0.1\textwidth}
\includegraphics[width=\columnwidth]{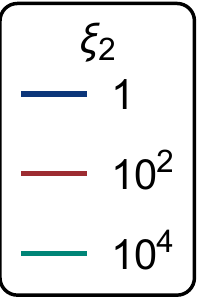}
\end{minipage}
}
\end{center}
\begin{center}
\small Palatini ($N=60$)\smallskip\\
\fbox{
\begin{minipage}{0.32\textwidth}
\includegraphics[width=\columnwidth]{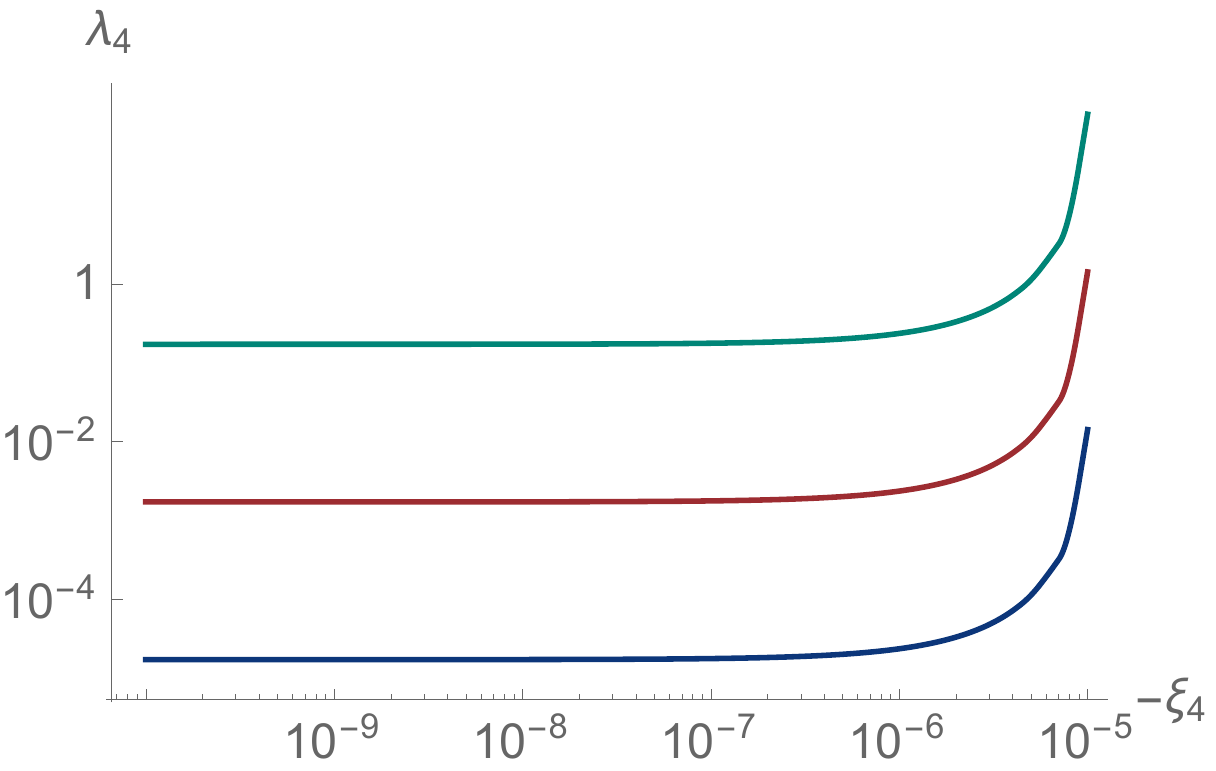}
\end{minipage}
\hskip 3mm
\begin{minipage}{0.32\textwidth}
\includegraphics[width=\columnwidth]{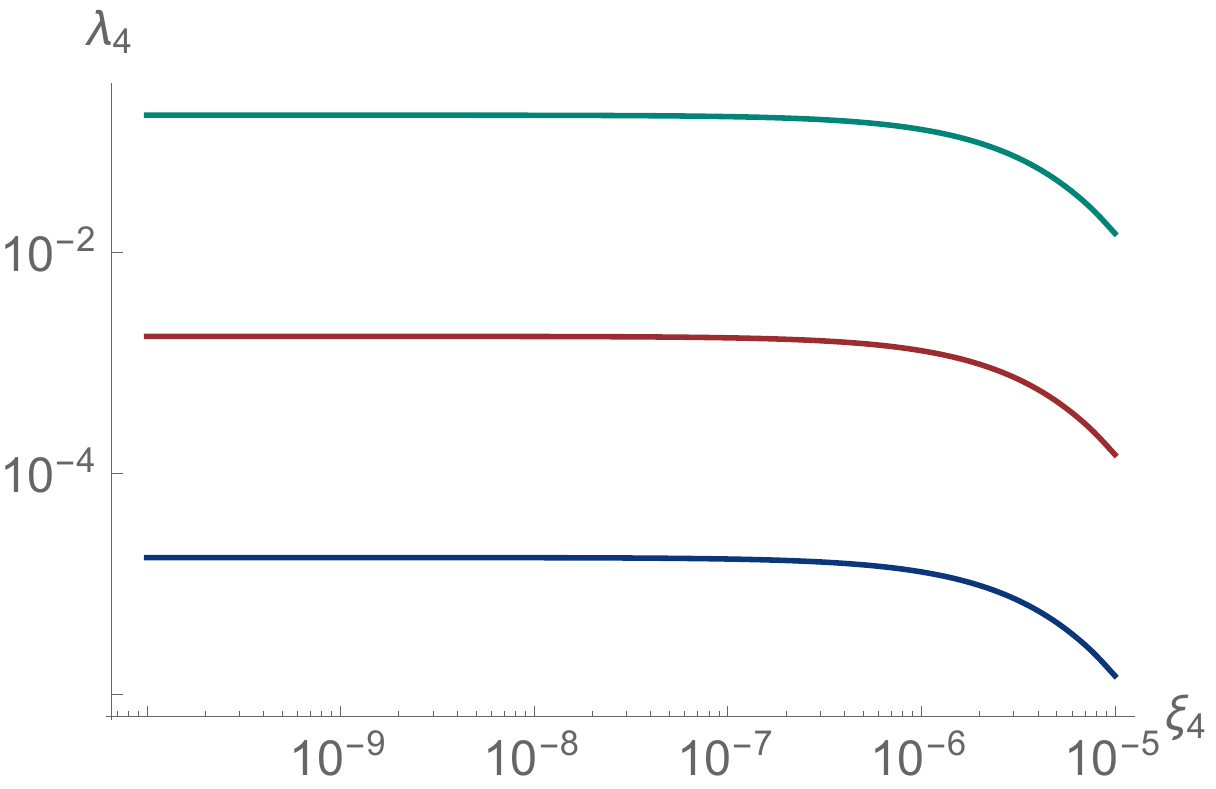}
\end{minipage}
\hskip 3mm
\begin{minipage}{0.1\textwidth}
\includegraphics[width=\columnwidth]{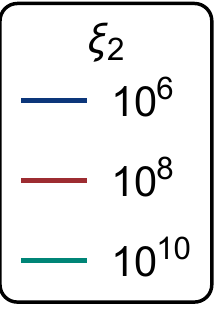}
\end{minipage}
}
\caption{\small
Behavior of $\lambda_4$ along constant $\xi_2$ slices in Fig.~\ref{fig:Nonminimal_nsr} for $N = 60$.
We show $\lambda_4$ as a function of $\ab{\xi_4}$ for $\xi_4<0$ (left) and $\xi_4>0$ (right).
The lines are $\xi_2 = 1$ (blue), $10^2$ (red), and $10^4$ (green) for the metric case,
while $\xi_2 = 10^6$ (blue), $10^8$ (red), and $10^{10}$ (green) for the Palatini case.
}
\label{fig:Nonminimal_lambda_60}
\end{center}
\end{figure}
%%%%%%%%%%

%%%%%%%%%%
\begin{figure}
\begin{center}
\small Metric ($N=60$)\smallskip\\
\fbox{
\begin{minipage}{0.37\textwidth}
\includegraphics[width=\columnwidth]{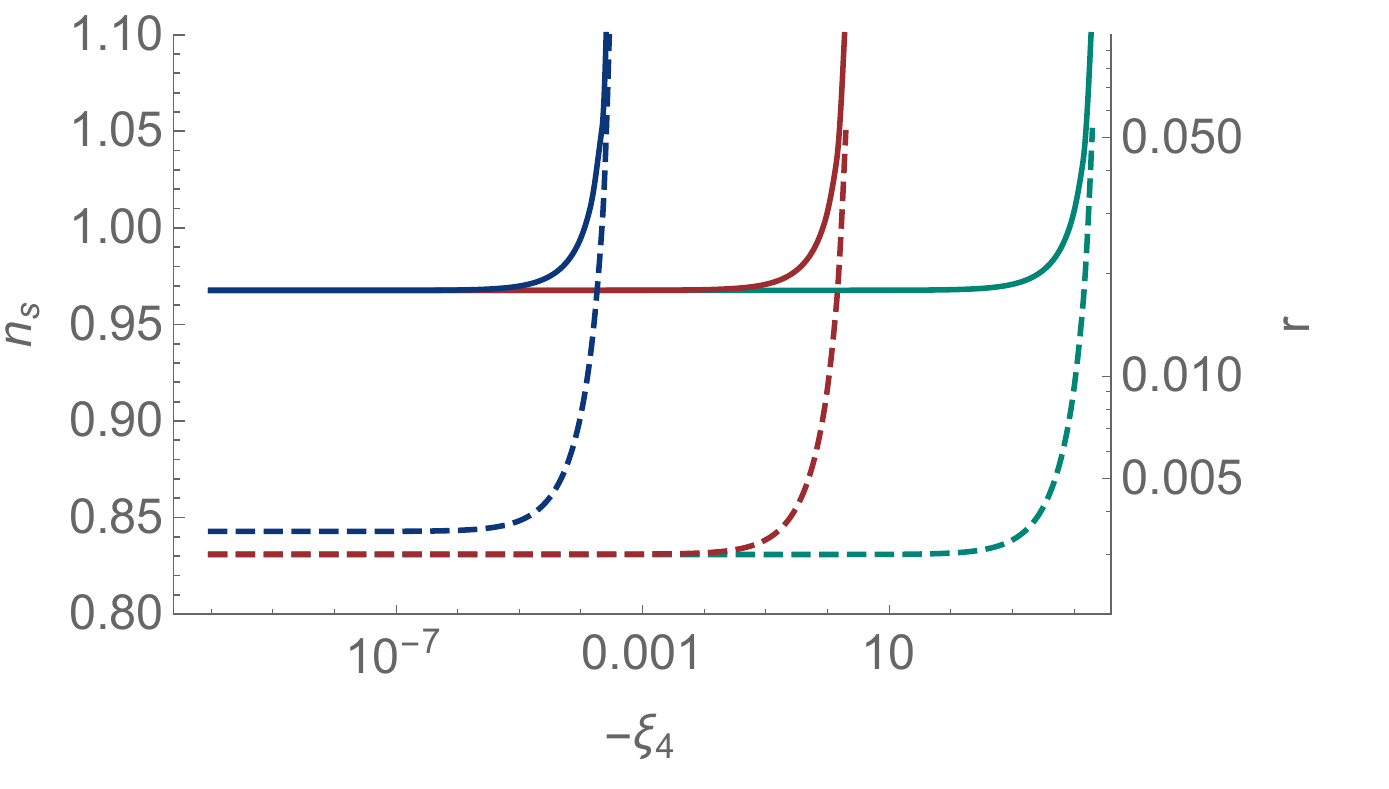}
\end{minipage}
\hskip 3mm
\begin{minipage}{0.37\textwidth}
\includegraphics[width=\columnwidth]{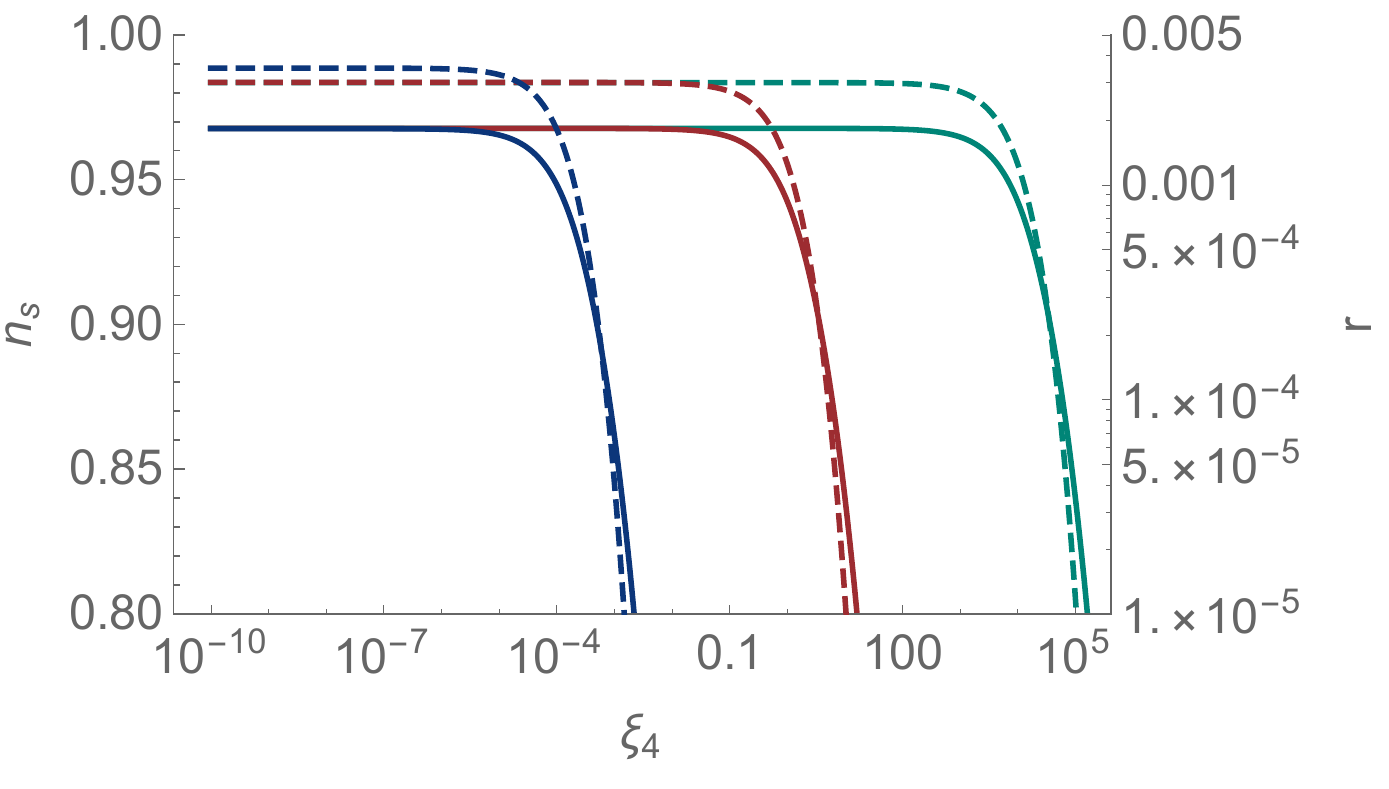}
\end{minipage}
\hskip 3mm
\begin{minipage}{0.1\textwidth}
\includegraphics[width=\columnwidth]{./figs/legNonminimalMetric.pdf}
\end{minipage}
}
\end{center}
\begin{center}
\small Palatini ($N=60$)\smallskip\\
\fbox{
\begin{minipage}{0.37\textwidth}
\includegraphics[width=\columnwidth]{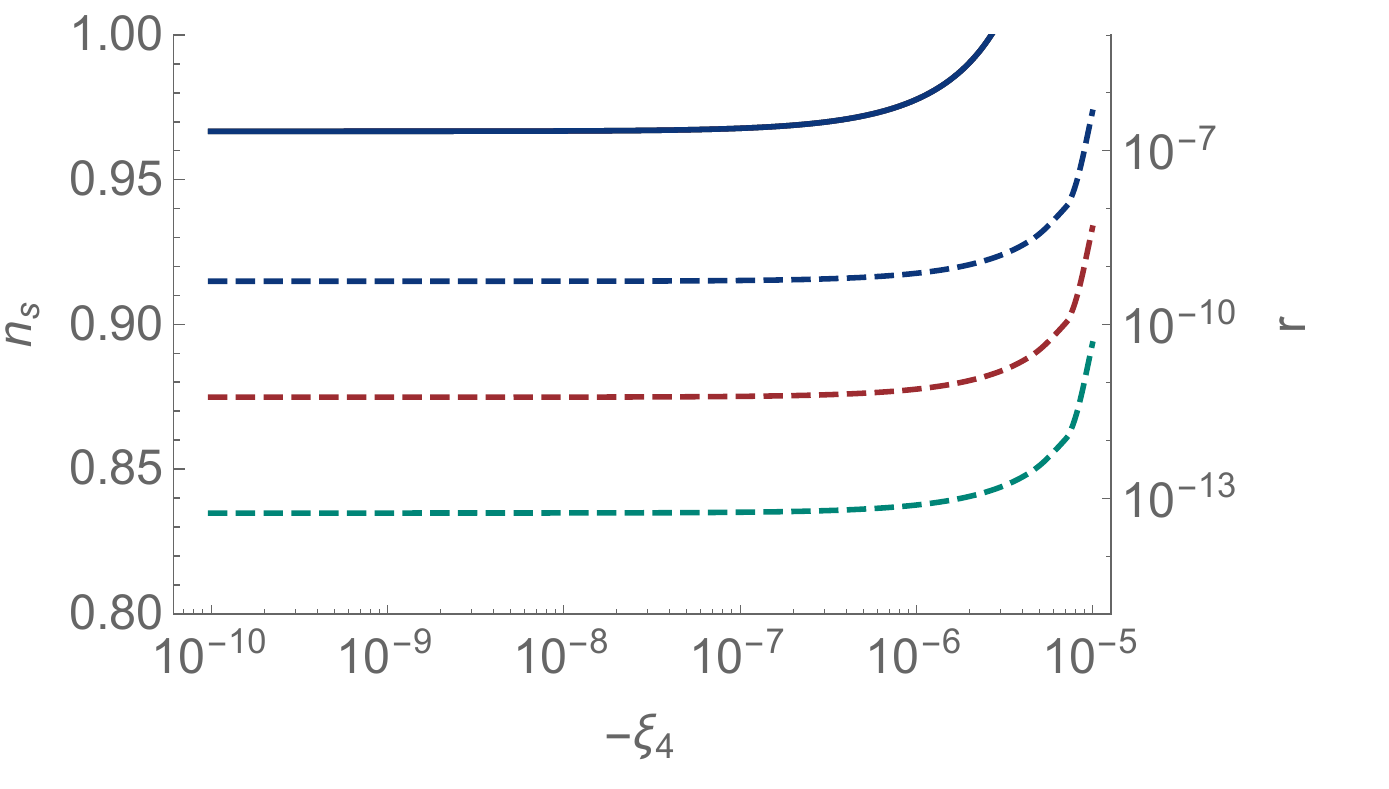}
\end{minipage}
\hskip 3mm
\begin{minipage}{0.37\textwidth}
\includegraphics[width=\columnwidth]{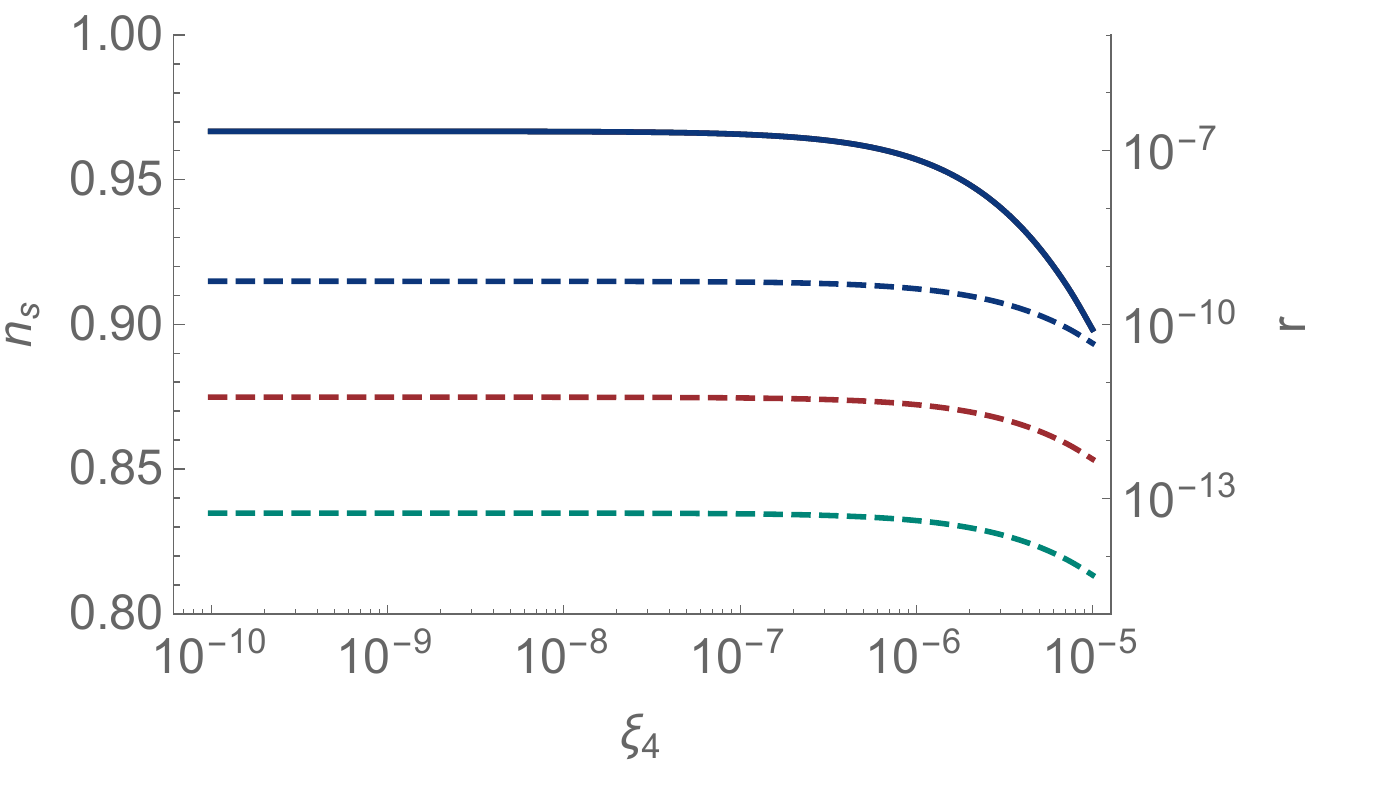}
\end{minipage}
\hskip 3mm
\begin{minipage}{0.1\textwidth}
\includegraphics[width=\columnwidth]{./figs/legNonminimalPalatini.pdf}
\end{minipage}
}
\caption{\label{fig:Nonminimal_nsr_60}\small
Behavior of $n_s$ (solid, left axis) and $r$ (dashed, right axis) along constant $\xi_2$ slices 
in Fig.~\ref{fig:Nonminimal_nsr} for $N = 60$.
All the solid lines are degenerate for the Palatini case.
}
\end{center}
\end{figure}
%%%%%%%%%%

\afterpage{\clearpage}

%%%%%%%%%%
\begin{figure}
\begin{center}
\small Metric ($N=50$)\smallskip\\
\fbox{
\begin{minipage}{0.32\textwidth}
\includegraphics[width=\columnwidth]{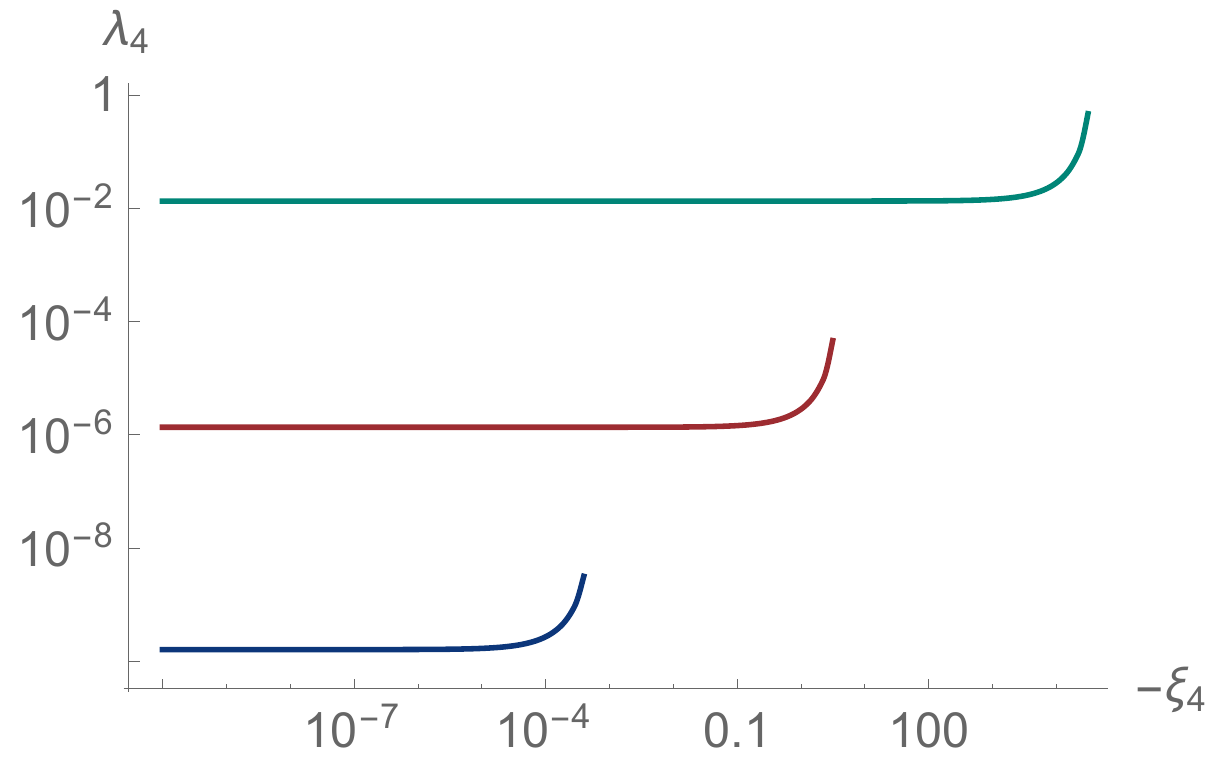}
\end{minipage}
\hskip 3mm
\begin{minipage}{0.32\textwidth}
\includegraphics[width=\columnwidth]{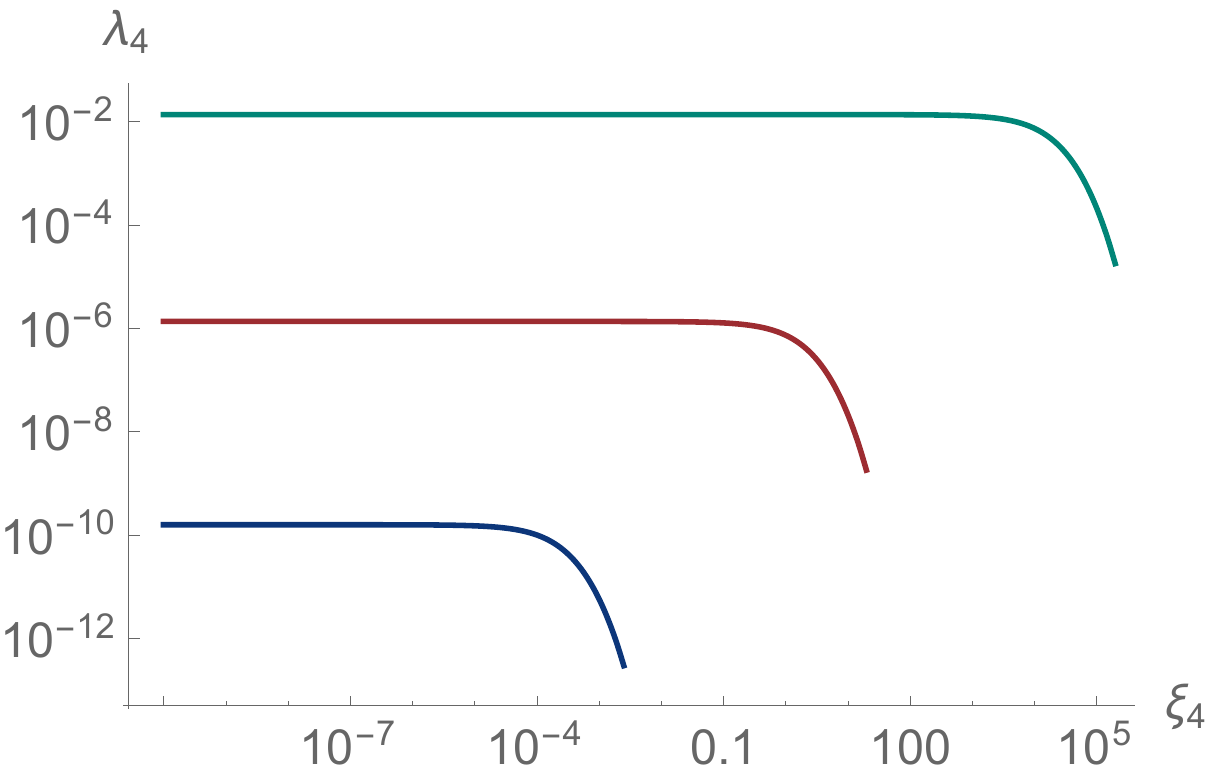}
\end{minipage}
\hskip 3mm
\begin{minipage}{0.1\textwidth}
\includegraphics[width=\columnwidth]{./figs/legNonminimalMetric.pdf}
\end{minipage}
}
\end{center}
\begin{center}
\small Palatini ($N=50$)\smallskip\\
\fbox{
\begin{minipage}{0.32\textwidth}
\includegraphics[width=\columnwidth]{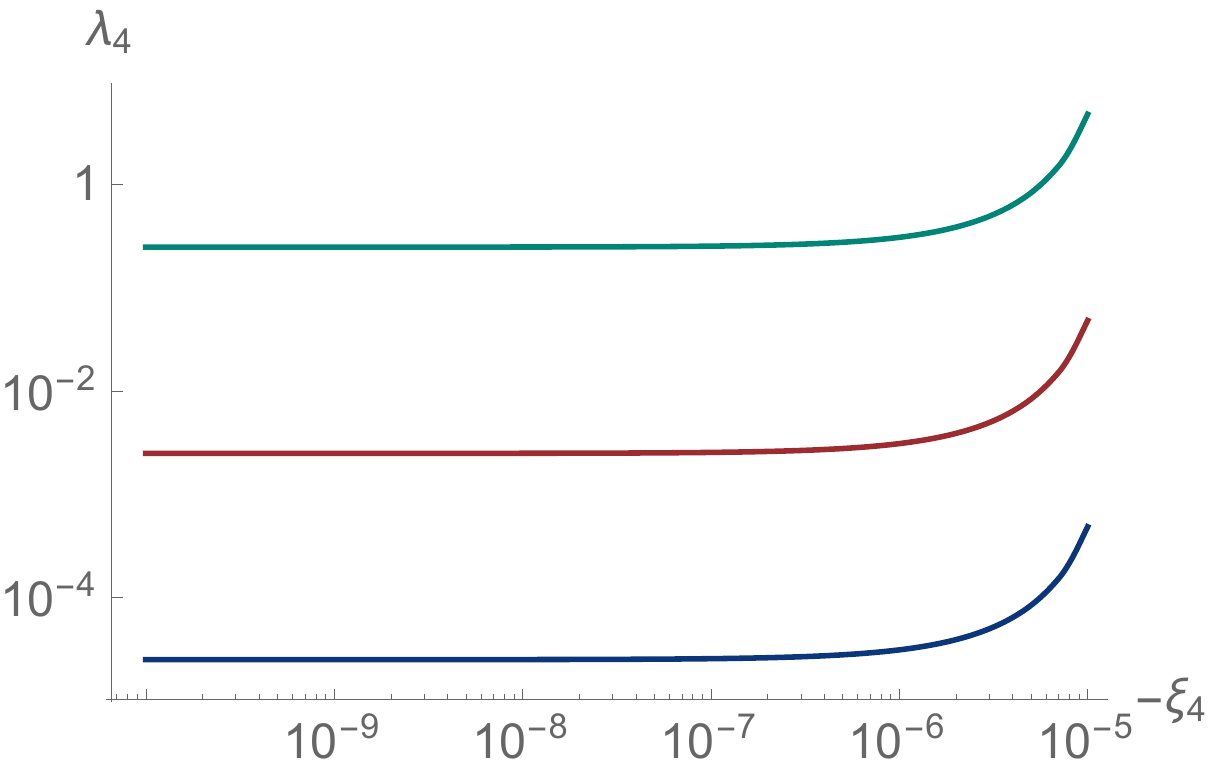}
\end{minipage}
\hskip 3mm
\begin{minipage}{0.32\textwidth}
\includegraphics[width=\columnwidth]{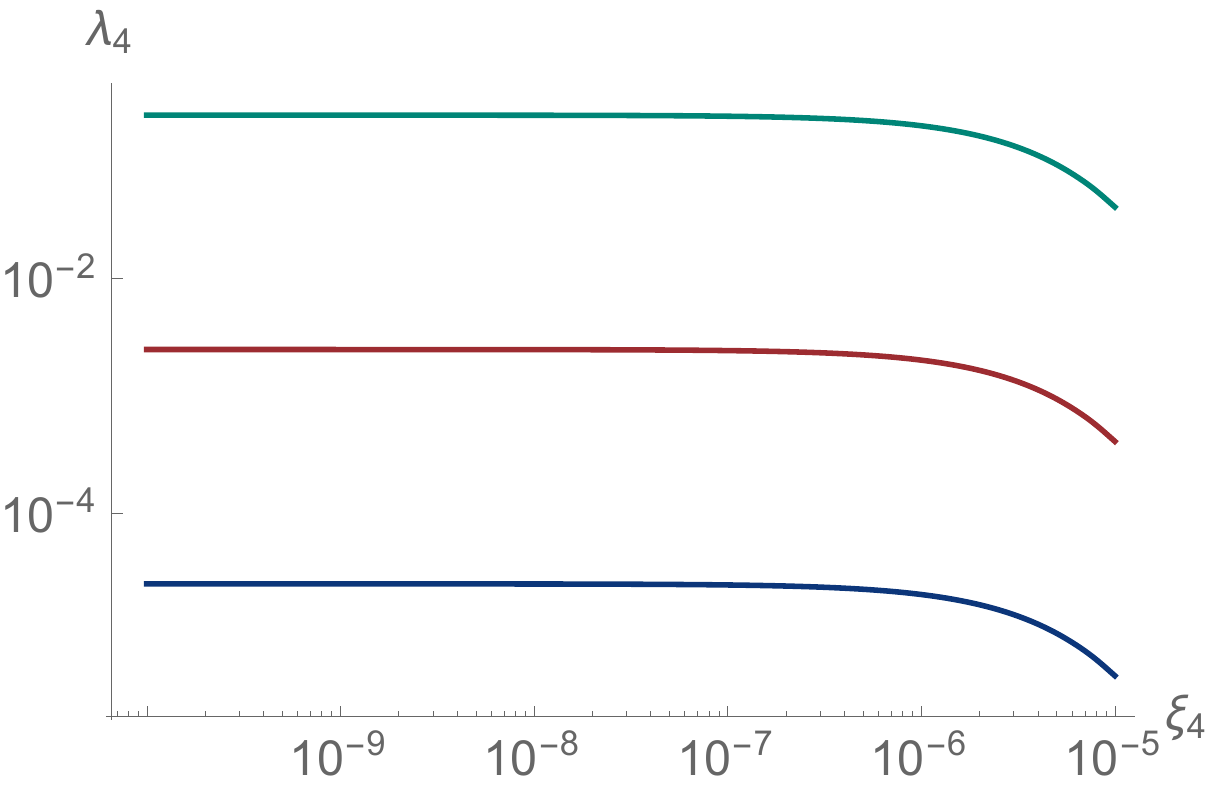}
\end{minipage}
\hskip 3mm
\begin{minipage}{0.1\textwidth}
\includegraphics[width=\columnwidth]{./figs/legNonminimalPalatini.pdf}
\end{minipage}
}
\caption{\small
The same as in Fig.~\ref{fig:Nonminimal_lambda_60} except for $N=50$.
}
\label{fig:Nonminimal_lambda_50}
\end{center}
\end{figure}
%%%%%%%%%%

%%%%%%%%%%
\begin{figure}
\begin{center}
\small Metric ($N=50$)\smallskip\\
\fbox{
\begin{minipage}{0.37\textwidth}
\includegraphics[width=\columnwidth]{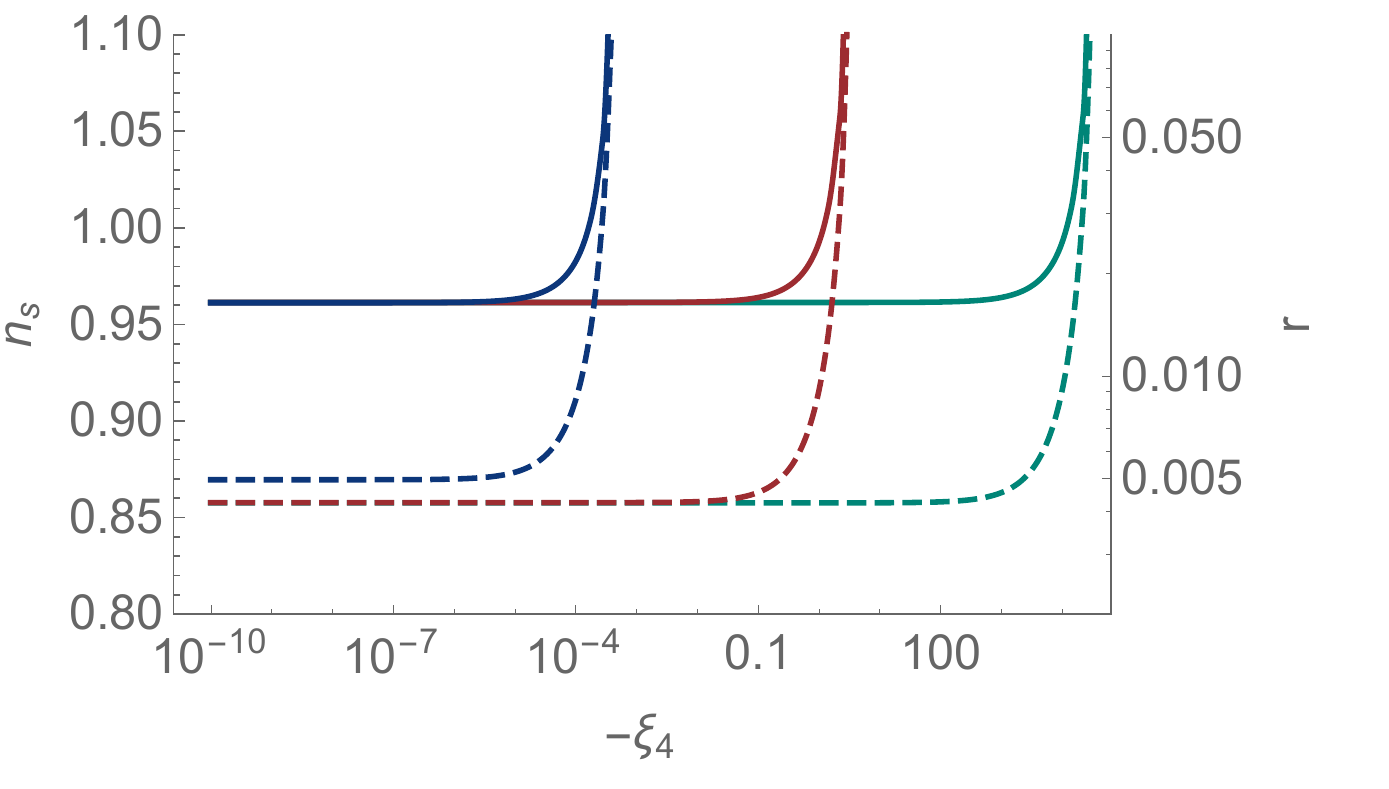}
\end{minipage}
\hskip 3mm
\begin{minipage}{0.37\textwidth}
\includegraphics[width=\columnwidth]{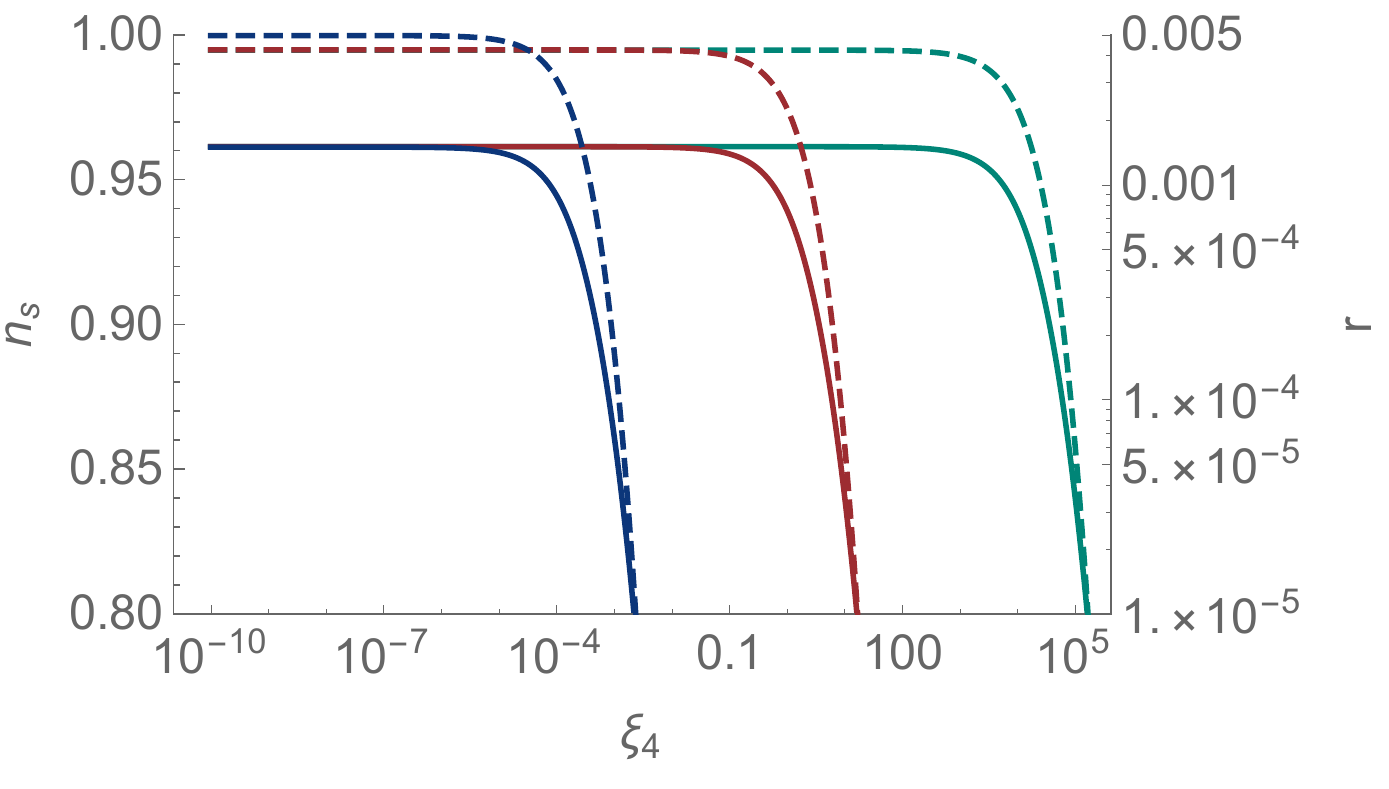}
\end{minipage}
\hskip 3mm
\begin{minipage}{0.1\textwidth}
\includegraphics[width=\columnwidth]{./figs/legNonminimalMetric.pdf}
\end{minipage}
}
\end{center}
\begin{center}
\small Palatini ($N=50$)\smallskip\\
\fbox{
\begin{minipage}{0.37\textwidth}
\includegraphics[width=\columnwidth]{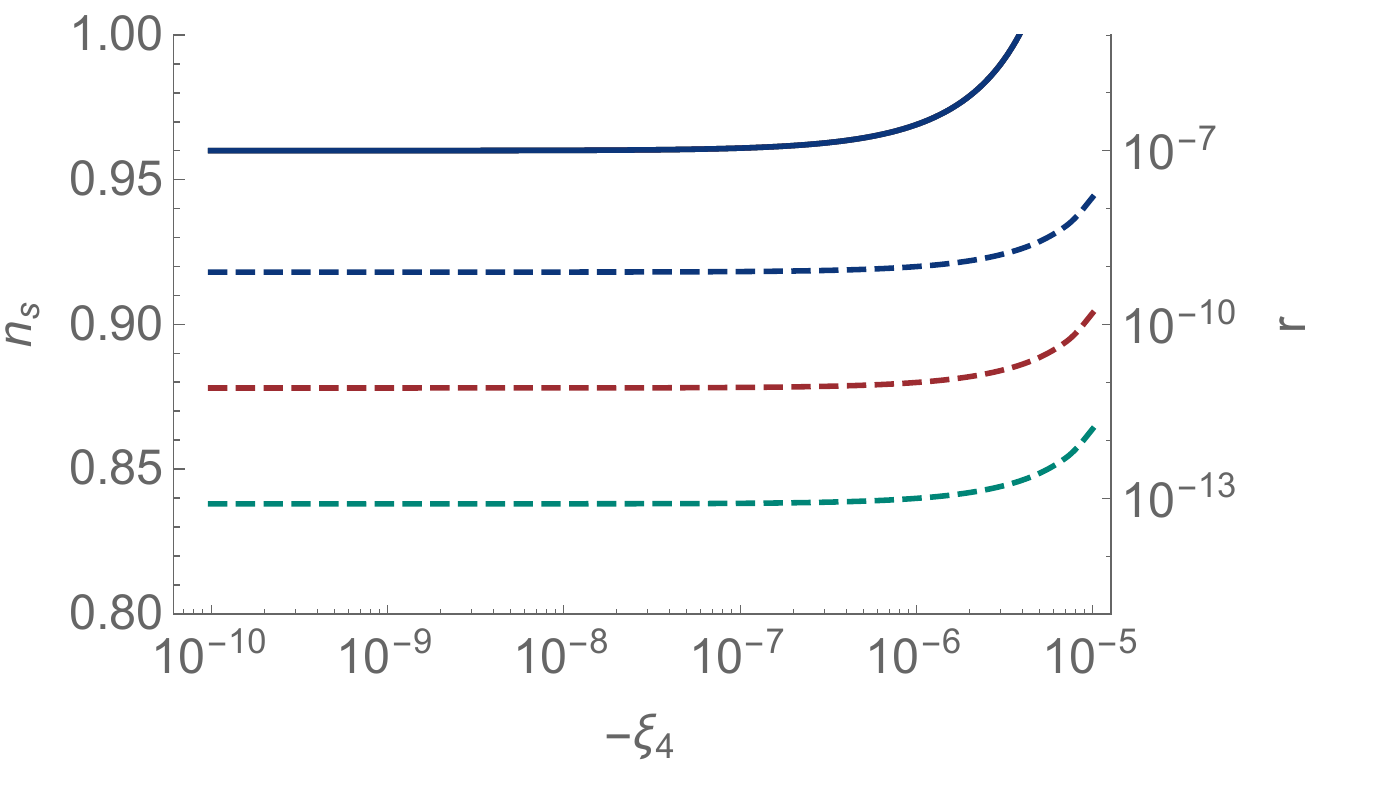}
\end{minipage}
\hskip 3mm
\begin{minipage}{0.37\textwidth}
\includegraphics[width=\columnwidth]{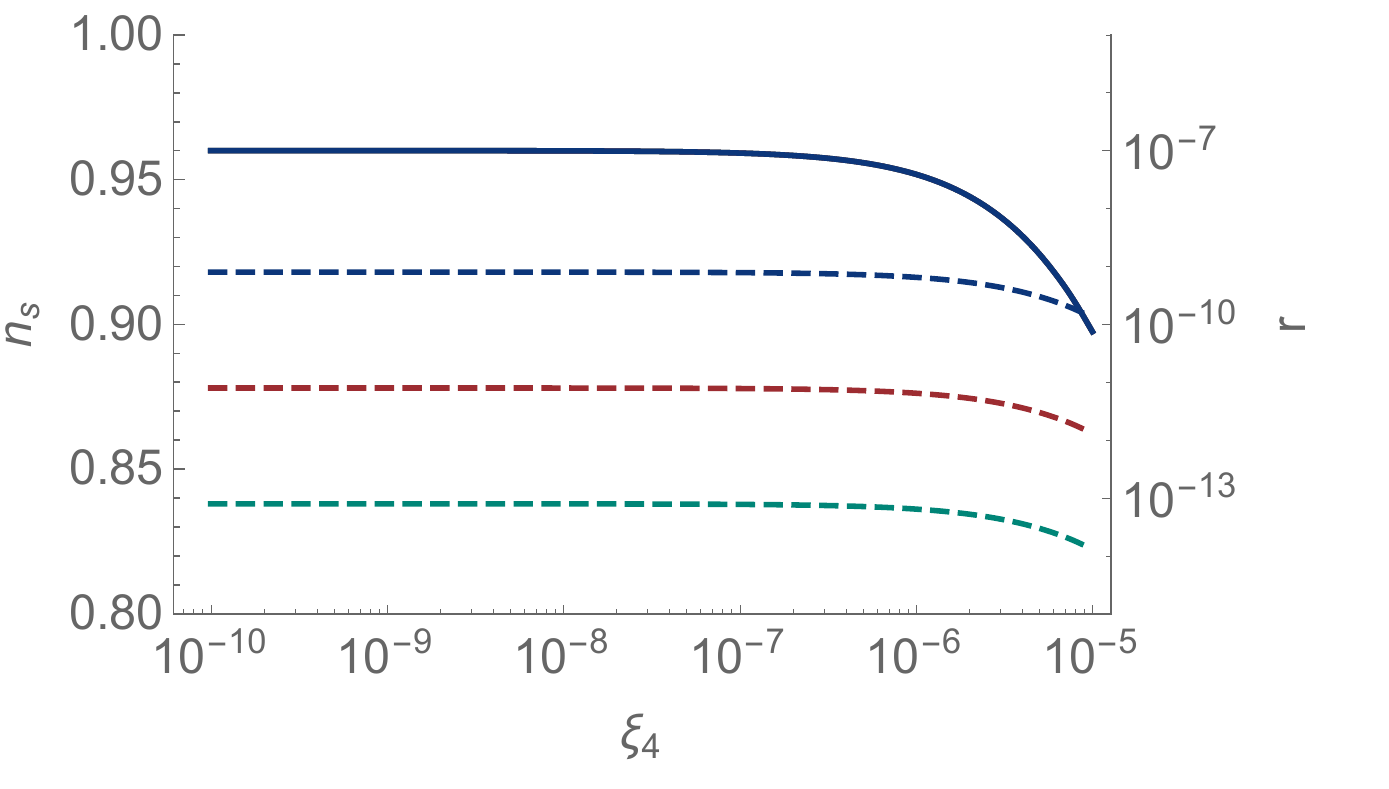}
\end{minipage}
\hskip 3mm
\begin{minipage}{0.1\textwidth}
\includegraphics[width=\columnwidth]{./figs/legNonminimalPalatini.pdf}
\end{minipage}
}
\caption{\small
The same as in Fig.~\ref{fig:Nonminimal_nsr_60} except for $N=50$.
}
\label{fig:Nonminimal_nsr_50}
\end{center}
\end{figure}
%%%%%%%%%%

\afterpage{\clearpage}

%%%%%%%%%%%%%%%%%%%%%%%%%%%%%%%%%%%%%%%%%%%%%%%%%%
\subsection{Results}
\label{subsec:Nonminimal_Results}
%%%%%%%%%%%%%%%%%%%%%%%%%%%%%%%%%%%%%%%%%%%%%%%%%%

In Fig.~\ref{fig:Nonminimal_nsr}, we show predictions of the metric (left) and Palatini (right) formalisms in the $n_s$-$r$ plane for $N = 60$ (top) and $50$ (bottom).
The blue contours show constant values of $\xi_2$,
while the red ones show constant values of $\xi_4$.
The value of the quartic coupling $\lambda_4$ is also presented as a density plot.
The smiley marker is the prediction of the quartic chaotic inflation,
while the star denotes the attractor point for $\xi_2 \gg 1$.
For $\xi_2 < 10^{-3}$ and $\xi_2 > 1$, the corresponding blue lines are almost degenerate with 
the upper and lower boundaries with the gray regions, respectively.
We see the following:
\begin{itemize}
\item 
For relatively small $\xi_2$ ($\lesssim 10^{-1}$),
even a small injection of $\ab{\xi_4}\gtrsim 10^{-5}$ drastically changes the inflationary predictions 
in both metric and Palatini formalisms.
\item
For relatively large $\xi_2$ ($\gtrsim 1$),
the inflationary predictions are stable against the injection of $\xi_4$ for the metric formalism (left):
we see that all the red lines converge to the attractor point.
On the other hand, the predictions are no more stable against the injection of $\ab{\xi_4}\gtrsim 10^{-5}$ 
for the Palatini formalism (right).
\end{itemize}

In Fig.~\ref{fig:Nonminimal_lambda_60}, 
we plot the behavior of $\lambda_4$ as a function of $\ab{\xi_4}$ for $\xi_4<0$ (left panel) and $\xi_4>0$ (right panel) 
with $N=60$ for fixed values of $\xi_2$.
In Fig.~\ref{fig:Nonminimal_nsr_60}, we plot $n_s$ (solid, left axis) and $r$ (dashed, right axis) similarly.
Figs.~\ref{fig:Nonminimal_lambda_50} and \ref{fig:Nonminimal_nsr_50} are the corresponding ones with $N=50$.
One immediately sees that the deviation from the attractor occurs around $\ab{\xi_4} \sim 10^{-4}\xi_2^2$ and 
around $\ab{\xi_4}\sim10^{-5}$ for the metric and Palatini cases, respectively.

%%%%%%%%%%%%%%%%%%%%%%%%%%%%%%%%%%%%%%%%%%%%%%%%%%
\subsection{Interpretation}
%%%%%%%%%%%%%%%%%%%%%%%%%%%%%%%%%%%%%%%%%%%%%%%%%%

Let us interpret our results, 
especially the threshold value of $\xi_4$ which gives deviation from the observationally allowed region.
In the following we use $N \gg 1$ and $\xi_2 \gg 1$ and keep the leading contribution
for each order in $\xi_4$ expansion when necessary.

In the metric formalism, let us first expand Eq.~(\ref{eq:Nonminimal_Metric_N}) by small $\xi_4$
around $\phi \simeq \phi_{\xi_4 = 0} \simeq \sqrt{4N/3\xi_2}$ (see Eq.~\eqref{eq:Metric_phiN}):
\begin{align}
N	
&=	
\left[
\frac{1 + 6\xi_2}{8} \phi^2 - \frac{3}{4} \ln (1 + \xi_2 \phi^2)
\right]
+
\xi_4
\left[
\frac{1 + 6\xi_2}{24} \phi^6
-
\frac{3}{4} \frac{\xi_2 \phi^6}{1 + \xi_2 \phi^2}
\right]
+
\cdots.
\end{align}
Substituting $\phi = \phi_{\xi_4 = 0}(1 + c \xi_4)$ and comparing leading terms in $\xi_4$ 
we find that the deviation of $\phi$ is given by
\begin{align}
\phi
&\simeq
\phi_{\xi_4 = 0}
\left(
1 - \frac{8N^2}{27\xi_2^2} \xi_4
\right).
\label{eq:Nonminimal_phidev_Metric}
\end{align}
We next expand $\epsilon$~\eqref{eq:Nonminimal_Metric_eps} 
and $\eta$~\eqref{eq:Nonminimal_Metric_eta} by small $\xi_4$ around the same point of $\phi$,
and then substitute Eq.~(\ref{eq:Nonminimal_phidev_Metric}):
\begin{align}
\epsilon
&\simeq
\frac{4}{3\xi_2^2 \phi^4}
\left(
1 - 2 \xi_4 \phi^4
\right)
\simeq 
\frac{3}{4N^2}
\left(
1 - \frac{64N^2}{27\xi_2^2} \xi_4
\right),
\\
\eta
&\simeq
- \frac{4}{3 \xi_2 \phi^2}
\left(
1 + \xi_4 \phi^4
\right)
\simeq 
- \frac{1}{N}
\left(
1 + \frac{64N^2}{27\xi_2^2} \xi_4
\right).
\end{align}
We see that the deviation becomes sizable for $\ab{\xi_4}\gtrsim \xi_2^2/N^2$.
Since $\epsilon$ and $\eta$ corresponds to the $r$ (vertical) and $n_s$ (horizontal) axes in Fig.~\ref{fig:Nonminimal_nsr}, respectively, the direction of deviation for positive and negative $\xi_4$ is also explained.

In the Palatini formalism, let us first expand Eq.~(\ref{eq:Nonminimal_Palatini_N}) by small $\xi_4$
around $\phi \simeq \phi_{\xi_4 = 0} \simeq \sqrt{8N}$ (see Eq.~\eqref{eq:Palatini_phiN}):
\begin{align}
N
&=
\frac{1}{8} \phi^2
+ \frac{1}{24} \xi_4 \phi^6 + \cdots.
\end{align}
Substituting $\phi = \phi_{\xi_4 = 0}(1 + c \xi_4)$ and comparing leading terms in $\xi_4$, we find
\begin{align}
\phi
&\simeq
\phi_{\xi_4 = 0}
\left(
1 - \frac{32N^2}{3} \xi_4
\right).
\label{eq:Nonminimal_phidev_Palatini}
\end{align}
We next expand $\epsilon$~\eqref{eq:Nonminimal_Palatini_eps} 
and $\eta$~\eqref{eq:Nonminimal_Palatini_eta} by small $\xi_4$ around the same point of $\phi$,
and then substitute Eq.~(\ref{eq:Nonminimal_phidev_Palatini}):
\begin{align}
\epsilon
&\simeq
\frac{8}{\xi_2 \phi^4}
\left(
1 - 2 \xi_4 \phi^4
\right)
\simeq 
\frac{1}{8N^2 \xi_2}
\left(
1 - \frac{256N^2}{3} \xi_4
\right),
\\
\eta
&\simeq
- \frac{8}{\phi^2}
\left(
1 + \xi_4 \phi^4
\right)
\simeq 
- \frac{1}{N}
\left(
1 + \frac{256N^2}{3} \xi_4
\right).
\end{align}
We see that the deviation becomes sizable for $\ab{\xi_4}\gtrsim 10^{-2}/N^2$ and 
that the direction of the deviation is the same as in the metric formalism.

To summarize, 
the expressions for the slow-roll parameters, especially for $\eta$, imply that 
the inflationary predictions (especially on $n_s$) are significantly affected for
\begin{align}
\ab{\xi_4}
&\gtrsim
\begin{cases}
\displaystyle
\frac{\xi_2^2}{N^2}
&~~
\tx{(metric)},
\\[3mm]
\displaystyle
\frac{10^{-2}}{N^2}
&~~
\tx{(Palatini)}.
\end{cases}
\label{eq:xi4_threshold}
\end{align}
This explains the behavior in 
Figs.~\ref{fig:Nonminimal_lambda_60}--\ref{fig:Nonminimal_nsr_50}.
After taking the curvature normalization into account,
these values read
\begin{align}
\ab{\xi_4}
&\gtrsim
\begin{cases}
\displaystyle
\frac{\xi_2^2}{N^2}
\sim
\frac{\lambda_4}{32\pi^2A_s}
\sim10^6\lambda_4
&~~
{\rm (metric)},
\\[3mm]
\displaystyle
\frac{10^{-2}}{N^2}
&~~
{\rm (Palatini)}.
\end{cases}
\label{eq:Nonminimal_Final}
\end{align}
Therefore, we see that the inflationary predictions in the metric formalism become stable for 
$\xi_2\gg1$ (unless $\lambda_4\ll10^{-6}$),
while they are spoiled by $\ab{\xi_4} \sim 10^{-5}$ 
independently of $\lambda_4$ or $\xi_2$ in the Palatini formalism.

In the case $\xi_4>0$, we note that in both the formalisms $\epsilon$ becomes zero at $\xi_4\phi^4=1$ 
and hence the $e$-folding diverges in the limit $\phi\to \xi_4^{-1/4}$. 
Then the necessary condition $\phi<\xi_4^{-1/4}$ at $\phi \sim \sqrt{4N/3\xi_2}$ (metric) 
and $\phi \sim \sqrt{8N}$ (Palatini) reads
\begin{align}
\xi_4
&\lesssim
\begin{cases}
\displaystyle{9\xi_2^2\ov16N^2}
&~~
\tx{(metric)},
\\[3mm]
\displaystyle{1\ov64N^2}
&~~
\tx{(Palatini)}.
\end{cases}
\end{align}
Note that we used the relation between $\phi$ and $N$ for $\xi_4=0$ 
(and for $\xi_2\gg1$ in metric formalism), and therefore these conditions are only approximate. 
We see that these conditions give comparable threshold values to Eq.~(\ref{eq:xi4_threshold}).

As we increase $\ab{\xi_4}$ from zero, the inflationary prediction starts to deviate at around the value in the right-hand side of Eq.~\eqref{eq:xi4_threshold}. We note that this occurs when the higher dimensional operator $\ab{\xi_4\phi^4}$ is still much smaller than the lower dimensional one $\xi_2\phi^2$. Indeed,
by substituting the approximate values $\phi\simeq\sqrt{4N/3\xi_2}$ (metric) and $\sqrt{8N}$ (Palatini),
the condition
\begin{align}
\ab{\xi_4\phi^4}
&\lesssim 
\xi_2\phi^2
\end{align}
becomes
\begin{align}
\ab{\xi_4}
&\lesssim	
\begin{cases}
\displaystyle{\xi_2^2\ov N}&\tx{(metric)},
\\[3mm]
\displaystyle10^{-1}{\xi_2\ov N}&\tx{(Palatini)},
\end{cases}
\end{align}
which is well satisfied at the value in the right-hand side of Eq.~\eqref{eq:xi4_threshold}.

\afterpage{\clearpage}

%%%%%%%%%%%%%%%%%%%%%%%%%%%%%%%%%%%%%%%%%%%%%%%%%%
\section{Sensitivity to corrections in the potential}
\setcounter{equation}{0}
\label{sec:Potential}
%%%%%%%%%%%%%%%%%%%%%%%%%%%%%%%%%%%%%%%%%%%%%%%%%%

Second, we consider a correction to the Jordan-frame potential:
\begin{align}
\Omega^2
&= 
1+\xi_2 \phi^2,
~~~~~~
V_\J
=
\lambda_4 \phi^4
+ \lambda_6 \phi^6.
\end{align}
Again we fix the signs of $\xi_2$ and $\lambda_4$ to be positive, 
while we allow both positive and negative signs for $\lambda_6$.
The Einstein-frame potential becomes
\begin{align}
V
&= 
\frac{\lambda_4 \phi^4 + \lambda_6 \phi^6}{(1 + \xi_2 \phi^2)^2}.
\end{align}
In this setup we have three parameters
\begin{itemize}
\item[]
\begin{center}
$(\lambda_4, \lambda_6, \xi_2)$,
\end{center}
\end{itemize}
but we can eliminate one since $A_s$ is fixed by the observation.

%%%%%%%%%%%%%%%%%%%%%%%%%%%%%%%%%%%%%%%%%%%%%%%%%%
\subsection{Predictions}
%%%%%%%%%%%%%%%%%%%%%%%%%%%%%%%%%%%%%%%%%%%%%%%%%%

Before discussing predictions in each formalism,
we note that the dependence of $\epsilon$, $\eta$, and $N$ on $\lambda_4$ and $\lambda_6$ shown below
appears only through the ratio $\lambda_6/\lambda_4$. 
In the following analysis, we fix the quartic coupling $\lambda_4$ by the overall normalization $A_s$
and adopt $\lambda_6/\lambda_4$ as an independent variable instead of $\lambda_6$.

%%%%%%%%%%%%%%%%%%%%%%%%%%%%%%%%%%%%%%%%%%%%%%%%%%
\subsubsection*{Metric formalism}
%%%%%%%%%%%%%%%%%%%%%%%%%%%%%%%%%%%%%%%%%%%%%%%%%%

In the metric formalism, the relation (\ref{eq:dchidphi}) between $\chi$ and $\phi$ is the same as
Eqs.~(\ref{eq:Metric_dchidphi})--(\ref{eq:Metric_chiphi}).
The slow-roll parameters and $e$-folding are calculated through Eqs.~(\ref{eq:epsGeneral})--(\ref{eq:NGeneral}):
\begin{align}
\epsilon
&=
\frac{2\paren{2 \lambda_4 + 3 \lambda_6 \phi^2 + \xi_2 \lambda_6 \phi^4}^2}
{\phi^2 \left[ 1 + \paren{\xi_2 + 6 \xi_2^2}\phi^2 \right] \paren{\lambda_4 + \lambda_6 \phi^2}^2},
\label{eq:Potential_Metric_eps}
\\
\eta
&= 
\frac{
2 \left[ 
6 \lambda_4 + \paren{2\xi_2 \lambda_4 + 24 \xi_2^2 \lambda_4 + 15 \lambda_6} \phi^2 
+ \cdots 
+ \paren{2 \xi_2^3 \lambda_6 + 12 \xi_2^4 \lambda_6} \phi^8
\right]
}
{\phi^2 \left[ 1 + \paren{\xi_2 + 6 \xi_2^2}\phi^2 \right]^2 \paren{\lambda_4 +  \lambda_6 \phi^2}},
\label{eq:Potential_Metric_eta}
\\
N
&= 
\int d\phi
~
\frac{\phi \left[ 1 + (\xi_2 + 6 \xi_2^2)\phi^2 \right] \paren{ \lambda_4 + \lambda_6 \phi^2}}
{2 \paren{1 + \xi_2 \phi^2} \paren{2 \lambda_4 + 3 \lambda_6 \phi^2 + \xi_2 \lambda_6 \phi^4}}.
\label{eq:Potential_Metric_N}
\end{align}
Here we do not show a complete expression for $\eta$ to avoid complications; see Appendix~\ref{app:Equations} for it.
Note that for $\lambda_6 < 0$ the $e$-folding diverges at the point where the potential derivative vanishes, namely, at
$\phi = \sqrt{-3 + \sqrt{9 - 8\xi_2 \lambda_4/\lambda_6}/2\xi_2}$.
The scalar power spectrum amplitude reads
\begin{align}
A_s
&=
{\lambda_4\phi^6\ov192\pi^2}
{
\displaystyle
\sqbr{1+\xi_2\phi^2\paren{1+6\xi_2}}
\paren{1+{\lambda_{6}\ov\lambda_4}\phi^2}^3
\ov
\displaystyle
\paren{1+\xi_2\phi^2}
\sqbr{1+{3\lambda_6\ov2\lambda_4}\phi^2\paren{1+{\xi_2\ov3}\phi^2}}^2
}.
\end{align}
%%

%%%%%%%%%%%%%%%%%%%%%%%%%%%%%%%%%%%%%%%%%%%%%%%%%%
\subsubsection*{Palatini formalism}
%%%%%%%%%%%%%%%%%%%%%%%%%%%%%%%%%%%%%%%%%%%%%%%%%%

In the Palatini formalism, the relation (\ref{eq:dchidphi}) between $\chi$ and $\phi$ is the same as
Eqs.~(\ref{eq:Palatini_dchidphi})--(\ref{eq:Palatini_chiphi}).
The slow-roll parameters and $e$-folding are calculated through Eqs.~(\ref{eq:epsGeneral})--(\ref{eq:NGeneral}):
\begin{align}
\epsilon
&=
\frac{2 \paren{2 \lambda_4 + 3 \lambda_6 \phi^2 + \xi_2 \lambda_6 \phi^4}^2}
{\phi^2 \paren{1 + \xi_2 \phi^2}\paren{\lambda_4 + \lambda_6 \phi^2}^2},
\label{eq:Potential_Palatini_eps}
\\
\eta
&= 
\frac{2 \left[ 6 \lambda_4 + \paren{-4 \xi_2 \lambda_4 + 15 \lambda_6} \phi^2 
+ 7 \xi_2 \lambda_6 \phi^4 + 2 \xi_2^2 \lambda_6 \phi^6 \right]}
{\phi^2 \paren{1 + \xi_2 \phi^2}\paren{ \lambda_4 +  \lambda_6 \phi^2}},
\label{eq:Potential_Palatini_eta}
\\
N
&= 
\int d\phi
~
\frac{\phi \paren{\lambda_4 +  \lambda_6 \phi^2}}
{2 \paren{2 \lambda_4 + 3 \lambda_6 \phi^2 + \xi_2 \lambda_6 \phi^4}}.
\label{eq:Potential_Palatini_N}
\end{align}
For $\lambda_6 < 0$, the $e$-folding diverges at the same point as the metric formalism,
where the potential derivative vanishes.
The scalar power spectrum amplitude reads
\begin{align}
A_s
&=
{\lambda_4\phi^6\ov192\pi^2}
{
\displaystyle
\paren{1+{\lambda_6\ov\lambda_4}\phi^2}^3
\ov
\displaystyle
\paren{1+\xi_2\phi^2}
\sqbr{1+{3\lambda_6\ov2\lambda_4}\phi^2\paren{1+{\xi_2\ov3}\phi^2}}^2
}.
\end{align}
%%

%%%%%%%%%%
\begin{figure}
\begin{center}
\includegraphics[width=0.42\columnwidth]{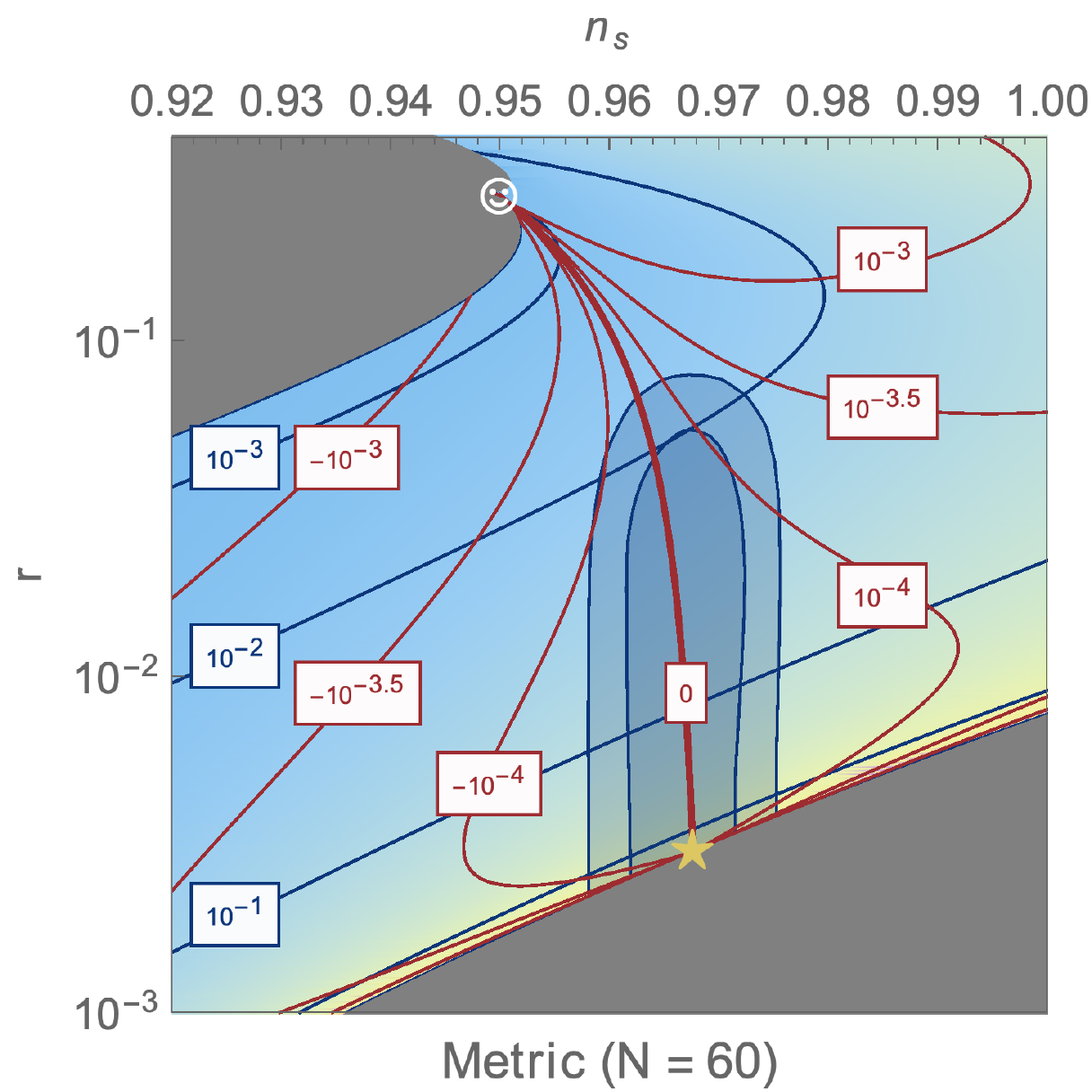}
\hskip 3mm
\includegraphics[width=0.42\columnwidth]{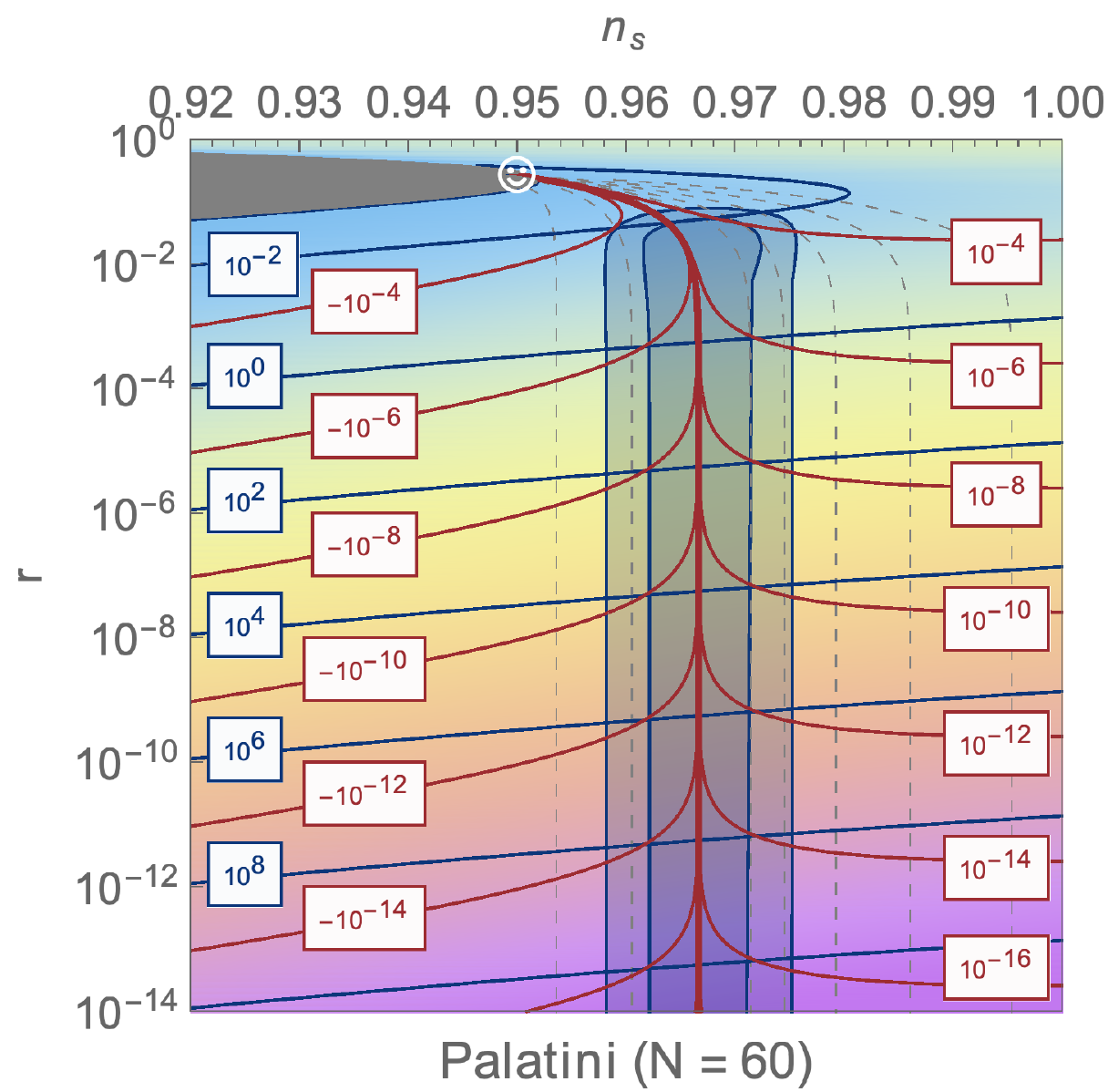}
\hskip 3mm
\includegraphics[width=0.105\columnwidth]{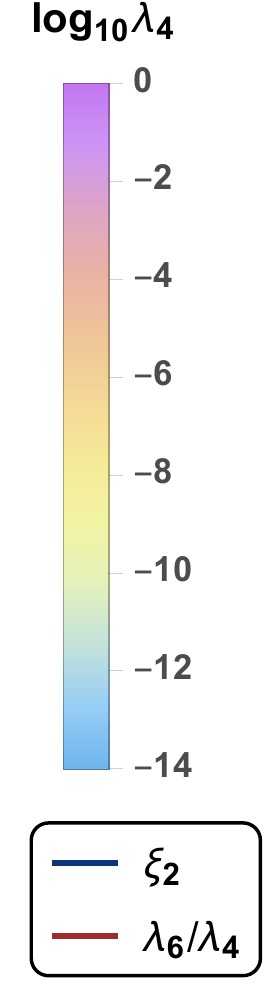}
\\
\vskip 4mm
\includegraphics[width=0.42\columnwidth]{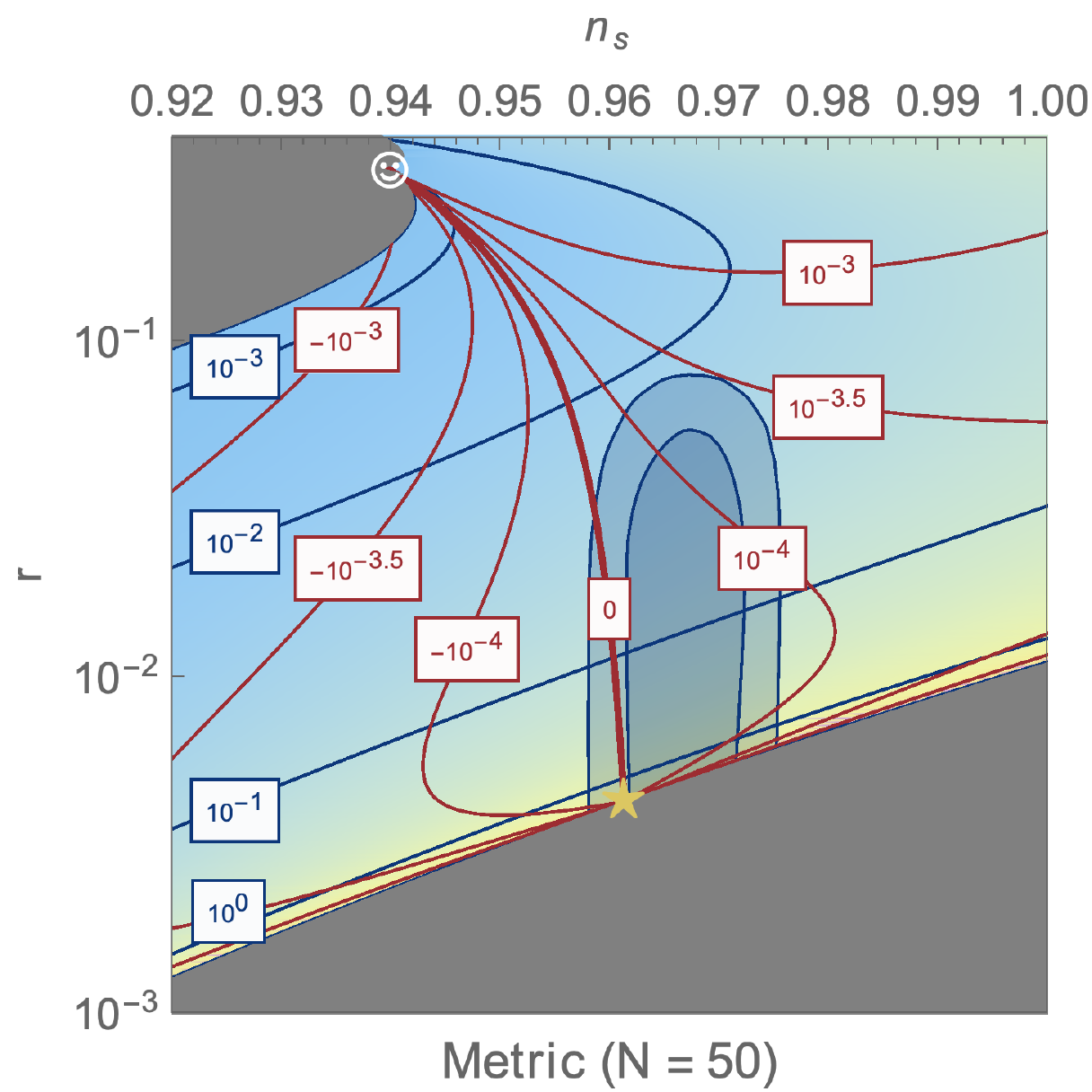}
\hskip 3mm
\includegraphics[width=0.42\columnwidth]{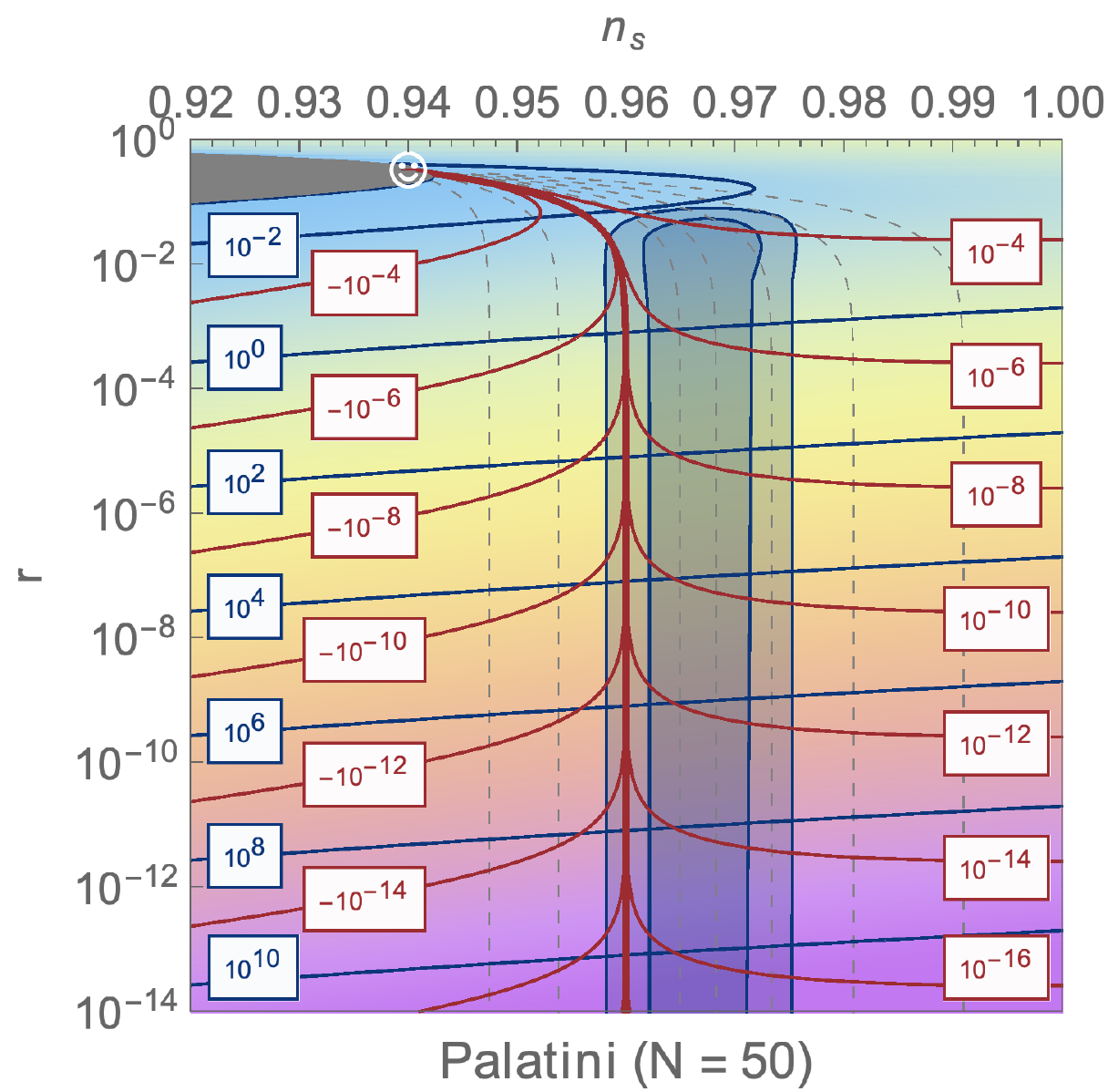}
\hskip 3mm
\includegraphics[width=0.105\columnwidth]{./figs/legPotential.pdf}
\caption{\small
Effects of the higher dimensional operator $\lambda_6\phi^6$ in the metric (left) and Palatini (right) formalisms with $N=60$ (top) and 50 (bottom).
We plot contours of fixed $\xi_2$ (blue, horizontal) and of fixed $\lambda_6/\lambda_4$ (red, vertical) 
in the $n_s$-$r$ plane.
The value of $\lambda_4$ is also shown as a density plot.
Allowed regions of $1\sigma$ and $2\sigma$ from the Planck experiment~\cite{Akrami:2018odb}
(TT,TE,EE+lowE+lensing+BK14+BAO) are also shown in the center.
Dashed lines in the right panels correspond to $\lambda_6=-10^{-16.25,\,-16.5}$ and $10^{-16.5,\,-16.25,\,-16,\,-15.75,\,-15.5}$ from left to right.
}
\label{fig:Potential_nsr}
\end{center}
\end{figure}
%%%%%%%%%%

\afterpage{\clearpage}

%%%%%%%%%%
\begin{figure}
\begin{center}
\small Metric ($N=60$)\smallskip\\
\fbox{
\begin{minipage}{0.32\textwidth}
\includegraphics[width=\columnwidth]{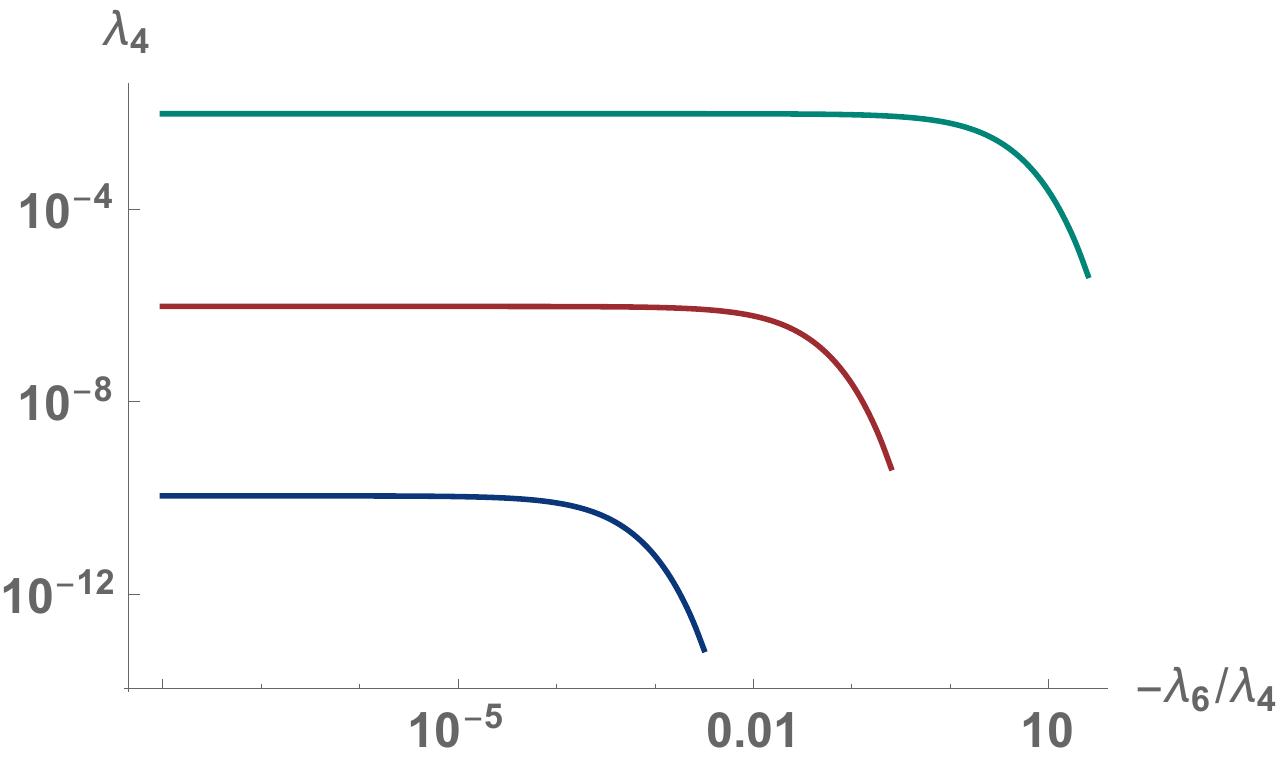}
\end{minipage}
\hskip 3mm
\begin{minipage}{0.32\textwidth}
\includegraphics[width=\columnwidth]{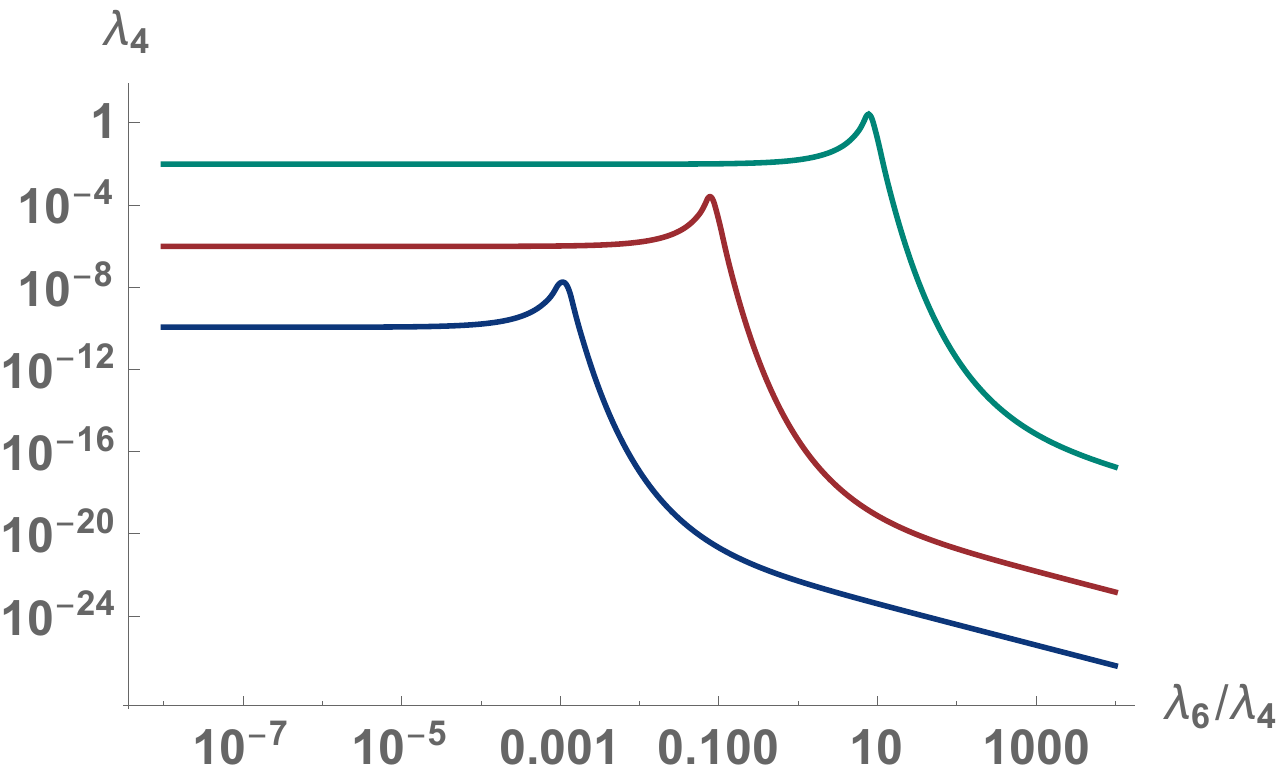}
\end{minipage}
\hskip 3mm
\begin{minipage}{0.1\textwidth}
\includegraphics[width=\columnwidth]{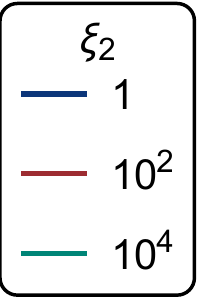}
\end{minipage}
}
\end{center}
\begin{center}
\small Palatini ($N=60$)\smallskip\\
\fbox{
\begin{minipage}{0.32\textwidth}
\includegraphics[width=\columnwidth]{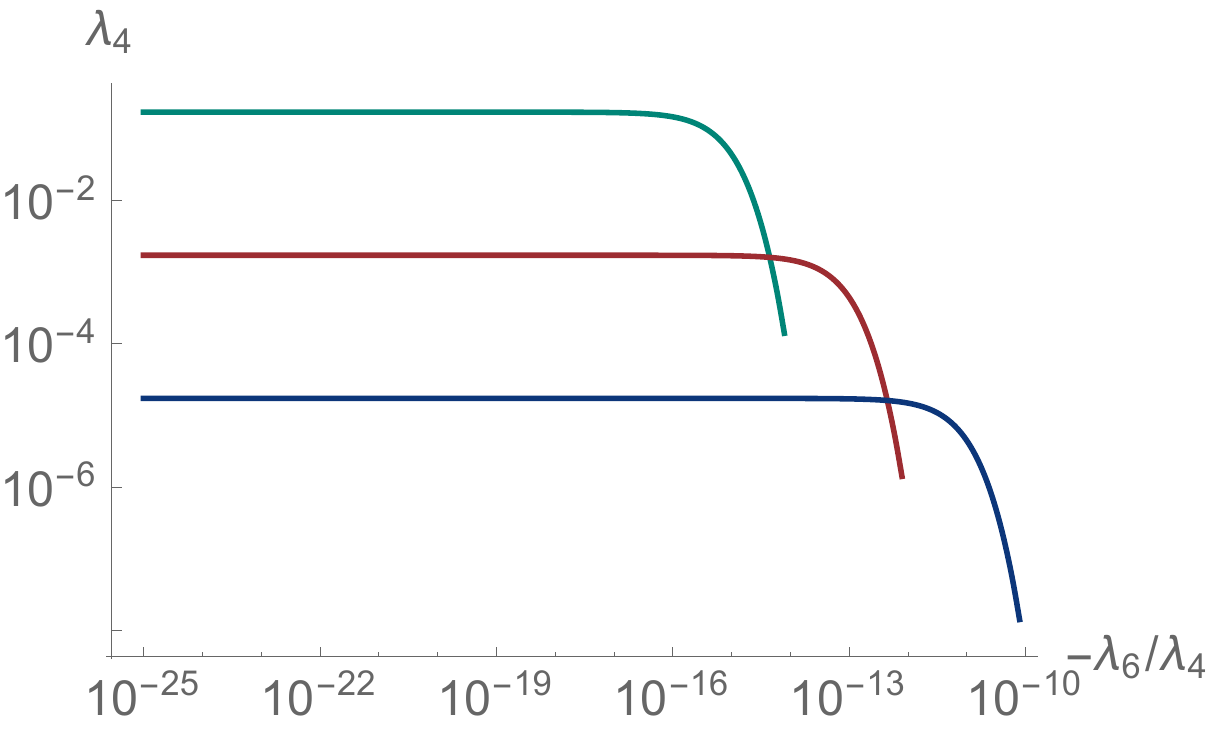}
\end{minipage}
\hskip 3mm
\begin{minipage}{0.32\textwidth}
\includegraphics[width=\columnwidth]{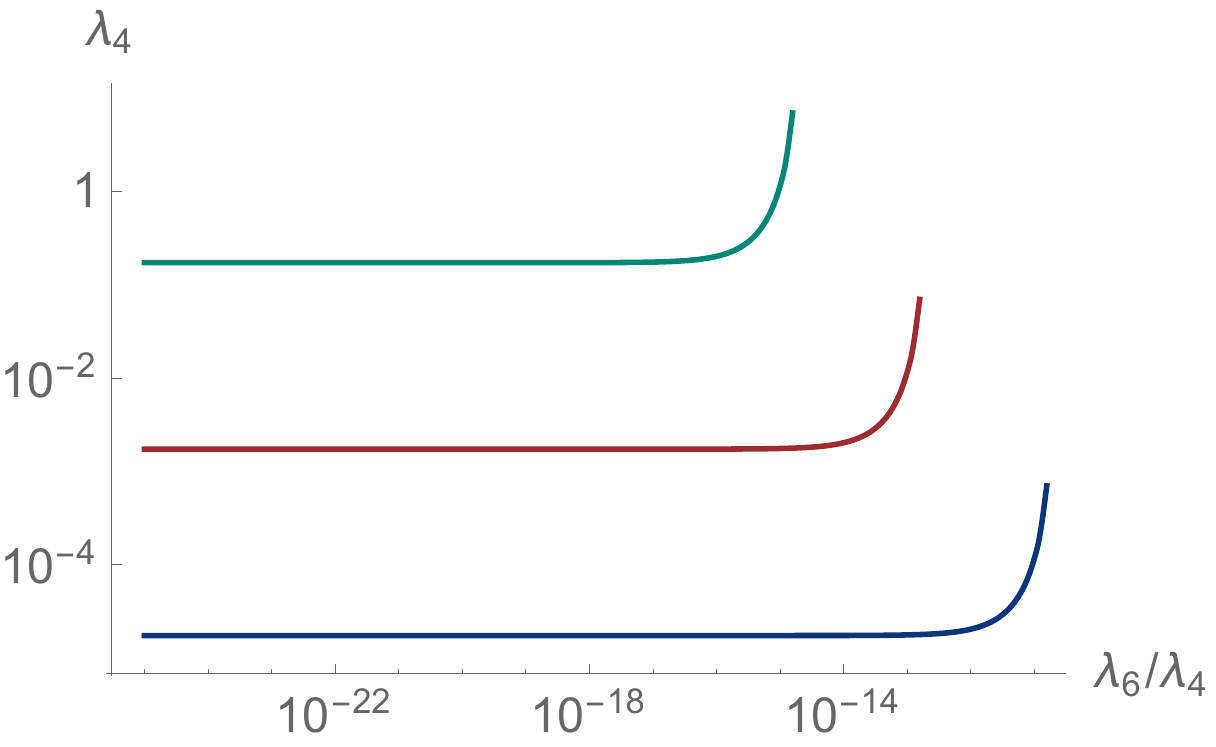}
\end{minipage}
\hskip 3mm
\begin{minipage}{0.1\textwidth}
\includegraphics[width=\columnwidth]{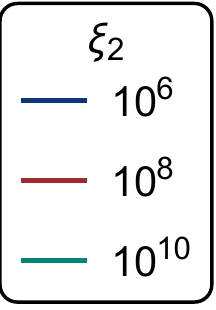}
\end{minipage}
}
\caption{\small
Behavior of $\lambda_4$ along constant $\xi_2$ slices in Fig.~\ref{fig:Potential_nsr} for $N = 60$.
We plot $\lambda_4$ as a function of 
$\ab{\lambda_6/\lambda_4}$ for $\lambda_6/\lambda_4 < 0$ (left) and $\lambda_6/\lambda_4 > 0$ (right).
The lines are $\xi_2 = 1$ (blue), $10^2$ (red), and $10^4$ (green) for the metric case,
while $\xi_2 = 10^6$ (blue), $10^8$ (red), and $10^{10}$ (green) for the Palatini case.
}
\label{fig:Potential_lambda_60}
\end{center}
\end{figure}
%%%%%%%%%%

%%%%%%%%%%
\begin{figure}
\begin{center}
\small Metric ($N=60$)\smallskip\\
\fbox{
\begin{minipage}{0.37\textwidth}
\includegraphics[width=\columnwidth]{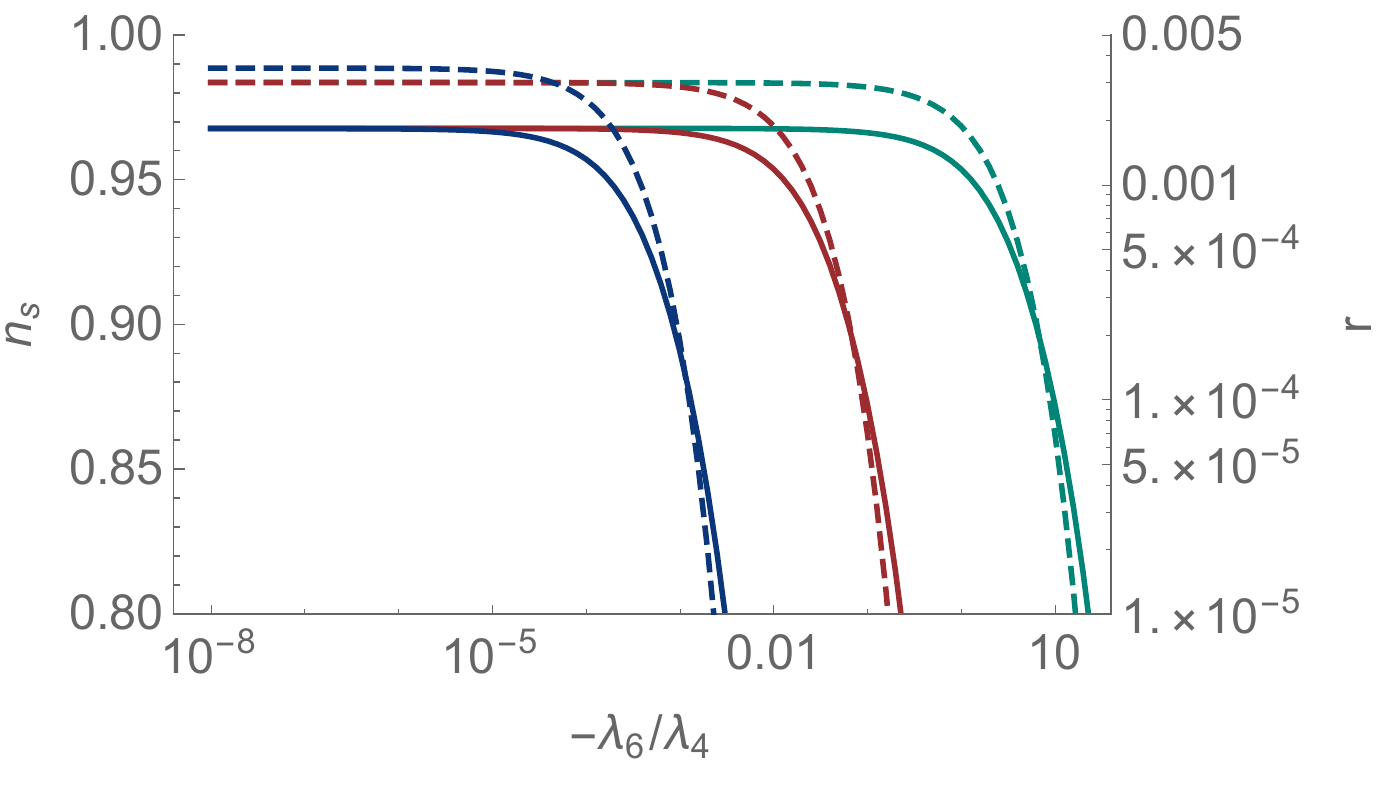}
\end{minipage}
\hskip 3mm
\begin{minipage}{0.37\textwidth}
\includegraphics[width=\columnwidth]{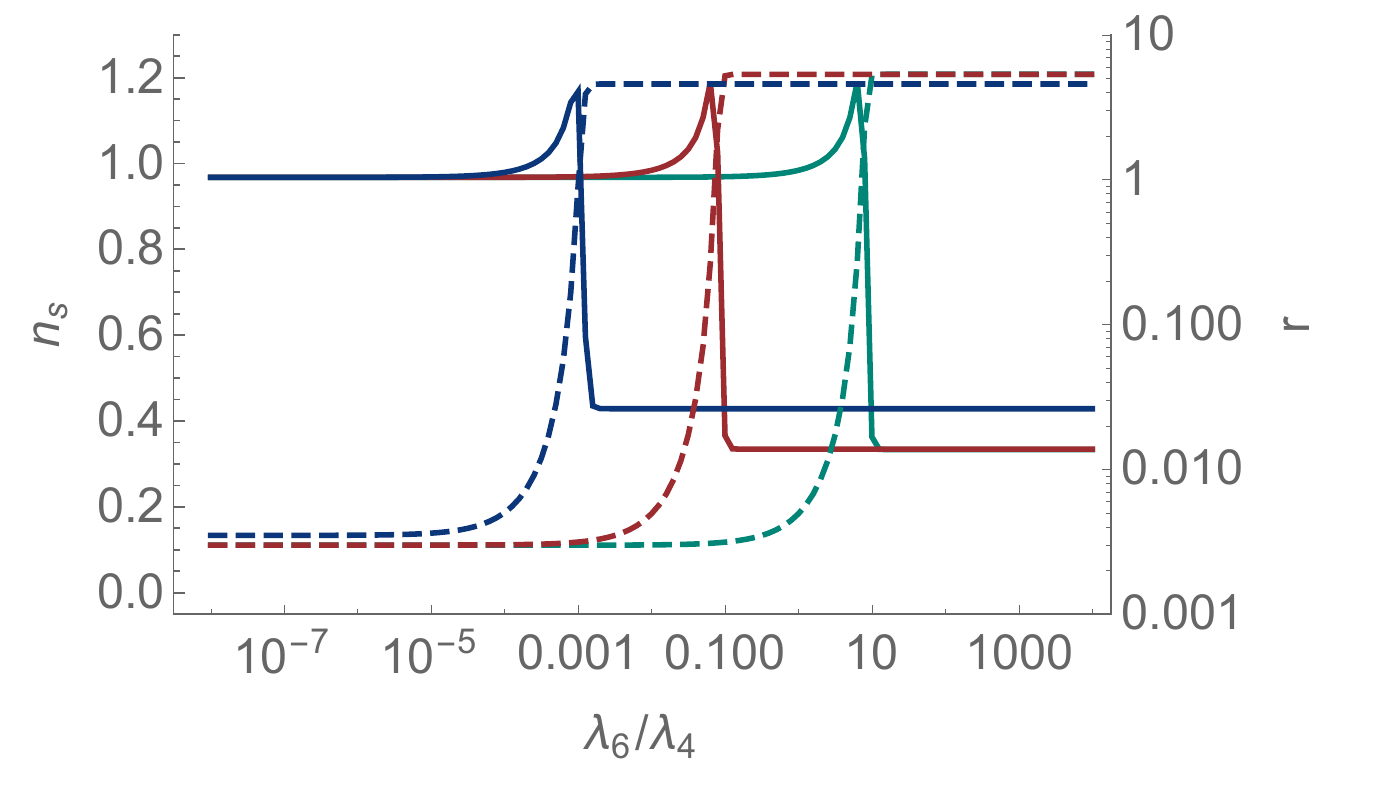}
\end{minipage}
\hskip 3mm
\begin{minipage}{0.1\textwidth}
\includegraphics[width=\columnwidth]{./figs/legPotentialMetric.pdf}
\end{minipage}
}
\end{center}
\begin{center}
\small Palatini ($N=60$)\smallskip\\
\fbox{
\begin{minipage}{0.37\textwidth}
\includegraphics[width=\columnwidth]{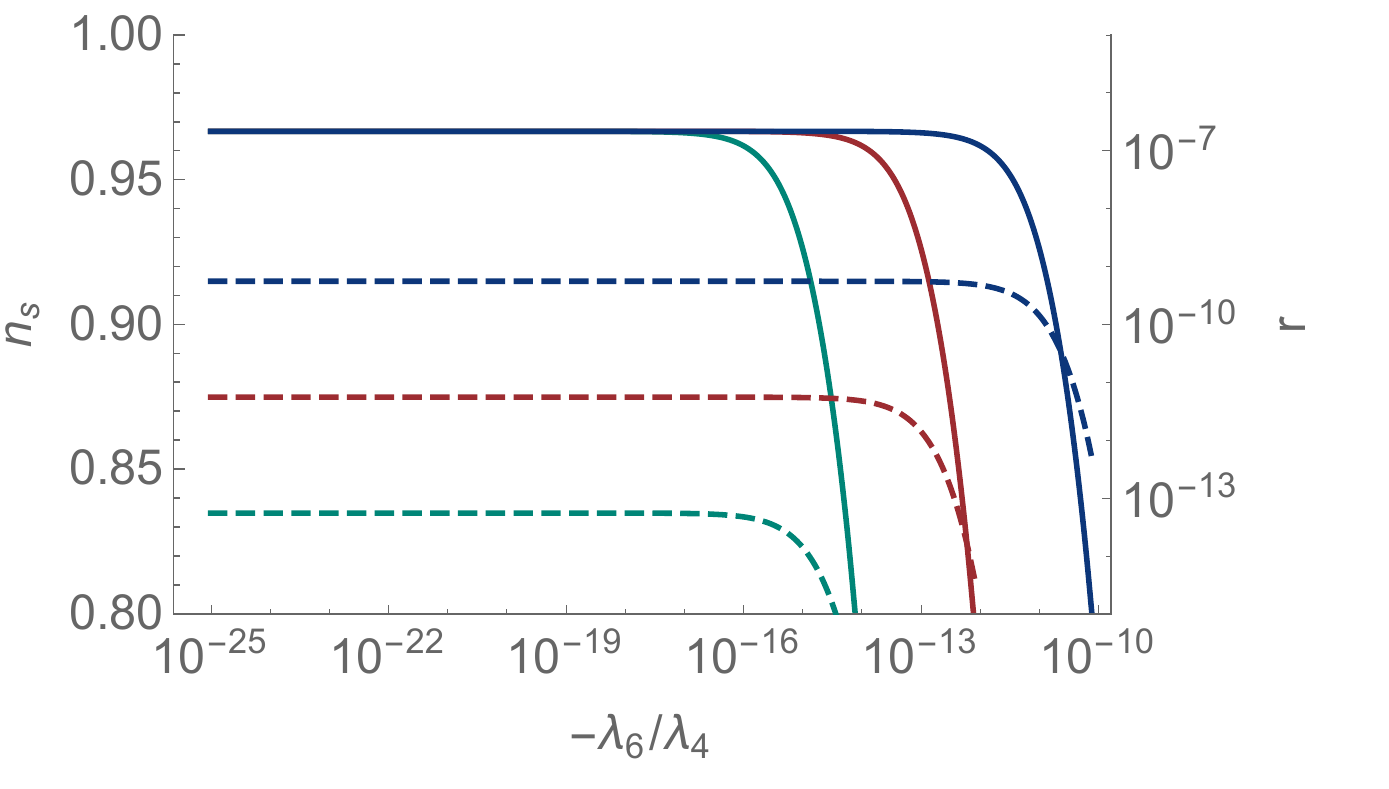}
\end{minipage}
\hskip 3mm
\begin{minipage}{0.37\textwidth}
\includegraphics[width=\columnwidth]{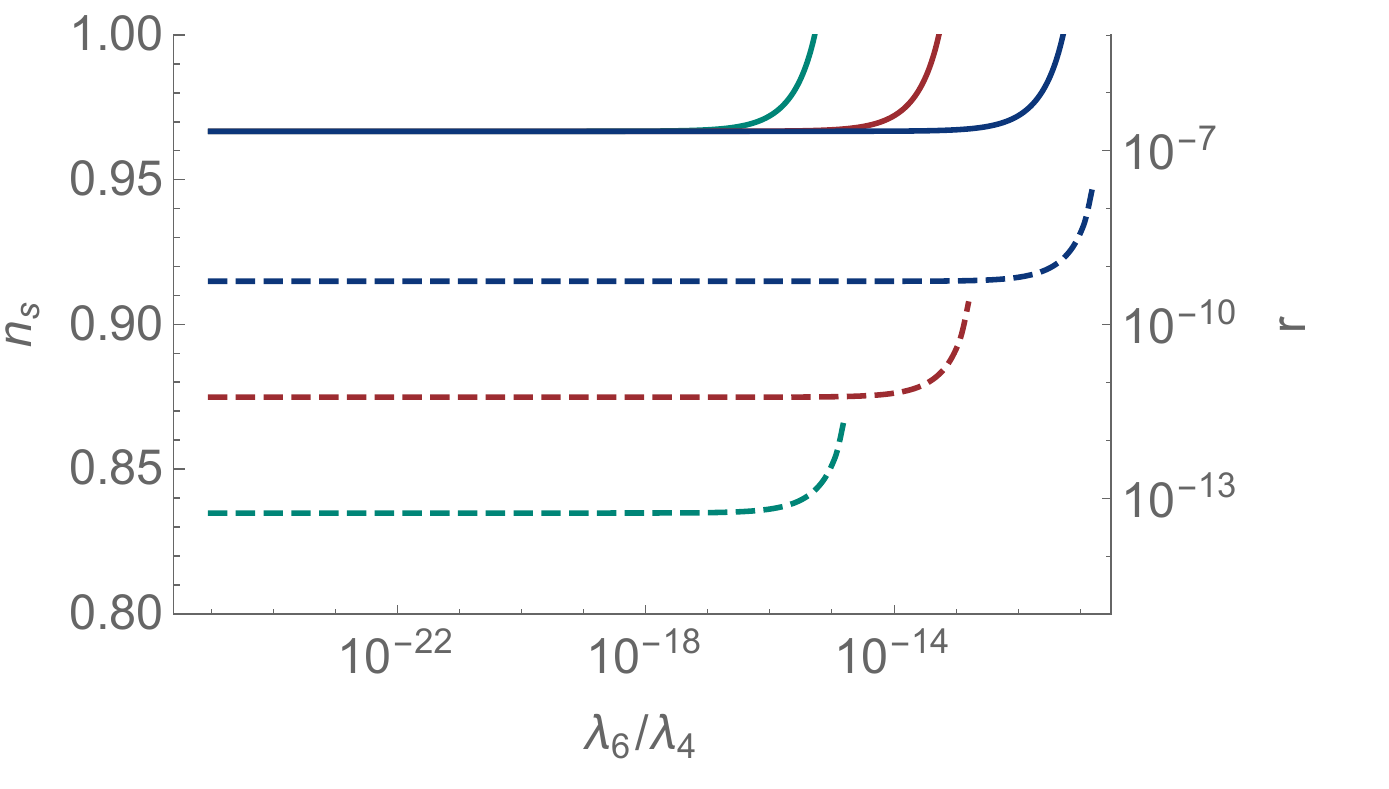}
\end{minipage}
\hskip 3mm
\begin{minipage}{0.1\textwidth}
\includegraphics[width=\columnwidth]{./figs/legPotentialPalatini.pdf}
\end{minipage}
}
\caption{\small
Behavior of $n_s$ (solid, left axis) and $r$ (dashed, right axis) along constant $\xi_2$ slices 
in Fig.~\ref{fig:Potential_nsr} for $N = 60$.
}
\label{fig:Potential_nsr_60}
\end{center}
\end{figure}
%%%%%%%%%%

\afterpage{\clearpage}

%%%%%%%%%%
\begin{figure}
\begin{center}
\small Metric ($N=50$)\smallskip\\
\fbox{
\begin{minipage}{0.32\textwidth}
\includegraphics[width=\columnwidth]{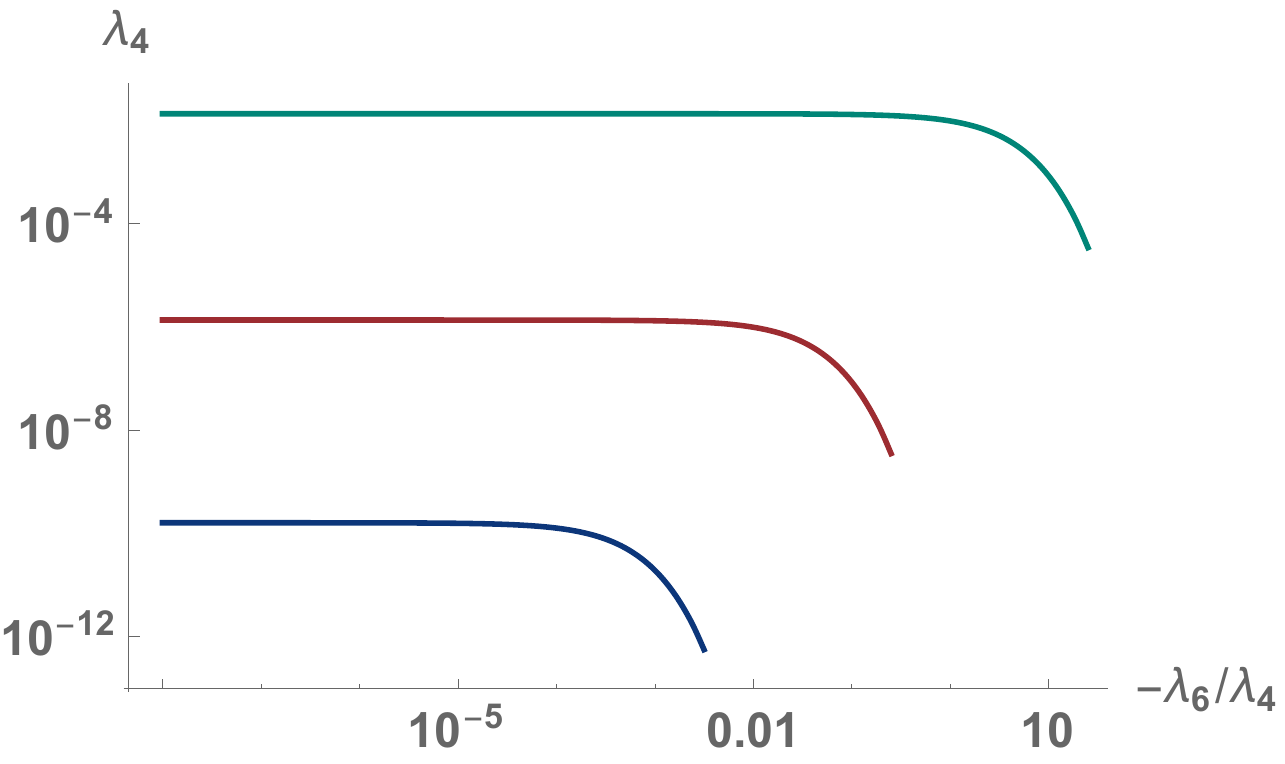}
\end{minipage}
\hskip 3mm
\begin{minipage}{0.32\textwidth}
\includegraphics[width=\columnwidth]{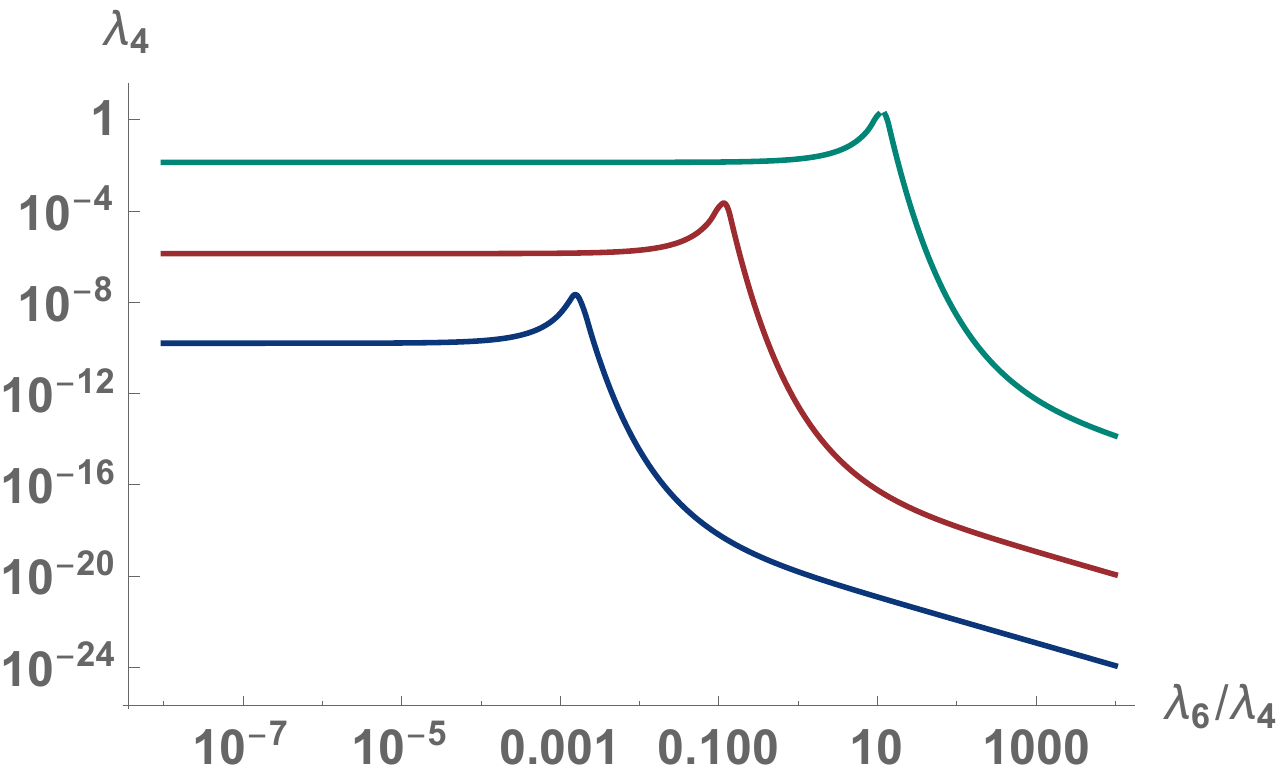}
\end{minipage}
\hskip 3mm
\begin{minipage}{0.1\textwidth}
\includegraphics[width=\columnwidth]{./figs/legPotentialMetric.pdf}
\end{minipage}
}
\end{center}
\begin{center}
\small Palatini ($N=50$)\smallskip\\
\fbox{
\begin{minipage}{0.32\textwidth}
\includegraphics[width=\columnwidth]{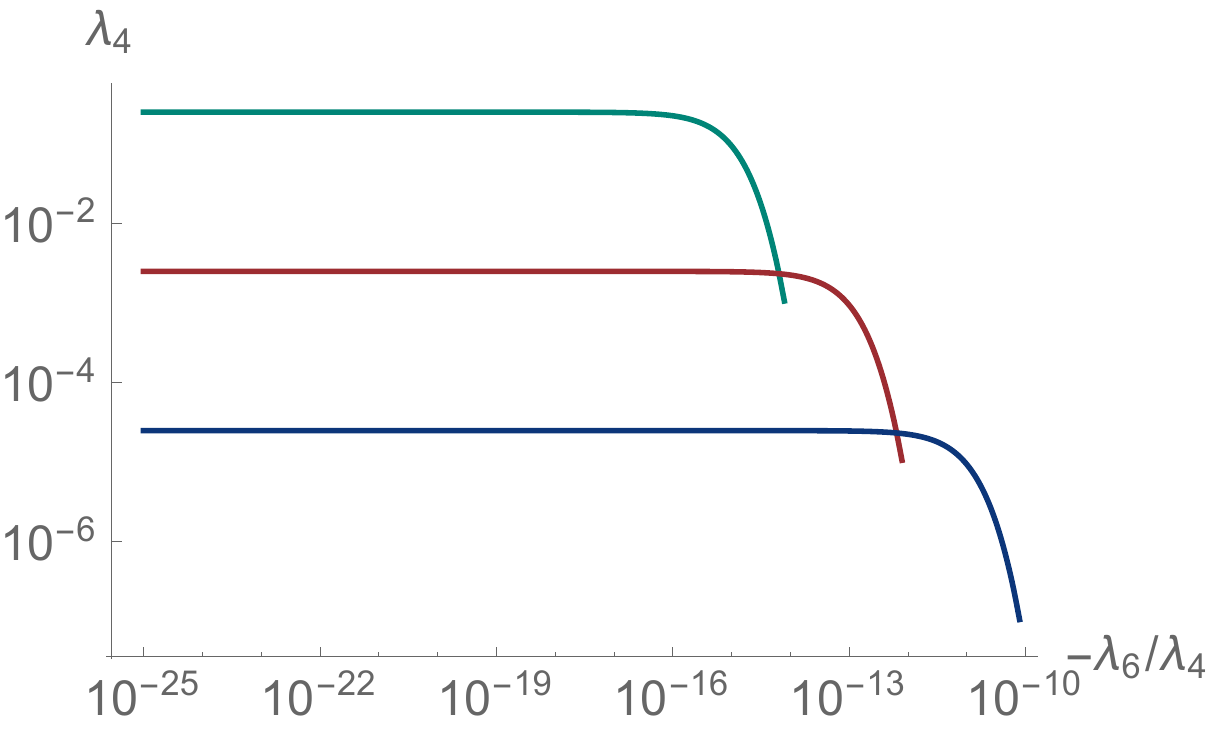}
\end{minipage}
\hskip 3mm
\begin{minipage}{0.32\textwidth}
\includegraphics[width=\columnwidth]{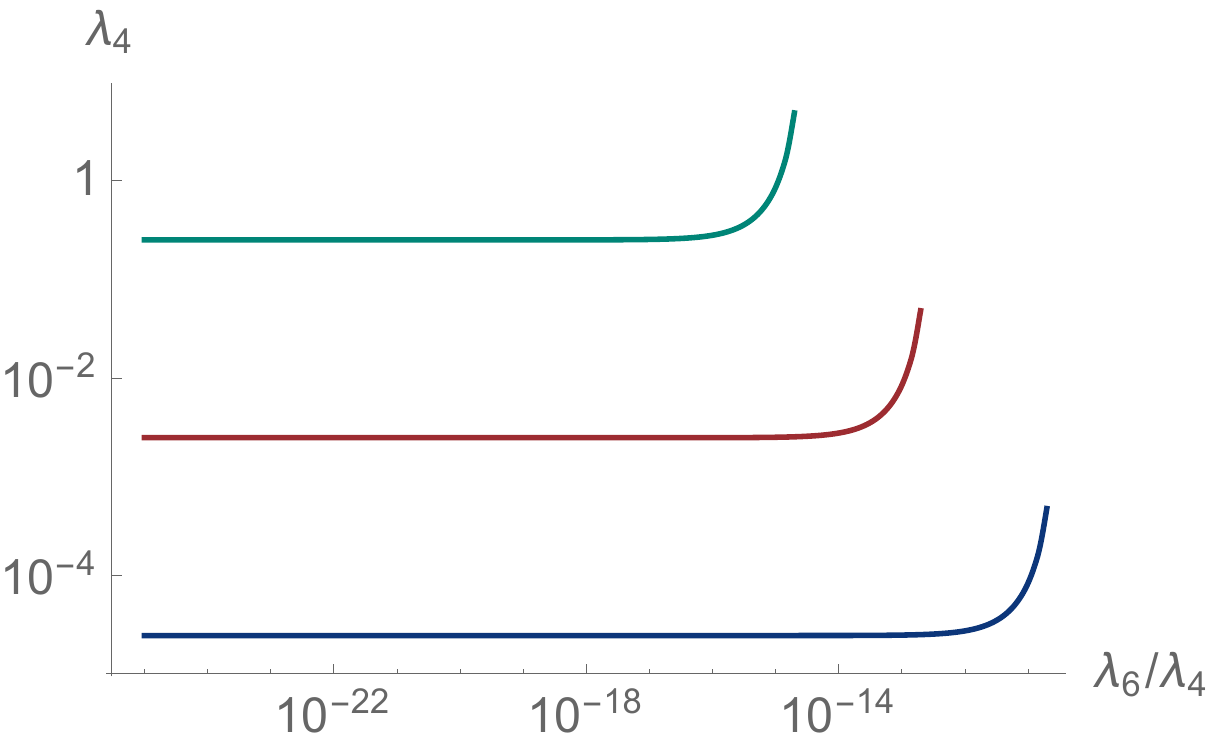}
\end{minipage}
\hskip 3mm
\begin{minipage}{0.1\textwidth}
\includegraphics[width=\columnwidth]{./figs/legPotentialPalatini.pdf}
\end{minipage}
}
\caption{\small
The same as in Fig.~\ref{fig:Potential_lambda_60} except for $N=50$.
}
\label{fig:Potential_lambda_50}
\end{center}
\end{figure}
%%%%%%%%%%

%%%%%%%%%%
\begin{figure}
\begin{center}
\small Metric ($N=50$)\smallskip\\
\fbox{
\begin{minipage}{0.37\textwidth}
\includegraphics[width=\columnwidth]{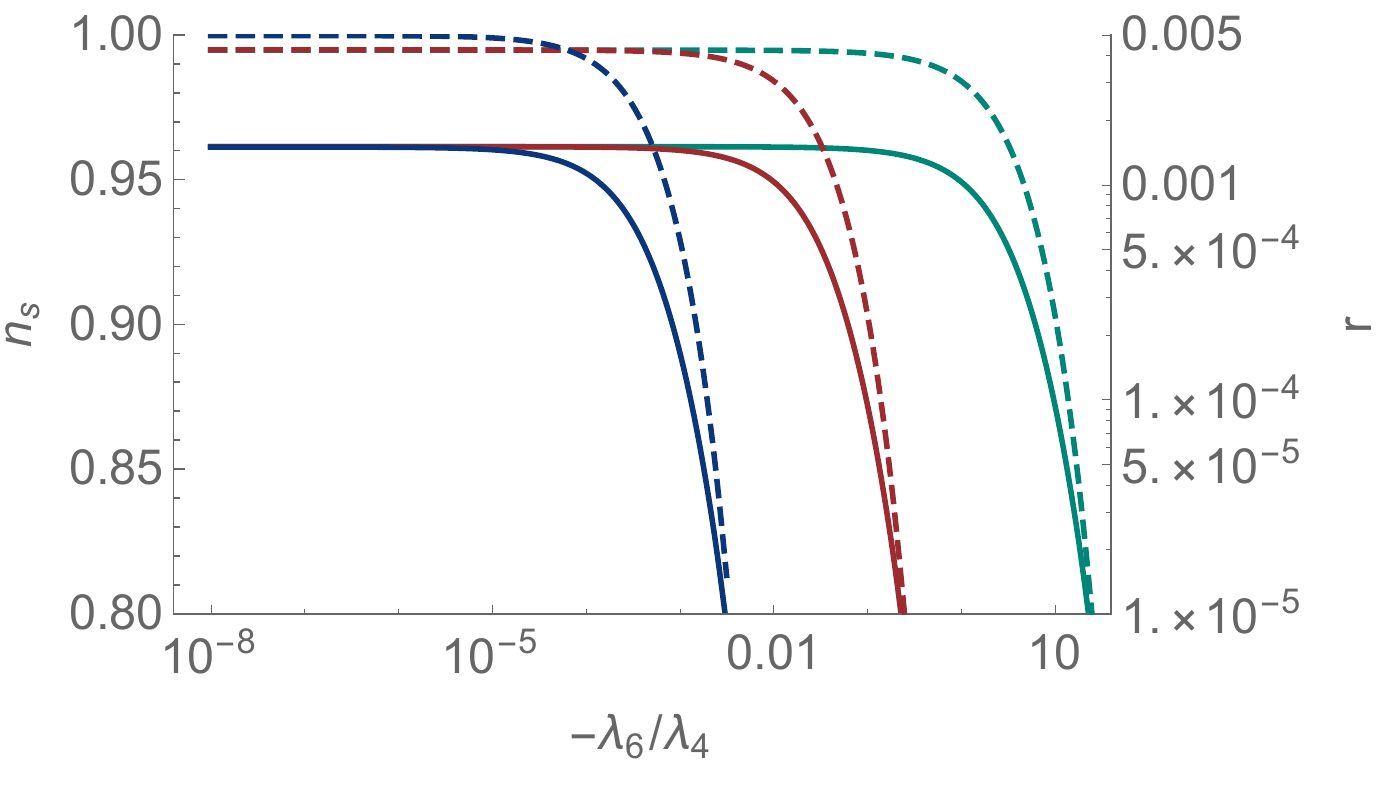}
\end{minipage}
\hskip 3mm
\begin{minipage}{0.37\textwidth}
\includegraphics[width=\columnwidth]{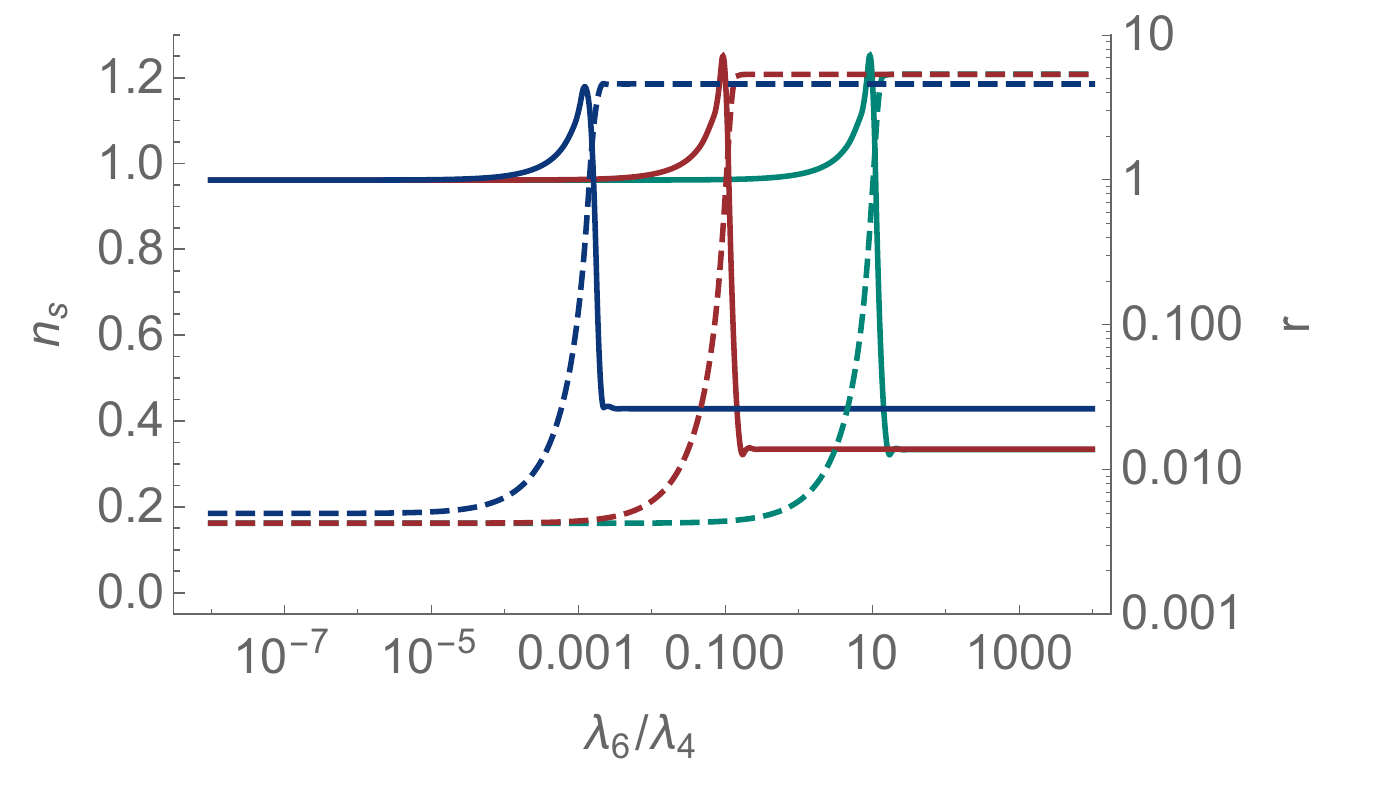}
\end{minipage}
\hskip 3mm
\begin{minipage}{0.1\textwidth}
\includegraphics[width=\columnwidth]{./figs/legPotentialMetric.pdf}
\end{minipage}
}
\end{center}
\begin{center}
\small Palatini ($N=50$)\smallskip\\
\fbox{
\begin{minipage}{0.37\textwidth}
\includegraphics[width=\columnwidth]{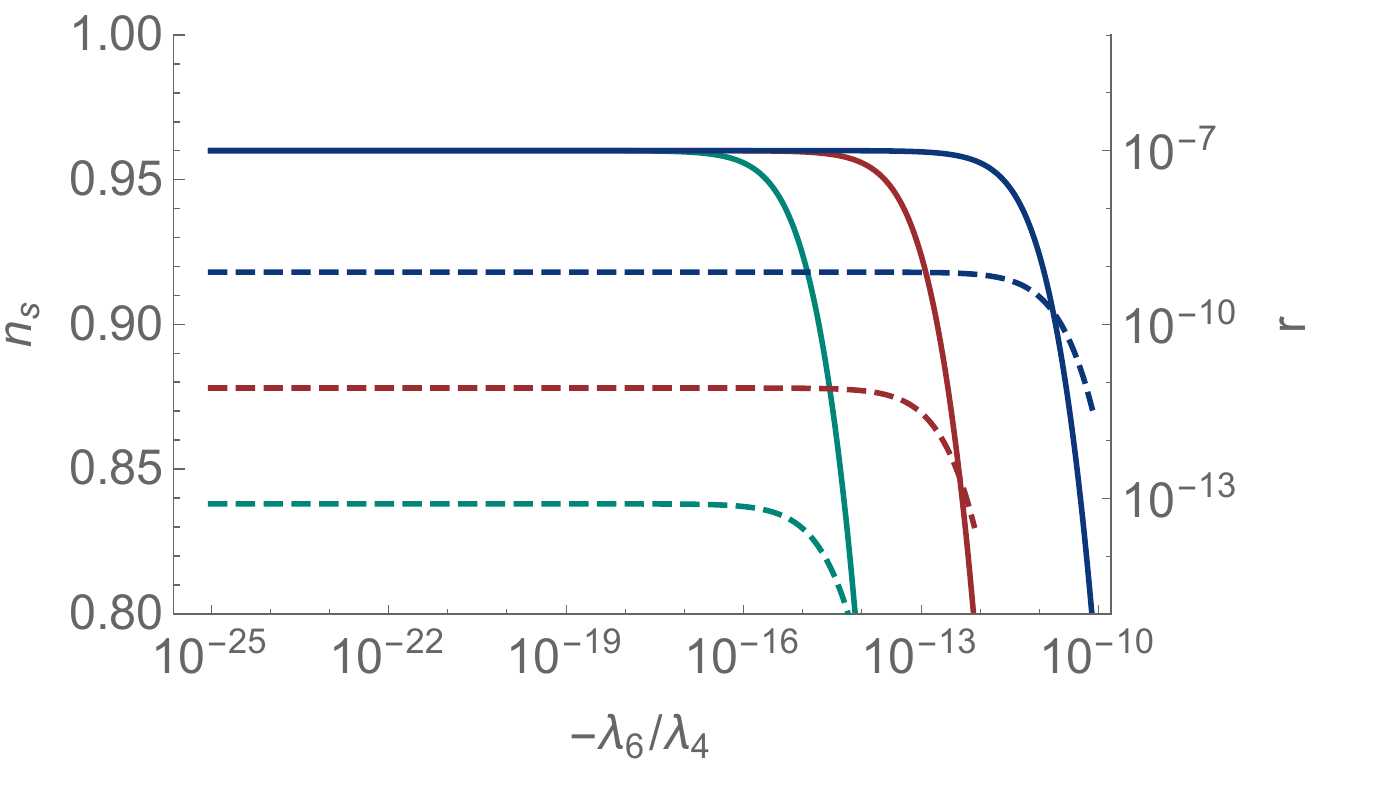}
\end{minipage}
\hskip 3mm
\begin{minipage}{0.37\textwidth}
\includegraphics[width=\columnwidth]{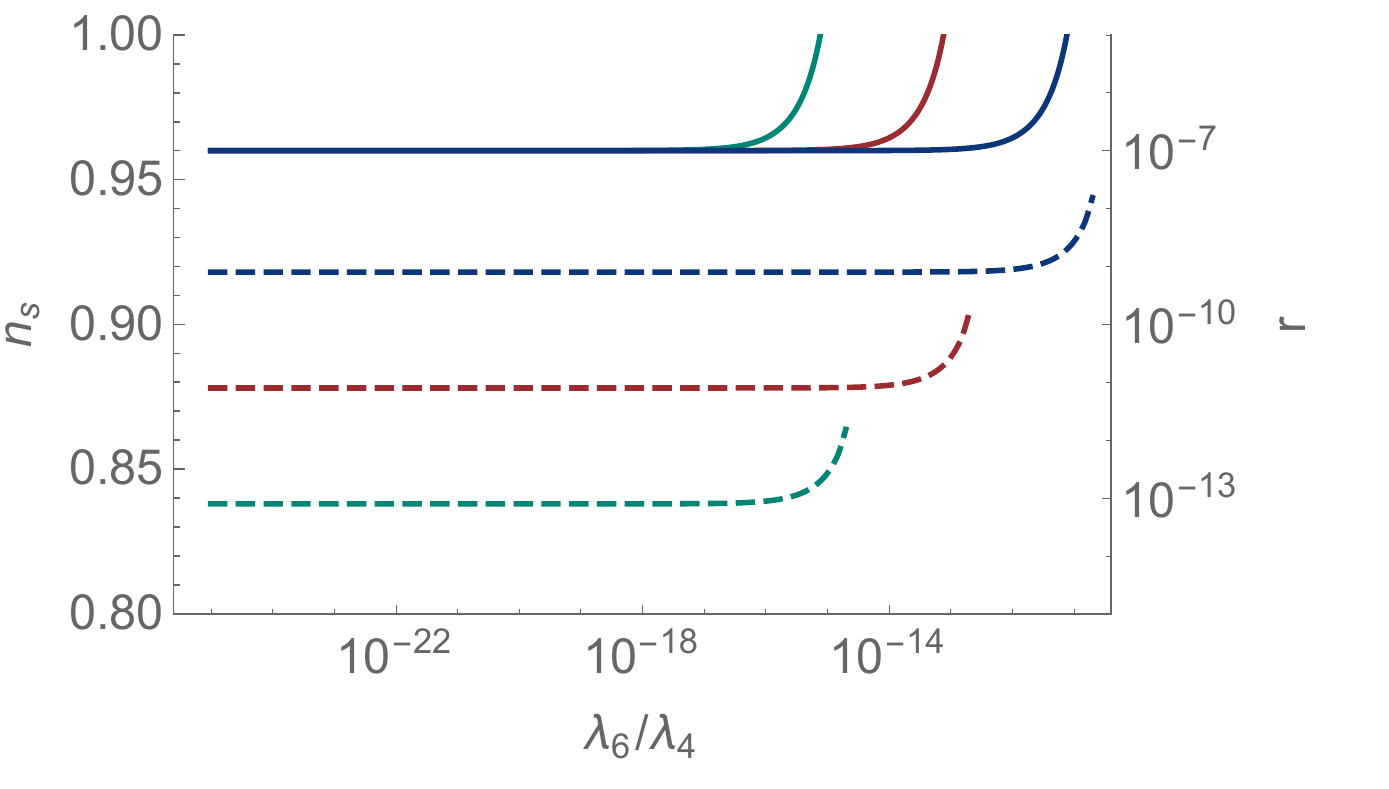}
\end{minipage}
\hskip 3mm
\begin{minipage}{0.1\textwidth}
\includegraphics[width=\columnwidth]{./figs/legPotentialPalatini.pdf}
\end{minipage}
}
\caption{\small
The same as in Fig.~\ref{fig:Potential_nsr_60} except for $N=50$.
}
\label{fig:Potential_nsr_50}
\end{center}
\end{figure}
%%%%%%%%%%

\afterpage{\clearpage}

%%%%%%%%%%%%%%%%%%%%%%%%%%%%%%%%%%%%%%%%%%%%%%%%%%
\subsection{Results}
\label{subsec:Potential_Results}
%%%%%%%%%%%%%%%%%%%%%%%%%%%%%%%%%%%%%%%%%%%%%%%%%%

First in Fig.~\ref{fig:Potential_nsr}, we show predictions of the metric (left) and Palatini (right) formalisms at $N=60$ (top) and 50 (bottom) in the $n_s$-$r$ plane.
The blue contours show constant values of $\xi_2$,
while the red ones show constant values of $\lambda_6/\lambda_4$.
The value of $\lambda_4$ is also shown as a density plot.
The smiley marker is the prediction of the quartic chaotic inflation,
while the star denotes the attractor point $\xi_2 \gg 1$.
For $\xi_2 < 10^{-3}$ and $\xi_2 > 1$, the corresponding blue lines are almost degenerate with 
the upper and lower boundaries with the gray regions, respectively.
We see the following:
\begin{itemize}
\item
For relatively small $\xi_2$ ($\lesssim 10^{-1}$),
even a small injection of $\ab{\lambda_6/\lambda_4}$ drastically changes the inflationary predictions
in both formalisms.
\item
For relatively large $\xi_2$ ($\gtrsim 1$),
the inflationary predictions are stable against the injection of $\lambda_6/\lambda_4$ for the metric formalism (left),
just as Sec.~\ref{sec:Nonminimal}.
On the other hand, the predictions are no more stable against the injection of 
$\ab{\lambda_6/\lambda_4} \gtrsim 10^{-5} \xi_2^{-1}$
for the Palatini formalism (right).
\end{itemize}

In Fig.~\ref{fig:Potential_lambda_60}, we plot the behavior of $\lambda_4$ for $N=60$ as a function of 
$\ab{\lambda_6/\lambda_4}$ for $\lambda_6/\lambda_4<0$ (left panel) and $\lambda_6/\lambda_4>0$ (right panel) 
while fixing the overall normalization $A_s$ as mentioned above.
Similarly, in Fig.~\ref{fig:Potential_nsr_60} we plot $n_s$ (left axis) and $r$ (right axis).
Figs.~\ref{fig:Potential_lambda_50} and \ref{fig:Potential_nsr_50} are the corresponding ones at $N=50$.
We note the following:
\begin{itemize}
\item
In the metric formalism (upper panels of Figs.~\ref{fig:Potential_lambda_60}--\ref{fig:Potential_nsr_50}), 
deviation from the attractor occurs at the threshold $\ab{\lambda_6/\lambda_4} \sim 10^{-3} \xi_2$. 
For $\lambda_6 < 0$
the behavior of $n_s$ and $r$ is relatively simple
(upper left panels of Figs.~\ref{fig:Potential_nsr_60} and \ref{fig:Potential_nsr_50}): 
as $\ab{\lambda_6/\lambda_4}$ increases, both $n_s$ and $r$ start to decrease at this threshold value.
From the relation $n_s \simeq 1 - 6\epsilon + 2\eta$ and $r \simeq 16\epsilon$,
we see that both $\epsilon$ and $\eta$ start to decrease at this threshold value
while keeping the relation $\ab{\epsilon} \ll \ab{\eta}$.
On the other hand, for $\lambda_6 > 0$
the behavior of $n_s$ and $r$ is more complicated
(upper right panels of Figs.~\ref{fig:Potential_nsr_60} and \ref{fig:Potential_nsr_50}):
as $\ab{\lambda_6/\lambda_4}$ increases, $r$ reaches to ${\mathcal O}(1)$ values 
while $n_s$ first increases and then decreases.
From the relation $n_s \simeq 1 - 6\epsilon + 2\eta$ and $r \simeq 16\epsilon$,
we see that as $\ab{\lambda_6/\lambda_4}$ increases 
$\epsilon$ reaches to ${\mathcal O}(0.1)$ values and exceeds $\eta$,
causing the decrease in $n_s$ observed in the upper right panels of 
Figs.~\ref{fig:Potential_nsr_60} and \ref{fig:Potential_nsr_50}.
\item
In the Palatini formalism (lower panels of Figs.~\ref{fig:Potential_lambda_60}--\ref{fig:Potential_nsr_50}), 
there is no attractor for $\xi_2 \gg 1$ and the deviation from the observationally allowed region occurs around
the threshold $\ab{\lambda_6/\lambda_4} \sim 10^{-5} \xi_2^{-1}$.
\end{itemize}
%%

%%%%%%%%%%%%%%%%%%%%%%%%%%%%%%%%%%%%%%%%%%%%%%%%%%
\subsection{Interpretation}
%%%%%%%%%%%%%%%%%%%%%%%%%%%%%%%%%%%%%%%%%%%%%%%%%%

Let us interpret our results and estimate the threshold value of $\lambda_6$ 
which causes significant deviation from the observationally allowed region.
In the following we use $N \gg 1$ and $\xi_2 \gg 1$ and keep the leading contribution
for each order in $\xi_4$ expansion when necessary.

In the metric formalism, let us first expand Eq.~(\ref{eq:Potential_Metric_N}) by small $\lambda_6/\lambda_4$
around $\phi \simeq \phi_{\lambda_6 = 0} \simeq \sqrt{4N/3\xi_2}$ (see Eq.~\eqref{eq:Metric_phiN}):
\begin{align}
N
&=
\int d\phi
~
\frac{\phi \cdot 6 \xi_2^2 \phi^2 \cdot \paren{ \lambda_4 + \lambda_6 \phi^2}}
{2 \cdot \xi_2 \phi^2 \cdot \paren{2 \lambda_4 + \xi_2 \lambda_6 \phi^4}}
+ \cdots
=
\frac{3}{4} \xi_2 \phi^2 
- \frac{1}{8} \frac{\lambda_6}{\lambda_4} \xi_2^2 \phi^6
+ \cdots.
\end{align}
Substituting $\phi = \phi_{\lambda_6 = 0}(1 + c (\lambda_6 / \lambda_4))$
and comparing leading terms in $\lambda_6 / \lambda_4$,
we find that the deviation of $\phi$ is given by
\begin{align}
\phi
&\simeq
\phi_{\lambda_6 = 0}
\left(
1 + \frac{4N^2}{27\xi_2} \frac{\lambda_6}{\lambda_4}
\right).
\label{eq:Potential_phidev_Metric}
\end{align}
We next expand $\epsilon$~\eqref{eq:Potential_Metric_eps} 
and $\eta$~\eqref{eq:Potential_Metric_eta} by small $\lambda_6/\lambda_4$ around the same point of $\phi$,
and then substitute Eq.~(\ref{eq:Potential_phidev_Metric}):
\begin{align}
\epsilon
&\simeq
\frac{4}{3\xi_2^2 \phi^4}
\left(
1 + \frac{\lambda_6}{\lambda_4} \xi_2 \phi^4
\right)
\simeq 
\frac{3}{4N^2}
\left(
1 + \frac{32N^2}{27\xi_2} \frac{\lambda_6}{\lambda_4}
\right),
\\
\eta
&\simeq
- \frac{4}{3 \xi_2 \phi^2}
\left(
1 - \frac{1}{2} \frac{\lambda_6}{\lambda_4} \xi_2 \phi^4
\right)
\simeq 
- \frac{1}{N}
\left(
1 - \frac{32N^2}{27\xi_2} \frac{\lambda_6}{\lambda_4}
\right).
\end{align}
We see that the deviation of $\eta$ (and hence of $n_s$) becomes sizable for 
$\ab{\lambda_6/\lambda_4}\gtrsim \xi_2/N^2$.
From the relation $n_s \simeq 1 - 6\epsilon + 2\eta$ and $r \simeq 16 \epsilon$,
the direction of deviation in Fig.~\ref{fig:Potential_nsr} for positive and negative $\lambda_6$ with a fixed $\xi_2$
can also be explained.

In the Palatini formalism, let us first expand Eq.~(\ref{eq:Potential_Palatini_N}) by small $\lambda_6/\lambda_4$
around $\phi \simeq \phi_{\lambda_6 = 0} \simeq \sqrt{8N}$ (see Eq.~\eqref{eq:Palatini_phiN}):
\begin{align}
N
&=
\int d\phi
~
\frac{\phi \cdot \paren{ \lambda_4 + \lambda_6 \phi^2}}{2 \cdot \paren{2 \lambda_4 + \xi_2 \lambda_6 \phi^4}}
+ \cdots
=
\frac{1}{8} \phi^2
- \frac{1}{48} \frac{\lambda_6}{\lambda_4} \xi_2 \phi^6
+ \cdots.
\end{align}
Substituting $\phi = \phi_{\lambda_6 = 0}(1 + c (\lambda_6 / \lambda_4))$
and comparing leading terms in $\lambda_6 / \lambda_4$, we find
\begin{align}
\phi
&\simeq
\phi_{\lambda_6 = 0}
\left(
1 + \frac{16N^2 \xi_2}{3} \frac{\lambda_6}{\lambda_4}
\right).
\label{eq:Potential_phidev_Palatini}
\end{align}
We next expand $\epsilon$~\eqref{eq:Potential_Palatini_eps} 
and $\eta$~\eqref{eq:Potential_Palatini_eta} by small $\lambda_6/\lambda_4$ around the same point of $\phi$,
and then substitute Eq.~(\ref{eq:Potential_phidev_Palatini}):
\begin{align}
\epsilon
&\simeq
\frac{8}{\xi_2 \phi^4}
\left(
1 + \frac{\lambda_6}{\lambda_4} \xi_2 \phi^4
\right)
\simeq 
\frac{1}{8N^2 \xi_2}
\left(
1 + \frac{128N^2 \xi_2}{3} \frac{\lambda_6}{\lambda_4}
\right),
\\
\eta
&\simeq
- \frac{8}{\phi^2}
\left(
1 - \frac{1}{2} \frac{\lambda_6}{\lambda_4} \xi_2 \phi^4
\right)
\simeq 
- \frac{1}{N}
\left(
1 - \frac{128N^2 \xi_2}{3} \frac{\lambda_6}{\lambda_4}
\right).
\end{align}
We see that the deviation of $\eta$ (and hence of $n_s$) becomes sizable for $\ab{\lambda_6/\lambda_4}\gtrsim 10^{-1}/N^2\xi_2$ 
and that the direction of the deviation is the same as in the metric formalism.

To summarize, the deviation of $\eta$ (and hence of $n_s$) becomes sizable for
\begin{align}
\ab{\lambda_6}
&\gtrsim
\begin{cases}
\displaystyle
{\lambda_4\xi_2\ov N^2}
&~~
\tx{(metric)},
\\[3mm]
\displaystyle 
10^{-1}{\lambda_4\ov N^2\xi_2}
&~~
\tx{(Palatini)},
\end{cases}
\label{eq:lambda6_threshold}
\end{align}
while $\epsilon$ (and hence $r$) is rather insensitive compared to $\eta$. 
If we substitute $\lambda_4$ with $A_s$, we get
\begin{align}
\ab{\lambda_6}
&\gtrsim
\begin{cases}
\displaystyle
10^2{A_s\xi_2^3\ov N^4}\sim 10^{-6}{\xi_2^3\ov N^4}
&~~
\tx{(metric)},
\\[3mm]
\displaystyle
{A_s\ov N^4}\sim 10^{-9}{1\ov N^4}
&~~
\tx{(Palatini)}.
\end{cases}
\end{align}
We see that near the attractor $\xi_2\gg1$ the metric formalism is stable against injection of $\lambda_6$ (unless $\xi_2\gg10^5$), while it is extremely sensitive to even tiny injection of $\lambda_6$ of order $10^{-16}$ in the Palatini formalism, regardless of $\xi_2$.

As we increase $\ab{\lambda_6}$ from zero, the inflationary prediction starts to deviate at around the value in the right-hand side of Eq.~\eqref{eq:lambda6_threshold}. We note that this occurs when the higher dimensional operator $\lambda_6\phi^6$ is still much smaller than the lower dimensional one $\lambda_4\phi^4$. Indeed, by substituting the approximate values $\phi\simeq\sqrt{4N/3\xi_2}$ (metric) and $\sqrt{8N}$ (Palatini),
the condition
\begin{align}
\ab{\lambda_6\phi^6}
&\lesssim 
\lambda_4\phi^4
\end{align}
becomes
\begin{align}
\ab{\lambda_6}
&\lesssim 
\begin{cases}
\displaystyle
{\lambda_4\xi_2\ov N}
&~~
\tx{(metric)},
\\[3mm]
\displaystyle
10^{-1}{\lambda_4\ov N}
&~~
\tx{(Palatini)},
\end{cases}
\end{align}
which is well satisfied at the value in the right-hand side of Eq.~\eqref{eq:lambda6_threshold}.

%%%%%%%%%%%%%%%%%%%%%%%%%%%%%%%%%%%%%%%%%%%%%%%%%%
\section{Summary and discussion}
\setcounter{equation}{0}
\label{sec:DC}
%%%%%%%%%%%%%%%%%%%%%%%%%%%%%%%%%%%%%%%%%%%%%%%%%%

The Higgs(-like) inflation with the quartic coupling $\lambda_4$ and the non-minimal coupling $\xi_2$ is 
one of the best models from the viewpoint of cosmic microwave background observations.
We have investigated the sensitivity of this model to higher dimensional operators 
in the Weyl rescaling factor ($\xi_4\phi^4$) or in the potential ($\lambda_6\phi^6$) 
both in the metric and Palatini formalisms.

We have found that, in the metric formalism,
injection of order $\ab{\xi_4}\sim 10^{-6}$ or $\ab{\lambda_6}\sim 10^{-16}$
makes the value of $n_s$ out of the observationally allowed region for $\xi_2\lesssim 1$,
while the inflationary predictions are relatively stable against $\lambda_6$ or $\xi_4$
for $\xi_2 \gg 1$ because of the existence of attractor.
On the other hand, in the Palatini formalism, we have found that large $\xi_2 \gg 1$ does not help the stability:
injection of order $\ab{\xi_4}\sim 10^{-6}$ or $\ab{\lambda_6}\sim 10^{-16}$ spoils 
the successful inflationary predictions regardless of the value of $\xi_2$.
We have also pointed out that these threshold values cannot be estimated from
a na\"ive comparison between $\xi_2 \phi^2$ and $\xi_4 \phi^4$ or 
between $\lambda_4 \phi^4$ and $\lambda_6 \phi^6$ at the $\phi$ value corresponding to
$e$-folding $N \sim 50 - 60$.
Our study underscores the theoretical challenge in realizing inflationary models 
with the nonminimal coupling in the Palatini formalism.

Our result shows that large $\xi_2$ is necessary not only for the realization of inflation itself 
but also for the robustness of the prediction against the injection of the tiny higher dimensional operators 
in the metric formalism.
This is interesting from theoretical point of view 
because it is unlikely to have such a large $\xi_2$ only in a single term in the field expansion 
in the low energy effective field theory.
This may indicate a principle beyond.\footnote{
See e.g.\ Ref.~\cite{odakinTalk}.
}\footnote{
Even in the vanilla model with only $\lambda_4$ and $\xi_2$, it is theoretically intriguing why we have $\lambda_2\ll\lambda_4$ and $1\ll\xi_2$; the former is nothing but the hierarchy problem for the Higgs mass-squred.
}

It is known that the radiative corrections to $\lambda_4$ becomes important for $\lambda_4\ll1$, 
namely for the critical Higgs inflation. This will be studied in a separate publication.

%%%%%%%%%%%%%%%%%%%%%%%%%%%%%%%%%%%%%%%%%%%%%%%%%%%%%%%
\subsection*{Acknowledgments}
%%%%%%%%%%%%%%%%%%%%%%%%%%%%%%%%%%%%%%%%%%%%%%%%%%%%%%%

The authors are grateful to Satoshi Yamaguchi for useful comments.
The work of RJ was supported by Grants-in-Aid for JSPS Overseas Research Fellow (No. 201960698).
The work of RJ was supported by IBS under the project code, IBS-R018-D1.
The work of SP was supported in part by the National Research Foundation of Korea (NRF) grant 
funded by the Korean government (MSIP) (NRF-2018R1A4A1025334) and (NRF-2019R1A2C1089334).

\appendix

%%%%%%%%%%%%%%%%%%%%%%%%%%%%%%%%%%%%%%%%%%%%%%%%%%
\section{Equations}
\setcounter{equation}{0}
\label{app:Equations}
%%%%%%%%%%%%%%%%%%%%%%%%%%%%%%%%%%%%%%%%%%%%%%%%%%

In this appendix we summarize expressions for the $\phi$-$\chi$ relations,
slow-roll parameters, and $e$-folding.
We consider $\xi_2 > 0$, $\lambda_4 > 0$, and $\phi > 0$ for Appendix~\ref{subapp:Nonminimal},
while $\xi_2 > 0$, $\lambda_4 > 0$, and $\phi > 0$ for Appendix~\ref{subapp:Potential}.

%%%%%%%%%%%%%%%%%%%%%%%%%%%%%%%%%%%%%%%%%%%%%%%%%%
\subsection{Correction to the nonminimal coupling}
\label{subapp:Nonminimal}
%%%%%%%%%%%%%%%%%%%%%%%%%%%%%%%%%%%%%%%%%%%%%%%%%%

%%%%%%%%%%%%%%%%%%%%%%%%%%%%%%%%%%%%%%%%%%%%%%%%%%
\subsubsection*{Metric formalism}
%%%%%%%%%%%%%%%%%%%%%%%%%%%%%%%%%%%%%%%%%%%%%%%%%%

%%

$\phi$-$\chi$ relation:
\begin{align}
\frac{d\chi}{d\phi}
&=
\frac{\sqrt{1 + \xi_2 \phi^2 + \xi_4 \phi^4 + 6\paren{\xi_2 \phi + 2\xi_4 \phi^3}^2}}{1 + \xi_2 \phi^2 + \xi_4 \phi^4}.
\end{align}
Slow-roll parameters and $e$-folding:
\begin{align}
\epsilon
&=
\frac{8(1 - \xi_4 \phi^4)^2}
{\phi^2 \left[ 1 + \paren{\xi_2 + 6\xi_2^2} \phi^2 + \paren{\xi_4 + 24\xi_2 \xi_4} \phi^4 + 24\xi_4^2 \phi^6 \right]},
\label{eq:app_Nonminimal_Metric_eps}
\\[0.6ex]
\eta
&= 
\frac{
4
\left[
\begin{matrix}
3 + \paren{\xi_2 + 12 \xi_2^2} \phi^2
+ (- 2 \xi_2^2 - 12 \xi_2^3 + 24 \xi_2 \xi_4 - 11 \xi_4) \phi^4 
\\[0.6ex]
- \paren{18 \xi_2 \xi_4 + 144 \xi_2^2 \xi_4} \phi^6
- \paren{2 \xi_2^2 \xi_4 + 12 \xi_2^3 \xi_4 + 11 \xi_4^2 + 384 \xi_2 \xi_4^2} \phi^8 
\\[0.6ex]
+ \paren{\xi_2 \xi_4^2 - 12 \xi_2^2 \xi_4^2 - 288 \xi_4^3} \phi^{10}
+ \paren{3 \xi_4^3 + 72 \xi_2 \xi_4^3} \phi^{12}
+ 96 \xi_4^4 \phi^{14}
\end{matrix}
\right]
}
{\phi^2 \left[1 + \paren{\xi_2 + 6 \xi_2^2} \phi^2 + \paren{\xi_4 + 24 \xi_2 \xi_4} \phi^4 + 24 \xi_4^2 \phi^6\right]^2},
\label{eq:app_Nonminimal_Metric_eta}
\\[0.6ex]
N
&= 
\int d\phi
~
\frac{\phi \left[ 1 + \paren{\xi_2 + 6\xi_2^2} \phi^2 + \paren{\xi_4 + 24 \xi_2 \xi_4} \phi^4 + 24 \xi_4^2 \phi^6 \right]}
{4(1 - \xi_4 \phi^4)(1 + \xi_2 \phi^2 + \xi_4 \phi^4)}
\nonumber \\
&=
\left\{
\begin{matrix}
\displaystyle
\frac{1 + 6\xi_2}{8\sqrt{\xi_4}}
~
\arctanh
\left[
\sqrt{\xi_4}\phi^2
\right]
-
\frac{3}{4}
\ln
\left[
\left(
1 - \xi_4 \phi^4
\right)
\left(
1 + \xi_2 \phi^2 + \xi_4 \phi^4
\right)
\right]
\\[2.4ex]
=
\displaystyle
\frac{1 + 6\xi_2}{16\sqrt{\xi_4}}
\ln
\left[
\frac{1 + \sqrt{\xi_4}\phi^2}{1 - \sqrt{\xi_4}\phi^2}
\right]
-
\frac{3}{4}
\ln
\left[
\left(
1 - \xi_4 \phi^4
\right)
\left(
1 + \xi_2 \phi^2 + \xi_4 \phi^4
\right)
\right]
&~~~~
(\xi_4 > 0~\&~\phi < \xi_4^{1/4}),
\\[2.4ex]
\displaystyle
\frac{1 + 6\xi_2}{8\sqrt{-\xi_4}}
~
{\rm arctan}
\left[
\sqrt{-\xi_4}\phi^2
\right]
-
\frac{3}{4}
\ln
\left[
\left(
1 - \xi_4 \phi^4
\right)
\left(
1 + \xi_2 \phi^2 + \xi_4 \phi^4
\right)
\right]
\\[2.4ex]
=
\displaystyle
\frac{1 + 6\xi_2}{16\sqrt{\xi_4}}
\ln
\left[
\frac{1 + \sqrt{\xi_4}\phi^2}{1 - \sqrt{\xi_4}\phi^2}
\right]
-
\frac{3}{4}
\ln
\left[
\left(
1 - \xi_4 \phi^4
\right)
\left(
1 + \xi_2 \phi^2 + \xi_4 \phi^4
\right)
\right]
&~~~~
(\xi_4 < 0).
\end{matrix}
\right.
\label{eq:app_Nonminimal_Metric_N}
\end{align}
%%

%%

%%%%%%%%%%%%%%%%%%%%%%%%%%%%%%%%%%%%%%%%%%%%%%%%%%
\subsubsection*{Palatini formalism}
%%%%%%%%%%%%%%%%%%%%%%%%%%%%%%%%%%%%%%%%%%%%%%%%%%

%%
$\phi$-$\chi$ relation:
\begin{align}
\frac{d\chi}{d\phi}
&=
\frac{1}{\sqrt{1 + \xi_2 \phi^2 + \xi_4 \phi^4}},
\\[0.6ex]
\chi
&= 
- i 
\sqrt{\frac{\xi_2 + \sqrt{\xi_2^2 - 4\xi_4}}{2\xi_4}}
~
F\fn{
i~\arcsinh
\left[
\sqrt{\frac{2\xi_4}{\xi_2 + \sqrt{\xi_2^2 - 4\xi_4}}}
~\phi
\right]
\,\Bigg|\,
- 1 
+ \frac{\xi_2 (\xi_2 + \sqrt{\xi_2^2 - 4\xi_4})}{2\xi_4}
},
\end{align}
where $F\fn{\varphi\,|\,m}$ is the elliptic integral of the first kind.
\\
Slow-roll parameters and $e$-folding:
\begin{align}
\epsilon
&= 
\frac{8(1 - \xi_4 \phi^4)^2}{\phi^2 (1 + \xi_2 \phi^2 + \xi_4 \phi^4)},
\label{eq:app_Nonminimal_Palatini_eps}
\\[0.6ex]
\eta
&= 
\frac{4(3 - 2\xi_2 \phi^2 - 14 \xi_4 \phi^4 - 2\xi_2 \xi_4 \phi^6 + 3 \xi_4^2 \phi^8)}
{\phi^2 (1 + \xi_2 \phi^2 + \xi_4 \phi^4)},
\label{eq:app_Nonminimal_Palatini_eta}
\\[0.6ex]
N
&= 
\int d\phi
~
\frac{\phi}{4(1 - \xi_4 \phi^4)}
\nonumber \\
&= 
\left\{
\begin{matrix}
\displaystyle
\frac{1}{8\sqrt{\xi_4}} 
~
\arctanh
\left[
\sqrt{\xi_4}\phi^2
\right]
=
\frac{1}{16\sqrt{\xi_4}} \ln \left[
\frac{1 + \sqrt{\xi_4}\phi^2}{1 - \sqrt{\xi_4}\phi^2}
\right]
&~~~~
(\xi_4 > 0~\&~\phi < \xi_4^{1/4}),
\\[2.4ex]
\displaystyle
\frac{1}{8\sqrt{-\xi_4}} 
~
{\rm arctan}
\left[
\sqrt{-\xi_4}\phi^2
\right]
=
\frac{1}{16\sqrt{\xi_4}} \ln \left[
\frac{1 + \sqrt{\xi_4}\phi^2}{1 - \sqrt{\xi_4}\phi^2}
\right]
&~~~~
(\xi_4 < 0).
\end{matrix}
\right.
\label{eq:app_Nonminimal_Palatini_N}
\end{align}
%%

%%

%%%%%%%%%%%%%%%%%%%%%%%%%%%%%%%%%%%%%%%%%%%%%%%%%%
\subsection{Correction to the potential}
\label{subapp:Potential}
%%%%%%%%%%%%%%%%%%%%%%%%%%%%%%%%%%%%%%%%%%%%%%%%%%

%%%%%%%%%%%%%%%%%%%%%%%%%%%%%%%%%%%%%%%%%%%%%%%%%%
\subsubsection*{Metric formalism}
%%%%%%%%%%%%%%%%%%%%%%%%%%%%%%%%%%%%%%%%%%%%%%%%%%

%%
$\phi$-$\chi$ relation:
\begin{align}
\frac{d\chi}{d\phi}
&=
\frac{\sqrt{1 + \xi_2 (1 + 6 \xi_2) \phi^2}}{1 + \xi_2 \phi^2},
\\[0.6ex]
\chi
&= 
\sqrt{\frac{1 + 6\xi_2}{\xi_2}}
\arcsinh
\left[
\sqrt{\xi_2 \paren{1 + 6\xi_2}} \phi
\right]
-
\sqrt{6}
\arctanh
\left[
\frac{\sqrt{6}\xi_2 \phi}{\sqrt{1 + \xi_2 \paren{1 + 6\xi_2} \phi^2}}
\right]
\nonumber \\[0.6ex]
&=
\sqrt{\frac{1 + 6\xi_2}{\xi_2}}
\ln
\left[
\sqrt{\xi_2 \paren{1 + 6\xi_2}} \phi
+
\sqrt{1 + \xi_2 \paren{1 + 6\xi_2} \phi^2}
\right]\nonumber\\
&\quad
-
\frac{\sqrt{6}}{2}
\ln
\left[
\frac{\sqrt{1 + \xi_2 \paren{1 + 6\xi_2} \phi^2} + \sqrt{6}\xi_2 \phi}{\sqrt{1 + \xi_2 \paren{1 + 6\xi_2} \phi^2} - \sqrt{6}\xi_2 \phi}
\right].
\end{align}
Slow-roll parameters and $e$-folding:
\begin{align}
\epsilon
&=
\frac{2(2 + 3 \lambda_{64} \phi^2 + \xi_2 \lambda_{64} \phi^4)^2}
{\phi^2 \left[ 1 + (\xi_2 + 6 \xi_2^2)\phi^2 \right] (1 + \lambda_{64} \phi^2)^2},
\label{eq:app_Potential_Metric_eps}
\\[0.6ex]
\eta
&= 
\frac{
2
\left[
\begin{matrix}
6 + (2 \xi_2 + 24 \xi_2^2 + 15 \lambda_{64}) \phi^2
+ (- 4 \xi_2^2 - 24 \xi_2^3 + 2 \xi_2 \lambda_{64} + 72 \xi_2^2 \lambda_{64}) \phi^4
\\[0.6ex]
+ (9 \xi_2^2 \lambda_{64} + 36 \xi_2^3 \lambda_{64}) \phi^6
+ (2 \xi_2^3 \lambda_{64} + 12 \xi_2^4 \lambda_{64}) \phi^8
\end{matrix}
\right]
}
{\phi^2 \left[ 1 + (\xi_2 + 6 \xi_2^2)\phi^2 \right]^2 (1 + \lambda_{64} \phi^2)},
\label{eq:app_Potential_Metric_eta}
\\[0.6ex]
N
&= 
\int d\phi
~
\frac{\phi \left[ 1 + (\xi_2 + 6 \xi_2^2)\phi^2 \right] (1 + \lambda_{64} \phi^2)}
{2(1 + \xi_2 \phi^2)(2 + 3 \lambda_{64} \phi^2 + \xi_2 \lambda_{64} \phi^4)},
\label{eq:app_Potential_Metric_N}
\end{align}
where we use
\al{
\lambda_{64} \equiv {\lambda_6 / \lambda_4}
}
for abbreviation.
%%

%%%%%%%%%%%%%%%%%%%%%%%%%%%%%%%%%%%%%%%%%%%%%%%%%%
\subsubsection*{Palatini formalism}
%%%%%%%%%%%%%%%%%%%%%%%%%%%%%%%%%%%%%%%%%%%%%%%%%%

%%
$\phi$-$\chi$ relation:
\begin{align}
\frac{d\chi}{d\phi}
&=
\frac{1}{\sqrt{1 + \xi_2 \phi^2}},
\\[0.6ex]
\chi 
&= 
\frac{1}{\sqrt{\xi_2}}
~
\arcsinh
\left[
\sqrt{\xi_2}\phi
\right]
=
\frac{1}{\sqrt{\xi_2}} \ln
\left[
\sqrt{\xi_2}\phi + \sqrt{1 + \xi_2 \phi^2}
\right].
\end{align}
Slow-roll parameters and $e$-folding:
\begin{align}
\epsilon
&=
\frac{2(2 + 3 \lambda_{64} \phi^2 + \xi_2 \lambda_{64} \phi^4)^2}
{\phi^2 (1 + \xi_2 \phi^2)(1 + \lambda_{64} \phi^2)^2},
\label{eq:app_Potential_Palatini_eps}
\\[0.6ex]
\eta
&= 
\frac{
2 \left[ 
6 + (- 4 \xi_2 + 15 \lambda_{64}) \phi^2 + 7 \xi_2 \lambda_{64} \phi^4 + 2 \xi_2^2 \lambda_{64} \phi^6 
\right]
}
{\phi^2 (1 + \xi_2 \phi^2)(1 + \lambda_{64} \phi^2)},
\label{eq:app_Potential_Palatini_eta}
\\[0.6ex]
N
&= 
\int d\phi
~
\frac{\phi (1 + \lambda_{64} \phi^2)}
{2 (2 + 3 \lambda_{64} \phi^2 + \xi_2 \lambda_{64} \phi^4)}.
\label{eq:app_Potential_Palatini_N}
\end{align}
%%

%%%%%%%%%%%%%%%%%%%%%%%%%%%%%%%%%%%%%%%%%%%%%%%%%%
\section{Difference from $f(R)$ Palatini theories}
\setcounter{equation}{0}
\label{app:Referee}
%%%%%%%%%%%%%%%%%%%%%%%%%%%%%%%%%%%%%%%%%%%%%%%%%%

It has been pointed out in Ref.~\cite{Iglesias:2007nv} that $f(R)$ theories in Palatini formalism 
are equivalent to the strong coupling limit of the Blans-Dicke theory.
This can be seen by rewriting the $f(R)$ action as
\begin{align}
S
&=
\int d^4x~\sqrt{-g}
f(R)
~~~~
\to
~~~~
S
=
\int d^4x~
\sqrt{-g}
\left[
f(\chi)
+
f'(\chi) (R - \chi)
\right].
\label{eq:fR}
\end{align}
This action is equivalent to the $w \to 0$ limit of the Brans-Dicke theory
\begin{align}
S
&=
\int d^4x~
\sqrt{-g}
\left[
\Phi R
- \frac{w}{\Phi} g^{\mu \nu} \nabla_\mu \Phi \nabla_\nu \Phi
- V(\Phi) 
\right],
\label{eq:BD}
\end{align}
with the identification 
$f'(\chi) = \Phi$ and $V(\Phi) = \chi f'(\chi) - f(\chi)$%{\color{red} for the Brans-Dicke scalar field $\Phi$ and its potential}
. The point here is that there is no kinetic term in the auxiliary field $\chi$.
In the Palatini formalism, no additional kinetic term of $\chi$ appears when moving to the Einstein frame.
The canonical field becomes $\Phi_{\rm canoncal} \sim \sqrt{w} \ln \Phi$ in that frame and  causes strong coupling problems in the $w \to 0$ limit.
This argument does not directly apply to our setup, since our starting action (\ref{eq:S}) has a kinetic term 
in the original frame.

Another issue mentioned in Ref.~\cite{Iglesias:2007nv} is about picking up a  particular theory from infinitely many possibilities.
In this reference, the relation between the connection and metric is imposed explicitly by a Lagrange multiplier.
If this multiplier is such that the connection becomes the Levi-Civita type
in a frame different from the original one (e.g. the Einstein frame),
a question would arise why a multiplier that gives the Levi-Civita connection in this specific frame 
is chosen despite infinitely many other possibilities.
However, in our setup we regard the connection as a variable which we take variation with.
Then the connection is uniquely  determined by the Levi-Civita form made from the Einstein-frame metric 
as a result of the variation principle, and therefore no such an ambiguity arises.

%%%%%%%%%%%%%%%%%%%%%%%%%%%%%%%%%%%%%%%%%%%%%%%%%%
\bibliography{biblio}
%%%%%%%%%%%%%%%%%%%%%%%%%%%%%%%%%%%%%%%%%%%%%%%%%%

%%%%%%%%%%%%%%%%%%%%%%%%%%%%%%%%%%%%%%%%%%%%%%%%%%
\end{document}